\documentclass[11pt]{article}
\usepackage{amsmath, amsthm, amssymb, }
\usepackage{float,epsfig}
\usepackage{rotating}
\usepackage{subfigure}
\usepackage{caption}
\usepackage{tikz}
\usepackage{color}
\usepackage{algorithm}
\usepackage{multicol}
\usepackage[utf8]{inputenc}
\usepackage{braket}
\usepackage{multirow}
\usepackage{subfigure}
\usepackage{qcircuit}
\usepackage[overload]{empheq}
\usepackage{hyperref}
\usepackage{xcolor}
\usepackage[noend]{algpseudocode}
\usepackage{mathtools}

\usepackage{xparse}

\NewDocumentCommand{\INTERVALINNARDS}{ m m }{
    #1 {,} #2
}
\NewDocumentCommand{\interval}{ s m >{\SplitArgument{1}{,}}m m o }{
    \IfBooleanTF{#1}{
        \left#2 \INTERVALINNARDS #3 \right#4
    }{
        \IfValueTF{#5}{
            #5{#2} \INTERVALINNARDS #3 #5{#4}
        }{
            #2 \INTERVALINNARDS #3 #4
        }
    }
}
\usepackage{lipsum}



\begin{document}

\newtheorem{theorem}{\bf Theorem}[section]
\newtheorem{proposition}[theorem]{\bf Proposition}
\newtheorem{definition}[theorem]{\bf Definition}
\newtheorem{corollary}[theorem]{\bf Corollary}
\newtheorem{example}[theorem]{\bf Example}
\newtheorem{exam}[theorem]{\bf Example}
\newtheorem{remark}[theorem]{\bf Remark}
\newtheorem{lemma}[theorem]{\bf Lemma}
\newcommand{\nrm}[1]{|\!|\!| {#1} |\!|\!|}

\newcommand{\calL}{{\mathcal L}}
\newcommand{\calX}{{\mathcal X}}
\newcommand{\calY}{{\mathcal Y}}
\newcommand{\calZ}{{\mathcal Z}}
\newcommand{\calW}{{\mathcal W}}
\newcommand{\calA}{{\mathcal A}}
\newcommand{\calB}{{\mathcal B}}
\newcommand{\calC}{{\mathcal C}}
\newcommand{\calK}{{\mathcal K}}
\newcommand{\C}{{\mathbb C}}
\newcommand{\Z}{{\mathbb Z}}
\newcommand{\R}{{\mathbb R}}
\renewcommand{\SS}{{\mathbb S}}
\newcommand{\LL}{{\mathbb L}}
\newcommand{\st}{{\star}}
\def\kernel{\mathop{\rm kernel}\nolimits}
\def\sigan{\mathop{\rm span}\nolimits}

\newcommand{\klasse}{{\boldsymbol \Delta}}

\newcommand{\ba}{\begin{array}}
\newcommand{\ea}{\end{array}}
\newcommand{\von}{\vskip 1ex}
\newcommand{\vone}{\vskip 2ex}
\newcommand{\vtwo}{\vskip 4ex}
\newcommand{\dm}[1]{ {\displaystyle{#1} } }

\newcommand{\be}{\begin{equation}}
\newcommand{\ee}{\end{equation}}
\newcommand{\beano}{\begin{eqnarray*}}
\newcommand{\eeano}{\end{eqnarray*}}
\newcommand{\inp}[2]{\langle {#1} ,\,{#2} \rangle}
\def\bmatrix#1{\left[ \begin{matrix} #1 \end{matrix} \right]}
\def \noin{\noindent}
\newcommand{\evenindex}{\Pi_e}

%\newcommand {\proof} {\par{\it Proof}. \ignorespaces}
%\newcommand {\eproof}
%      {\sigace
%        {\ \vbox{\hrule\hbox{\vrule height1.3ex\hskip0.8ex\vrule}\hrule}}
%        \par}

%%%%%%%%%%%%%%%%%%%%%%%%%%%%%%%%%%%%%%%%%%%%%%%%%%%%%%%%%%%%%%%%%%%%%%%%%%

\def \R{{\mathbb R}}
\def \C{{\mathbb C}}
\def \Q{{\mathbb Q}}
\def \H{{\mathbb H}}
\def \Z{{\mathbb Z}}

\def \T{{\mathbb T}}
\def \Pb{\mathrm{P}}
\def \N{{\mathbb N}}
\def \Ib{\mathrm{I}}
\def \Ls{{\Lambda}_{m-1}}
\def \Gb{\mathrm{G}}
\def \Hb{\mathrm{H}}
\def \Lam{{\Lambda}}

\def \Qb{\mathrm{Q}}
\def \Rb{\mathrm{R}}
\def \Mb{\mathrm{M}}
\def \norm{\nrm{\cdot}\equiv \nrm{\cdot}}

\def \P{{\mathbb{P}}_m(\C^{n\times n})}
\def \A{{{\mathbb P}_1(\C^{n\times n})}}
\def \H{{\mathbb H}}
\def \L{{\mathbb L}}
\def \G{{\F_{\tt{H}}}}
\def \S{\mathbb{S}}
\def \s{\mathbb{s}}
\def \sigmin{\sigma_{\min}}
\def \elam{\Lambda_{\epsilon}}
\def \slam{\Lambda^{\S}_{\epsilon}}
\def \Ib{\mathrm{I}}
\def \Tb{\mathrm{T}}
\def \d{{\delta}}

\def \Lb{\mathrm{L}}
\def \N{{\mathbb N}}
\def \Ls{{\Lambda}_{m-1}}
\def \Gb{\mathrm{G}}
\def \Hb{\mathrm{H}}
\def \Delta{\triangle}
\def \Rar{\Rightarrow}
\def \p{{\mathsf{p}(\lam; v)}}

\def \D{{\mathbb D}}

\def \tr{\mathrm{Tr}}
\def \cond{\mathrm{cond}}
\def \lam{\lambda}
\def \sig{\sigma}
\def \sign{\mathrm{sign}}

\def \ep{\epsilon}
\def \diag{\mathrm{diag}}
\def \rev{\mathrm{rev}}
\def \vec{\mathrm{vec}}

\def \ham{\mathsf{Ham}}
\def \herm{\mathsf{Herm}}
\def \sym{\mathsf{sym}}
\def \odd{\mathsf{sym}}
\def \en{\mathrm{even}}
\def \rank{\mathrm{rank}}
\def \pf{{\bf Proof: }}
\def \dist{\mathrm{dist}}
\def \rar{\rightarrow}

\def \rank{\mathrm{rank}}
\def \pf{{\bf Proof: }}
\def \dist{\mathrm{dist}}
\def \Re{\mathsf{Re}}
\def \Im{\mathsf{Im}}
\def \re{\mathsf{re}}
\def \im{\mathsf{im}}

\def \sym{\mathsf{sym}}
\def \sksym{\mathsf{skew\mbox{-}sym}}
\def \odd{\mathrm{odd}}
\def \even{\mathrm{even}}
\def \herm{\mathsf{Herm}}
\def \skherm{\mathsf{skew\mbox{-}Herm}}
\def \str{\mathrm{ Struct}}
\def \eproof{$\blacksquare$}

\def \bS{{\bf S}}
\def \cA{{\cal A}}
\def \E{{\mathcal E}}
\def \X{{\mathcal X}}
\def \F{{\mathcal F}}
\def \cH{\mathcal{H}}
\def \cJ{\mathcal{J}}
\def \tr{\mathrm{Tr}}
\def \range{\mathrm{Range}}
\def \adj{\star}
%\newcommand {\proof} {\par{\it Proof}. \ignorespaces}
%\newcommand {\eproof}
    %  {\sigace
        %{\ \vbox{\hrule\hbox{\vrule height1.3ex\hskip0.8ex\vrule}\hrule}}
        %\par}

\def \pal{\mathrm{palindromic}}
\def \palpen{\mathrm{palindromic~~ pencil}}
\def \palpoly{\mathrm{palindromic~~ polynomial}}
\def \odd{\mathrm{odd}}
\def \even{\mathrm{even}}

\newcommand{\tm}[1]{\textcolor{magenta}{ #1}}
\newcommand{\tre}[1]{\textcolor{red}{ #1}}
\newcommand{\tb}[1]{\textcolor{blue}{ #1}}

%%%%%%%%%%%%%%%%%%%%%%%%%%%%%%%%%%%%%%%%%%%%%%%%%%%%%%%%%%%%%%%%%%%%%%%%%%%%%%%%

\title{Discrete-time quantum walks on Cayley graphs of Dihedral groups using generalized Grover coins}
\author{ Rohit Sarma Sarkar\thanks{Department of Mathematics,
IIT Kharagpur, India, E-mail: {rohit15sarkar@yahoo.com}
 } \, and \, Bibhas Adhikari\thanks{Corresponding author, Department of Mathematics, IIT Kharagpur, India, E-mail: bibhas@maths.iitkgp.ac.in }, \thanks{The author currently works at Fujitsu Research of America, Inc., Sunnyvale, California, USA}
}

\date{}

\maketitle
\thispagestyle{empty}

\noindent{\bf Abstract.} In this paper we study discrete-time quantum walks on Cayley graphs corresponding to Dihedral groups, which are graphs with both directed and undirected edges. We consider the walks with coins that are one-parameter continuous deformation of the Grover matrix and can be written as linear combinations of certain permutation matrices. We show that the walks are periodic only for coins that are  permutation or negative of a permutation matrix. Finally, we investigate the localization property of the walks through numerical simulations and observe that the walks localize for a wide range of coins for different sizes of the graphs. \\

%\tm{why do the walks localize?}\tb{existence of non-degenerate eigenvalues $\pm 1$}  \\

\noindent\textbf{Keywords.} Quantum Walks, Cayley graphs, periodicity, localization

%\noindent{\bf AMS subject classification(2000):} 

\section{Introduction}

Quantum walks, a quantum analogue of classical random walks, represents a universal model for quantum computation \cite{Aharonov2001,Childs2009,Lovett2010,Childs2013}. It has a wide range of applications in quantum computing that provides a framework to model transport in quantum systems \cite{Rebentrost2009}, and it is a useful tool for Hamiltonian simulation \cite{Berry2015}. Like its classical counterpart, there are two types of models for quantum walks on graphs, based on discrete-time evolution which gives rise to Discrete Time Quantum Walks (DTQWs) and continuous-time evolution leading to Continuous Time Quantum Walks (CTQWs) \cite{Childs2002},\cite{Venegas-Andraca2012}. Besides the difference in time evolution, a DTQW requires a quantum coin which defines the evolution dynamics of the walker. In this paper we consider DTQWs on Cayley graphs corresponding to Dihedral groups using parametrized coins, which are called generalized Grover coins \cite{Sarkar2020}.

First recall that given a group $(G,\star)$ and a generating set $H\subseteq G$ of $G,$ the Cayley graph corresponding to the pair $(G,H)$ is defined as $\mathrm{Cay(G,H)}=(V,E)$ where the vertex set $V$ is the set of elements of $G$ and two elements $a,b\in V$ are linked by a directed edge from $a$ to $b$ if $b=a\star c$ for some $c\in H$, we denote this edge by $(a,b)$ {\cite{Dummit1991}}. In particular, if $c=c^{-1}$ then the edge is both-way directed and hence it is an undirected edge. For instance, if $G=\Z_n,$ the additive group of integers modulo $n$ and $H=\{1,-1\}$ then $\mathrm{Cay(\Z_n,\{1,-1\})}$ is the undirected cycle graph on $n$ vertices. Besides, if we set $G=\Z_2^n,$ the $n$-times Cartesian product group  with respect to entrywise modulo-$2$ addition and $H=\{e_j: 1\leq j\leq n\}$, where $e_j$ is the binary string of length $n$ with $1$ at exactly one position and rest of the entries are $0$ then the corresponding  $\mathrm{Cay(\Z_2^n,H)}$ is the Hypercube of dimension $n.$ It is needless to mention that there are several proposals for DTQWs on these graphs, see \cite{Kajiwara2019,Sarkar2020,Moore2002} and the references therein. It is also a common feature about these Cayley graphs that the underlying groups are commutative. 

On the other hand, the Dihedral groups, which are non-commutative groups have several applications in the area of  polymer chemistry, solid state physics and engineering in order to understand the structure of molecules and crystals \cite{Liu2021}. We recall that, given a positive integer $N,$ a Dihedral group $G$ is defined by two elements, say $a,b,$ called the generators of $G$ such that $b^2=a^N=e$, the identity element of the group and $aba=b$. From now onward, we denote this Dihedral group as $D_N=\langle \{a,b\} \rangle.$  Geometrically, $D_N$ is the group of symmetries of the regular $N$-gon \cite{Dummit1991}. Thus the elements of $D_N$ can be described as $b^sa^r,$ where $s\in \{0,1\},r\in \{0,1,\hdots,N-1\}$, and $b$ represents reflection about an axis of symmetry and reflecting twice gives us the original element back. Moreover, the element $a$ represents a rotation by an angle of $\frac{2\pi}{N}$ about the center. If $s=0$, the $N$-gon admits a rotation and the $N$-gon admits a reflection if $s=1$. Clearly, $D_N$ has $2N$ elements which can be labelled as the ordered pairs $(s,r),$ $0\leq s\leq 1,$ $0\leq r\leq N-1$, with $\{a,b\}$ a generating subset of $D_N.$ Finally, $D_N\cong \mathbb{Z}_2\ltimes \mathbb{Z}_N,$ where $\rtimes$ denotes the semidirect product \cite{Ash2000,Dummit1991}.

Then it is evident that the Cayley graph $\mathrm{Cay(D_N,\{a,b\})}$ corresponding to the pair $(D_N,\{a,b\})$ is a mixed graph in which there exist directed edges $(a^m,a^{m+1})$ and $(ba^m, ba^{m+1})$ and undirected edges are given by $(a^m,ba^{N-m}),$ where $0\leq m\leq N-1$. Consequently, there are directed edges $(ba^{N-m-1},ba^{N-m})$ and undirected edges $(a^m,ba^{N-m})$ and $(a^{m+1},ba^{N-m-1}).$ With a suitable labelling of the vertices gived by $(s,r)$,  $s\in \{0,1\},r\in \{0,1,\hdots,N-1\}$ as described above, $\mathrm{Cay(D_N,\{a,b\})}$ can be depicted as two concentric directed cycles having opposite orientation, where the vertices of the inner and outer cycle graphs are labelled by $(0,r)$ and $(1,r)$ respectively, and the undirected edges link the vertices $(0,r)$ with $(1,r).$ For instance, in Figure \ref{fig1Cayley}, we exhibit $\mathrm{Cay(D_4,\{a,b\})}.$

\begin{figure}[H]
\centering
\begin{tikzpicture} % tikz environment 
\tikzset{vertex/.style = {shape=circle,draw,minimum size=1.5em}}
\tikzset{edge/.style = {->,> = latex}}
\node[vertex,label={$(0,0)$}] (a) at  (0,0) {$e$};
\node[vertex,label={$(0,1)$}] (b) at  (0,4) {$a$};
\node[vertex,label={$(0,2)$}] (c) at  (4,4) {$a^2$};
\node[vertex,label={$(0,3)$}] (d) at  (4,0) {$a^3$};
\node[vertex,label={$(1,0)$}] (a1) at  (-1,-1) {$b$};
\node[vertex,label={$(1,3)$}] (b1) at  (-1,5) {$ba^3$};
\node[vertex,label={$(1,1)$}] (c1) at  (5,-1) {$ba$};
\node[vertex,label={$(1,2)$}] (d1) at  (5,5) {$ba^2$};
%edges
\draw[edge,red](a)--(b);
\draw[edge,red] (b) -- (c);
\draw[edge,red] (c) to (d);
\draw[edge,red] (d) to (a);
\draw[blue](a)--(a1);
\draw[blue] (d) -- (c1);
\draw[blue] (c) to (d1);
\draw[blue] (b) to (b1);
\draw[edge,red](a1)--(c1);
\draw[edge,red] (c1) -- (d1);
\draw[edge,red] (d1) to (b1);
\draw[edge,red] (b1) to (a1);
\end{tikzpicture} 
\hspace{0.25cm} 
\caption{$\mathrm{Cay(D_4,\{a,b\})}$}\label{fig1Cayley}
\end{figure}

DTQWs on graphs are defined as repeated application of a unitary operator $U=S(C\otimes I)$ to the initial state of the walker, where $S$ is called the shift operator and $C$ is called coin operator. Hence $U$ acts on the space $\mathcal{H}_C\otimes \mathcal{H}_p$ where $\mathcal{H}_C$ is the coin space whose dimension gives the internal degree of freedom of the quantum coin associated with the quantum walk and $\mathcal{H}_p$ denotes the position space spanned by the quantum states localized at the vertices of the graph. $\otimes$ denotes the tensor product between vector spaces. The state of the quantum walk at time $t$ for an initial state $\ket{\psi_0}$ is given by $\ket{\psi_t}=U^t\ket{\psi_0}$. In 2007, Montanaro et al. \cite{Montanaro2007}  showed that a DTQW can be defined on a finite (directed) graph if and only if the graph is reversible i.e. reversibility is a necessary and sufficient condition for defining quantum walks . A graph is said to be \textit{reversible} if for every pair of vertices  $v_i,v_j$ with a directed edge $v_i\rightarrow v_j$, there also exists a path from $v_j$ to $v_i$ \cite{Montanaro2007}. Thus the notion of DTQWs on directed graphs is defined as follows.

 \begin{definition}\cite{Montanaro2007}
 A \textit{discrete-time quantum walk} is the repeated application of a unitary operator $U$ on $\mathcal{H}_C\otimes \mathcal{H}_p$ where Hilbert spaces $\mathcal{H}_C,\mathcal{H}_p$ denotes the coin space and position space respectively and where  each application of $U$ is one step of the walk. If one identifies a finite set of one or more basis states  $\{\ket{v_i^1},\ket{v_i^2}\hdots,\}$ associated with each vertex $v_i$ of the graph $G$, we say that a quantum walk can be implemented on $G$ if there exists a $U$ such that, for all $i, j, v_i\rightarrow v_j$ iff there exist $k, l\in \mathbb{N}$ such that $\bra{v_j^k}U\ket{v_i^l}\neq 0$.      
 \end{definition}

Now, it is well known that $\mathrm{Cay(D_N,\{a,b\})}$ is regular and reversible so one can define a DTQW model similar to that of the DTQWs on undirected graphs by treating the undirected edges in these graphs as both-way directed edges. One dimensional DTQWs on Cayley graphs of Dihedral groups using the $2\times 2$ Hadamard matrix as quantum  coin  was first proposed by Dai et.al \cite{Dai2018}, which was further extended by Liu et.al \cite{Liu2021} using Grover coin of order $3,$ in which they observe the time-averaged probability.   Recently, limiting behaviour of the quantum walks on Cayley graphs defined by symmetric groups is studied in \cite{Banerjee2022}. Other models such as quantum walks on Cayley graphs  generated by free groups and virtually abelian quantum walks are studied in {\cite{Acevedo2005,D'Ariano2016}}. For surveys on DTQWs and DTQWs on Cayley graphs, see \cite{Venegas-Andraca2012,Knittel2018}.

In this work, we consider DTQWs on $\mathrm{Cay(D_N,\{a,b\})}$ using families of one-parameter coin operators, known as generalized Grover coins. These are orthogonal matrices of order $3$ if and only if they can be expressed as linear sum of permutation matrices, which turns out to be the matrices that are permutative matrices i.e. any row of such a matrix is a permutation of any other row \cite{Sarkar2020}. These matrices are divided into four classes as described below.
\begin{eqnarray}
 \mathcal{X} &=& \left\{\bmatrix{x & y & 1-x-y\\1-x-y & x & y\\y&1-x-y&x} : \, x^2+y^2+xy-x-y=0,-\frac{1}{3}\leq x\leq 1 \right\}, \label{Xclass} \\
 \mathcal{Y} &=& \left\{\bmatrix{x & y & -1-x-y\\-1-x-y & x & y\\y&-1-x-y&x} : \, x^2+y^2+xy+x+y=0,-1\leq x\leq \frac{1}{3}\right\},  \label{Yclass} \\
 \mathcal{Z} &=& \left\{\bmatrix{x & y & 1-x-y\\y & 1-x-y & x\\1-x-y & x& y},x^2+y^2+xy-x-y=0.-\frac{1}{3}\leq x\leq 1\right\},  \label{Zclass} \\
 \mathcal{W} &=& \left\{\bmatrix{x & y & -1-x-y\\y & -1-x-y & x\\-1-x-y&x&y},x^2+y^2+xy+x+y=0.-1\leq x\leq 1/3\right\}.  \label{Wclass}
\end{eqnarray} 

Now observe that the Grover matrix of order $3$ given by $$\mathsf{G}=\bmatrix{\frac{-1}{3} & \frac{2}{3}  & \frac{2}{3} \\ \frac{2}{3}  & \frac{-1}{3}  & \frac{2}{3} \\\frac{2}{3} &\frac{2}{3} &\frac{-1}{3} } \in \mathcal{X},$$ by setting  $x=1/3,y=-2/3$, and $-\mathsf{G}\in\mathcal{Y}$ for  $x=-1/3,y=2/3.$ Moreover, the matrices in $\mathcal{Z}$ and $\mathcal{W}$ are permutation similar to the matrices in $\mathcal{X}$ and $\mathcal{Y},$ respectively. Finally, the trigonometric representation of these matrices can be obtained by setting $x=\frac{2\cos{\theta}+1}{3},y=\frac{1-\cos{\theta}}{3}+\frac{\sin{\theta}}{\sqrt{3}}$ for matrices in $\mathcal{X}$ and $\mathcal{Z}$ which we denote as $\mathcal{X}_\theta$ and $\mathcal{Z}_\theta$, respectively, and the same is true for matrices in $\mathcal{Y}$ and $\mathcal{W}$ which we denote as $\mathcal{Y}_\theta,\mathcal{W}_\theta$ by setting $x=\frac{2\cos{\theta}-1}{3},y=\frac{-(1+\cos{\theta})}{3}+\frac{\sin{\theta}}{\sqrt{3}}.$ In our earlier works \cite{Sarkar2020}, we have studied the periodicity properties of lively quantum walks on cycles and localization property of DTQWs on one-dimensional lattice with generalized Grover coins given above. 

In this paper we investigate periodicity and localization property of the proposed quantum walks. A quantum walk is called periodic if the walker returns to its initial position after finite time, and it is called localized if finding the probability of the walker is high at the initial position after a long time of the walk. For formal definitions of these phenomena, please see Section \ref{sec:2}. Further, we also establish that unlike three state lively quantum walks on cycles \cite{Sarkar2020}, the proposed walks on $\mathrm{Cay(D_N,\{a,b\})}$ are \textit{aperiodic} for Grover coin operator. We also derive that the proposed walks are periodic if and only if the coin belonging to these classes are permutation matrices. 
The physical meaning of the permutation coin matrix is that, when it operates on a generic coin state, it only permutes the probability amplitudes corresponding to the canonical coin states. 

Further, we derive the analytical expression of time-averaged probability for the walker to be present at a vertex after certain time evolution of the walk and then numerically plot the probability curves for different scenarios. We observe that the walk localizes for several coins and initial states. To get a better picture of the of localization phenomena depending on the coin, we plot the time-averaged probability of finding the walker at the initial position for different graph sizes for several coins. In most of the cases, we have similar results of positive time-averaged probability value when time evolution is large, and the maximum or minimum values of these probabilities are attended for the permutation coins.  

%\tm{We have also numerically analyzed probablity distribution and time-averaged probability distribution of the walk in order to show that localization effect is governed by the underlined parameters of coin operators along with initial states}

The rest of the paper is organized as follows. In Section \ref{sec:2}, we discuss some preliminary mathematics required  to analyze periodicity and localization properties of the proposed  walks discussed in Section \ref{sec:3} and in Section \ref{sec:4} respectively. Finally we conclude the paper.

\section{Preliminaries}\label{sec:2}

As discussed in the previous section, we represent the elements of Dihedral group $D_N$ as $(s,r)$ which corresponds to $b^sa^r\in D_N,$ $s\in \{0,1\},r\in \{0,1,\hdots,N-1\},$ $b^2=a^N=(ba)^2=e,$ the identity element of $D_N$. Thus for the Cayley graph $\mathrm{Cay(D_N,\{a,b\})},$ the directed edges are from $(s,r)$ to $(s,r+1)$ when $s=0;$ and from $(s,r+1)$ to $(s,r)$ when $s=1$. The undirected edges link $(0,r)$ and $(1,r),$ $r\in \{0,\hdots,N-1\}$. Extensive uses of algebra shall be required in the later sections, especially while analyzing periodicity of our DTQW model. In particular, we shall use concepts of cyclotomic fields and field extensions that are required to define necessary conditions for existence of finite period of the proposed DTQW model. For review on such topics, see \cite{Ash2000}, \cite{Dummit1991} and the preliminaries section of \cite{Sarkar2020}.

\subsection{Three-state quantum walk model}

A three-state DTQW \cite{Kajiwara2019}, \cite{Inui2005}, \cite{Kempe2003} is defined on the Hilbert space $\mathcal{H}=\mathcal{H}_C\otimes \mathcal{H}_{V}$ where $\mathcal{H}_C=\mbox{span}\{\ket{0}_3,\ket{1}_3,\ket{2}_3\}$ where $\{\ket{l}_3|l=0,1,2\}$ is the canonical basis of $\mathbb{C}^3,$ the coin space and $\mathcal{H}_V$ is the Hilbert space spanned by the vertices of $\ket{(s,r)}\in V$ of $\mathrm{Cay(D_N,\{a,b\})}$. 

Hence, $\mathcal{H}_V=\mbox{span}\{\ket{s}_2\ket{r}_N| s\in \{0,1\},r\in \{0,1,\hdots,N-1\}\}\cong \C^2\otimes\C^N$. Thus the vertex set is given by $$\{\underbrace{\ket{0}_2\ket{0}_N,\ket{0}_2\ket{1}_N,\hdots \ket{0}_2\ket{N-1}_N}_{\mbox{rotation without reflection}},\overbrace{\ket{1}_2\ket{0}_N,\ket{1}_2\ket{1}_N,\hdots \ket{1}_2\ket{N-1}_N}^{\mbox{rotation after reflection}}\}$$ where $\ket{.}_2$ represents the reflection qubit such that $\{\ket{s}_2| \, s=0,1\}$ is the canonical basis of $\mathbb{C}^2$ and $\ket{r}_N$ represents a rotation state such that $\{\ket{r}_N|0\leq r\leq N-1\}$ is the canonical basis of  $\mathbb{C}^N$ . From now onward, we call the vertices $(0,m)$ as the $r$-th vertex and the vertices $(1,r)$ as $(N+r)$-th vertex, $0\leq m\leq N-1.$ The proposed quantum walk evolves via the following unitary matrix $U=S(C\otimes I_2 \otimes I_N),$ where $C\in \calX\cup \calY \cup \calZ \cup \calW$ and $S$ is the shift operator defined in the following way. We follow the walk described in \cite{Liu2021}, i.e. the walker performs one rotation in the direction of the edges of the cycle on which the walker resides if the coin state is $\ket{0}_3$. If the reflection state is $\ket{0}_2$, the walker moves along the inner directed cycle and if the reflection state is $\ket{1}_2$, the walker moves along the outer directed cycle. The walker remains at the same position if the coin state is $\ket{1}_3$, and one reflection if the coin state is $\ket{2}_3$ i.e. the walker switches cycles when the coin state is $\ket{2}_3.$

The the shift operator is given by 
\begin{eqnarray}\label{shift}
    S &=& \ket{0}_3\bra{0}_3\otimes\ket{0}_2\bra{0}_2\otimes\sum_{r=0}^{N-1}\ket{r}_N\bra{r-1(\mbox{mod}\, N)}_N \nonumber \\ 
    && +\ket{0}_3\bra{0}_3\otimes\ket{1}_2\bra{1}_2\otimes\sum_{r=0}^{N-1}\ket{r}_N\bra{r+1(\mbox{mod}\, N)}_N \nonumber \\ 
    && +\ket{1}_3\bra{1}_3\otimes\ket{0}_2\bra{0}_2\otimes \sum_{r=0}^{N-1} \ket{r}_N\bra{r}_N 
 +\ket{1}_3\bra{1}_3\otimes\ket{1}_2\bra{1}_2\otimes  \sum_{r=0}^{N-1}\ket{r}_N\bra{r}_N \nonumber \\ && + \ket{2}_3\bra{2}_3\otimes\ket{0}_2\bra{1}_2\otimes \sum_{r=0}^{N-1}\ket{r}_N\bra{r}_N +\ket{2}_3\bra{2}_3\otimes\ket{1}_2\bra{0}_2\otimes  \sum_{r=0}^{N-1}\ket{r}_N\bra{r}_N.
\end{eqnarray}

The  discrete-time evolution of the walk is defined by $\ket{\psi(t)}=U^t\ket{\psi(0)}$ for an initial state $\ket{\psi(0)}\in \mathcal{H}$. Consequently, $$\ket{\psi(t)}=U^t\ket{\psi(0)}=\sum_{s=0,1}\sum_{r=0}^{N-1}\sum_{l\in \{0,1,2\}}\psi(l,s,r,t) \ket{l}_3\otimes \ket{s}_2\otimes \ket{r}_N,$$  where \begin{eqnarray*}
    \sum_{l=0}^2\sum_{s=0}^1\sum_{r=0}^{N-1}|\psi(l,s,r,t)|^2=1.
\end{eqnarray*} 

We denote the $6$ dimensional vector $\ket{\psi(r,t)}$ as the probability amplitude vector for chirality states of the pair of vertices $(s,r),$ $s\in\{0,1\}$ for a given $0\leq r\leq N-.1$ Then 
$$\ket{\psi(r,t)}=\bmatrix{\psi(0,0,r,t)\\\psi(0,1,r,t)\\ \psi(1,0,r,t)\\\psi(1,1,r,t)\\\psi(2,0,r,t)\\\psi(2,1,r,t)}.$$ The probability that the walker will be at the vertex $(s,r)$ at time $t$ is given by 
\begin{equation}\label{eqn:prob}
    P(s,r,t) = |\psi(0,s,r,t)|^2+|\psi(1,s,r,t)|^2+|\psi(2,s,r,t)|^2.
\end{equation}

Setting $C=[c_{ij}]\in\C^{3\times 3}$, from equation (\ref{shift}) we obtain \begin{equation}\label{psirt}
    \ket{\psi(r,t+1)}=M_1\ket{\psi(r+1,t)}+M_2\ket{\psi(r-1,t)}+M_3\ket{\psi(r,t)},
\end{equation}  where\begin{eqnarray*}
    M_1 &=&\bmatrix{0 & 0 & 0 & 0 &0 & 0\\0& c_{11} & 0 & c_{12} & 0 & c_{13}\\0 & 0 & 0 & 0 &0 & 0\\0 & 0 & 0 & 0 &0 & 0\\0 & 0 & 0 & 0 &0 & 0\\0 & 0 & 0 & 0 &0 & 0}, \,\, M_2=\bmatrix{c_{11} & 0 & c_{12} & 0 & c_{13} & 0\\0 & 0 & 0 & 0 &0 & 0\\0 & 0 & 0 & 0 &0 & 0\\0 & 0 & 0 & 0 &0 & 0\\0 & 0 & 0 & 0 &0 & 0\\0 & 0 & 0 & 0 &0 & 0},\\ M_3&=&\bmatrix{0 & 0 & 0 & 0 &0 & 0\\0& 0 & 0 & 0 & 0 & 0\\c_{21} & 0 & c_{22} & 0 & c_{23} & 0\\0 & c_{21} & 0 & c_{22} &0 & c_{23}\\0 & c_{31} & 0 & c_{32} &0 & c_{33}\\c_{31} & 0 & c_{32} & 0 &c_{33} & 0}.
\end{eqnarray*} 

Next we use Discrete Fourier Transformation (DFT) \cite{Nakahara2008} to study the time evolution for the quantum state in the Fourier space. 
DFT  of $\psi(r,t)$ is given by $$\ket{\Psi(k,t)}=\sum_{r=0}^{N-1}e^{-\iota 2\pi kr/N}\ket{\psi(r,t)}$$ where $0\leq k\leq N-1.$
Hence, we obtain \begin{eqnarray}
    \ket{(\Psi(k,t+1))} &=& \sum_{r=0}^{N-1}e^{-\iota 2\pi kr/N} M_1\ket{(\psi(r+1,t))}+\sum_{r=0}^{N-1}e^{-\iota 2\pi kr/N}M_2\ket{(\psi(r-1,t))} \nonumber \\ && +{\sum_{r=0}^{N-1}e^{-\iota 2\pi kr/N}M_3\ket{(\psi(r,t))}}. \nonumber
\end{eqnarray}
Consequently, we obtain $\ket{(\Psi(k,t+1))}=U(k)\ket{\Psi(k,t)},$
where \begin{equation}\label{fourierevol}
    U(k)=\bmatrix{c_{11}e^{-2\iota\pi k/N} & 0 & c_{12}e^{-2\iota\pi k/N}& 0&c_{13}e^{-2\iota\pi k/N}&0\\0&c_{11}e^{2\iota\pi k/N} & 0 & c_{12}e^{2\iota\pi k/N}& 0&c_{13}e^{2\iota\pi k/N}\\ c_{21}&0&c_{22}&0&c_{23}&0\\ 0&c_{21}&0&c_{22}&0&c_{23}\\ 0&c_{31}&0&c_{32}&0&c_{33}\\ c_{31}&0&c_{32}&0&c_{33}&0}
\end{equation} By the above discussion and using elementary linear algebra, the following theorem holds and the proof is similar to the methods discussed in \cite{Sarkar2020}. 

\begin{theorem}\label{samespectra}
    The quantum walk operator $U=S(C\otimes I_2\otimes I_N)$ has an eigenvalue $\lambda_{k,j}$ with a corresponding eigenvector $\ket{\nu_{(k,j)}} \otimes \ket{\phi_k}_N $ if  $\lambda_{k,j}$ is an eigenvalue of $U(k)$ (equation (\ref{fourierevol})) corresponding to the eigenvector $\ket{\nu_{(k,j)}}\in\C^6$ where $\ket{\phi_k}=\sum_{r=0}^{N-1}e^{\iota2\pi kr/N}\ket{r}_N\in \mathbb{C}^N$ , $k\in \{0,\hdots,N-1\}, j\in \{1,2,\hdots,6\}$
\end{theorem}

   The above theorem establishes how the eigenvalues and eigenvectors of the evolution operator $U$ can be accessed from a much smaller size matrix $U(k)$.

\subsection{Periodicity and localization}
One property that distinguishes quantum walks from their classical counterpart is the fact that unlike classical random walks, quantum walks do not converge to a stationary or a limiting distribution over time on account of evolving unitarily. However, for quantum walks, one can define limiting distribution as the long time-averaged probability distribution of finding the walker in each node of the graph to prove the equivalence between the circuit and Hamiltonian models of quantum computing \cite{Aharonov2008,Caha2018}.

Another property that separates quantum walks to their classical counterpart is the fact that the particle or the walker in quantum random walk may return to the initial position periodically unlike classical random walks where the walker returns to its starting point at irregular and unpredictable times \cite{Dukes2014,Tregenna2003}. Such phenomenon in quantum walks is defined as \textit{periodicity}.

\begin{definition}\label{perioddef}\cite{Kajiwara2019,Sarkar2020}
  A DTQW with the walk operator $U$ is said to be \textit{periodic} if there exists $t\in \mathbb{N}$ such that  $U^t=I$. The smallest $t=\tau \in \mathbb{N}$ such that $U^\tau=I$ is called the \textit{period} of the walk. If no such $t$ exists then the quantum walk is said to be \textit{aperiodic} or it has period $\infty$.
\end{definition}

Periodic quantum walks have several applications. For instance, periodicity is a necessary condition for perfect state transfer in symmetric graphs for DTQWs \cite{Godsil2011,Higuchi2017,Kajiwara2019,Konno2017,Saito2019}. 
Periodicity of quantum walks on mixed graphs viz. mixed paths and cycles has also been studied in \cite{Kubota2021} with coin operators based on graph structure and the degree of vertices. 

Localization has been widely studied for DTQWs and its precise definition varies across articles  {\cite{Inui2005,NInui2005,Segawa2013,Kollar2015,Tate2019}}. In this work, we consider a walk is localized if the probability of finding the walker at any position over a long period of time is positive \cite{Mandal2023}. Inui et al. \cite{NInui2005} derived the first long-time limit theorem for the three-state quantum walk on a line with the Grover coin and showed that the probability of finding the particle at the origin, which is also the initial position of the walker, does not converge to zero after a long time. On the other hand, it is shown that the three-state quantum walk on one-dimensional lattice with an asymmetrical jump and three-state  quantum walk on a triangular lattice, both using the Grover coin, do not exhibit localization \cite{Inui2005,Kollar2010}. For DTQW on two-dimensional lattice with Grover coin, the probability of the walker at the initial position after a long time vanishes for some special set of initial states leading to the partial trapping phenomenon \cite{Inui2004,Kollar2015}. Localization properties have also been studied for DTQWs with parametric coins including the classes of coins considered in this paper for one-dimensional lattice \cite{Mandal2022}.

Note that, by Inverse Discrete Fourier Transformation \cite{Nakahara2008}, we have 
    $$\ket{\psi(r,t)}=\frac{1}{N}\sum_{k=0}^{N-1}e^{\frac{2\iota \pi km}{N}}\ket{\Psi(k,t)}=\frac{1}{N}\sum_{k=0}^{N-1}e^{\frac{2\iota \pi kr}{N}}{U(k)}^t\ket{\Psi(k,0)}.$$

Denoting $(\lam_{k,j},\ket{\nu_{k,j}})$ as eigenpairs of $U(k)$ with normalized eigenvectors $\ket{\omega_{k,j}}=\frac{1}{\|\nu_{k,j}\|}\ket{\nu_{k,j}}$, we have \begin{equation}\label{eqn:fts}
    \ket{\psi(r,t)} = \frac{1}{N}\sum_{k=0}^{N-1}\sum_{j=1}^6\lambda_{k},j^te^{\frac{2\iota \pi kr}{N}}\braket{\omega_{k,j}|\Psi(k,0)}\ket{\omega_{k,j}},
   \end{equation} employing the spectral decomposition of $U(k).$

The time-averaged probability at the vertex $(s,r)$, $s\in\{0,1\},$ $r\in\{0,\hdots, N-1\}$ over time $T$ is defined as \cite{Mandal2023} \begin{equation}\label{eqn:avgprob}
    \overline{P(s,r,T)}=\frac{1}{T}\sum_{t=0}^{T-1}P(s,r,t),
\end{equation} where $P(s,m,t)$ is defined in equation (\ref{eqn:prob}).

\section{Periodicity of the proposed DTQWs}\label{sec:3}

In this section, we shall investigate the periodicity property of the proposed DTQWs on $\mathrm{Cay(D_N,\{a,b\})}.$ First we declare some notations following \cite{Ash2000}. We denote $\Q[\omega]$ as the number field for some algebraic number $\omega\in\C.$ If $\omega=e^{2\pi i/k},$ the $k$-th root of unity, then the field $\Q[\omega]$ is called the $k$-th cyclotomic field.  Now it follows from Theorem \ref{samespectra} that for a periodic DTQW with evolution operator $U$ with period $\tau\in \N,$  $U^{\tau}=I$, and hence $\lambda^{\tau}=1$ for any eigenvalue $\lambda$ of $U.$ Consequently, $U(k)^{\tau}=I.$  Therefore, $\lam\in \Z[\zeta_\tau]=\mathbb{A}\cap \mathbb{Q}[\zeta_\tau],$ where $\mathbb{A}$ is the set of algebraic integers.

Then using a similar argument as in \cite{Kajiwara2019}, we have the following lemmas.

\begin{lemma}\label{neccon1period}
Let $U$ be the evolution operator of the proposed quantum walk on $\mathrm{Cay(D_N,\{a,b\})}$, and $\lam_j(k),$ $j=1,\hdots,6$ are the eigenvalues of $U(k),$ $k\in \{0,\hdots, N-1\}.$ Then $\sum_{i=1}^6\lambda_j(k)\in  \mathbb{Z}[\zeta_\tau]$.

\end{lemma}

\begin{lemma}\label{neccon2period}
 Let $C=[c_{ij}]\in\C^{3\times 3}$ be a coin operator for the proposed walk on $\mathrm{Cay(D_N,\{a,b\})}.$ Suppose $\tau$ is the period of the walk with the evolution operator $U$. Then  $U^\tau\neq I_{6N}$ if $c_{11} \not\in \frac{1}{2N}\mathbb{Z}[\zeta_{lcm(N,\tau)}]$ or $c_{22}\not\in \frac{1}{2N}\mathbb{Z}[\zeta_{\tau}].$  
\end{lemma}

\pf From Lemma \ref{neccon1period}, we see that if $U^\tau=I_{6N}$, it implies that $\sum_{i=1}^6\lambda_j(k)\in  \mathbb{Z}[\zeta_\tau]$ where $\lambda_i(k)$'s are eigenvalues of $U(k)$, $j\in \{1,2\hdots,6\}$ and $k\in \{0,\hdots,N-1\}$. Hence, from Cayley-Hamilton Theorem and equation (\ref{fourierevol}), we obtain \begin{eqnarray*}
    c_{11}e^{\iota 2\pi k/N}+2c_{22}+c_{11}e^{-\iota 2\pi k/N} \in \mathbb{Z}[\zeta_\tau].
\end{eqnarray*}

Now $e^{\pm\iota 2\pi k/N}\mathbb{Z}[\zeta_\tau]\subset \mathbb{Z}[\zeta_{lcm(N,\tau)}]$. Hence\begin{eqnarray*}
   && \sum_{k=0}^{N-1}(c_{11}e^{\iota 2\pi k/N}+2c_{22}+c_{11}e^{-\iota 2\pi k/N})\in \mathbb{Z}[\zeta_\tau] \implies 2Nc_{22} \in \mathbb{Z}[\zeta_\tau],\\
    && \sum_{k=0}^{N-1}e^{-\iota 2\pi k/N}(c_{11}e^{\iota 2\pi k/N}+2c_{22}+c_{11}e^{-\iota 2\pi k/N})\in \mathbb{Z}[\zeta_{lcm(N,\tau)}]] \implies Nc_{11} \in \mathbb{Z}[\zeta_{lcm(N,\tau)}],\\
   && \sum_{k=0}^{N-1}e^{\iota 2\pi k/N}(c_{11}e^{\iota 2\pi k/N}+2c_{22}+c_{11}e^{-\iota 2\pi k/N})\in \mathbb{Z}[\zeta_{lcm(N,\tau)}] \implies Nc_{11} \in \mathbb{Z}[\zeta_{lcm(N,\tau)}].
    \end{eqnarray*}
    Finally, 
    \begin{eqnarray*}
   && \sum_{k=0}^{N-1}e^{-\iota 2\pi k/N}(c_{11}e^{\iota 2\pi k/N}+2c_{22}+c_{11}e^{-\iota 2\pi k/N}) \\
   && +\sum_{k=0}^{N-1}e^{\iota 2\pi k/N}(c_{11}e^{\iota 2\pi k/N}+2c_{22}+c_{11}e^{-\iota 2\pi k/N})\in \mathbb{Z}[\zeta_{lcm(N,\tau)}] 
    \implies 2Nc_{11}\in \mathbb{Z}[\zeta_{lcm(N,\tau)}].
    \end{eqnarray*}
    The rest of the proof follows immediately. \hfill{$\square$}

    Now we investigate the periodicity properties of the proposed walk when the coin operator $C$ belongs to either of the sets $\mathcal{X}, \mathcal{Y},  \mathcal{Z}, \mathcal{W}$ as given by equations (\ref{Xclass}) - (\ref{Wclass}).

    \subsection{$C\in \mathcal{X}\cup \mathcal{Y}$}
    In this subsection, we analyze the periodicity of the time evolution operator $U$ when the coin is chosen from $\mathcal{X}$ and $\mathcal{Y}$. First we have the following theorem. 
    
    \begin{theorem}\label{eigenX}
        Let $C\in \mathcal{X}.$ Then the eigenvalues of $U(k)$ are
\begin{eqnarray*}
   && \lambda_{k,1}=-1, \,\, \lambda_{k,2}= 1, \,\,   \lambda_{k,3},\lambda_{k,4}=\frac{(x+1)}{2} \pm \iota \frac{\sqrt{(1-x)(x+3)}}{2}, \\
   && \lambda_{k,5}, \lambda_{k,6}=\frac{(x-1+2x\cos{(2\pi k/N)})}{2} \pm \iota \frac{\sqrt{(1+x+2x\cos{(2\pi k/N)})(3-x-2x\cos{(2\pi k/N)})}}{2},
\end{eqnarray*} with corresponding eigenvectors 
            $$\ket{\nu_{k,j}}=\begin{cases}
            \bmatrix{0&0&0&0&-1&1}^T  \mbox{ if } x=1,y=0,j=1\\
            \bmatrix{0&0&1&0&0&0}^T,  \mbox{ if } x=1,y=0,j=2\\  
        \bmatrix{0&0&0&1&0&0}^T,  \mbox{ if } x=1,y=0,j=3\\
        \bmatrix{0&0&0&0&1&1}^T,  \mbox{ if } x=1,y=0,j=4\\
        \bmatrix{0&1&0&0&0&0}^T,  \mbox{ if } x=1,y=0,j=5\\
        \bmatrix{1&0&0&0&0&0}^T,  \mbox{ if } x=1,y=0,j=6\\
        \bmatrix{0&0&-1&-1&1&1}^T,   \mbox{ if } x=\frac{-1}{3},y=\frac{2}{3},j=1\\
                \bmatrix{\frac{\lambda_{k,j}(\lambda_{k,j}(1-x-y)+y)}{1-\lambda_{k,j}x-\lambda_{k,j}xe^{i\frac{2\pi k}{N}}+\lambda_{k,j}^2xe^{i\frac{2\pi k}{N}}}\\ \frac{(\lambda_{k,j}(1-x-y)+y)}{x-\lambda_{k,j}x-\lambda_{k,j}xe^{-i\frac{2\pi k}{N}}+\lambda_{k,j}^2e^{-i\frac{2\pi k}{N}}} \\ 
            \frac{\lambda_{k,j}((1-x-y)+ \lambda_{k,j}ye^{i\frac{2\pi k}{N}})}{1-\lambda_{k,j}x-\lambda_{k,j}xe^{i\frac{2\pi k}{N}}+\lambda_{k,j}^2xe^{i\frac{2\pi k}{N}}}  \\ 
            \frac{((1-x-y)+ \lambda_{k,j}ye^{-i\frac{2\pi k}{N}})}{x-\lambda_{k,j}x-\lambda_{k,j}xe^{-i\frac{2\pi k}{N}}+\lambda_{k,j}^2e^{-i\frac{2\pi k}{N}}}\\
            \frac{\lambda_{k,j}(x-\lambda_{k,j}x-\lambda_{k,j}xe^{i\frac{2\pi k}{N}}+\lambda_{k,j}^2e^{i\frac{2\pi k}{N}})}{1-\lambda_{k,j}x-\lambda_{k,j}xe^{i\frac{2\pi k}{N}}+\lambda_{k,j}^2xe^{i\frac{2\pi k}{N}}}\\ 1}, \mbox{ otherwise,} 
            \end{cases} $$
where $-1/3\leq x\leq 1$, $k\in\{0,1,\hdots,N-1\}.$  
\end{theorem}

    \pf From (\ref{fourierevol}), $U(k)$ is given by 
    $$\bmatrix{xe^{-2\iota\pi k/N} & 0 & ye^{-2\iota\pi k/N}& 0&(1-x-y)e^{-2\iota\pi k/N}&0\\0&xe^{2\iota\pi k/N} & 0 & ye^{2\iota\pi k/N}& 0&(1-x-y)e^{2\iota\pi k/N}\\ (1-x-y)&0&x&0&y&0\\ 0&(1-x-y)&0&x&0&y\\ 0&y&0&(1-x-y)&0&x\\ y&0&(1-x-y)&0&x&0},$$ where $x^2+y^2+xy-x-y=0,-1/3\leq x\leq 1$. The matrix has the characteristic equation as \begin{eqnarray}
    \chi(z)=z^6-2x(1+\cos{(2\pi k/N)})z^5+(2(x^2+x)\cos{(2\pi k/N)} +x^2)z^4\\ \nonumber-(2(x^2+x)\cos{(2\pi k/N)} +x^2)z^2+2x(1+\cos{(2\pi k/N)})z-1
\end{eqnarray}
Clearly all complex roots come in conjugate pairs. Hence, $-1$ becomes a root. Further, other real root can either be $1$ or $-1$. However, since the determinant of the matrix is seen to be $-1$ from the characteristic equation, hence $1$ is the other real root. Hence $\chi(z)=(z^2-1)\eta(z)$ where $\eta(z)$ is a polynomial with degree $ 4$ given by  $\eta(z)=(z^4-2x(1+\cos{(2\pi k/N)(z^3+z)})+(2(x^2+x)\cos{2\pi k/N}+x^2+1)z^2+1).$ Then the desired result follows by calculating the roots of $\eta(z)$ following Ferrari's method \cite{Dummit1991}. This completes the proof. \hfill{$\square$}

Now we consider the one-parametric trigonometric representation of the matrices in $\mathcal{X}$ that enable us to derive the periodicity results. By setting $x=\frac{2\cos\theta+1}{3}$ and $y=\frac{1-\cos\theta}{3} + \frac{\sin\theta}{\sqrt{3}}$ the set $\mathcal{X}$ is given by  \begin{eqnarray*}
   \mathcal{X}_{\theta} =\left\{\bmatrix{\frac{2\cos{\theta}+1}{3}  &  \frac{1-\cos{\theta}}{3}+\frac{\sin{\theta}}{\sqrt{3}} & \frac{1-\cos{\theta}}{3}-\frac{\sin{\theta}}{\sqrt{3}}\\\frac{1-\cos{\theta}}{3}+\frac{\sin{\theta}}{\sqrt{3}}&\frac{2\cos{\theta}+1}{3}&\frac{1-\cos{\theta}}{3}+\frac{\sin{\theta}}{\sqrt{3}} \\ \frac{1-\cos{\theta}}{3}+\frac{\sin{\theta}}{\sqrt{3}} & \frac{1-\cos{\theta}}{3}-\frac{\sin{\theta}}{\sqrt{3}}    &  \frac{2\cos{\theta}+1}{3} }|-\pi\leq \theta \leq \pi  \right\}.
\end{eqnarray*} Clearly, for $\theta = \pi$, we obtain the Grover coin $\mathsf{G}\in \mathcal{X}_{\theta}.$ Then we have the following result.

\begin{theorem}\label{Xperiod}
Let $C\in \mathcal{X}_{\theta}.$ Then the period of the proposed walk on $\mathrm{Cay(D_N,\{a,b\})}$ is given by 
$$\begin{cases}
    lcm\{2,c_kp_k:1\leq k\leq N-1\},\mbox{ where }\frac{2k}{N}=\frac{m_k}{p_k},gcd(m_k,p_k)=1 \,\mbox{with}\, c_k=1 \mbox{ if } m_k \mbox{ is even and } \\ \hfill{ c_k=2 \mbox{ if } m_k \mbox{ is odd } ,\mbox{ if } C = I(\theta= 0)}  \vspace{0.05in}\\
    6, \mbox{ if } \theta\in \{\frac{2\pi}{3},\frac{4\pi}{3}\}\\
    \infty, \mbox{ otherwise. }
\end{cases}$$
\end{theorem}

\pf From Theorem \ref{eigenX}, and by putting $x=\frac{2\cos{\theta}+1}{3}$, we get the eigenvalues of $U$ in terms of $\theta$ and $k,N$. Hence, by putting $\theta=0$, we get the coin operator as identity matrix of order 3  and $U$ in this case has eigenvalues to be $-1,1,1,1,e^{\pm\iota 2\pi k/N}$, $k\in \{0,\hdots,N-1\}$. For $k=0$, the eigenvalues become $-1,1,1,1,1,1$. Obviously in this case the period is $2$ since $\lambda_{k,j}^2=1 $ for $k=0,2\leq j\leq 6$. For $k>1$, however, let $\frac{2K}{N}=\frac{m_k}{p_k}$ such that $gcd(m_k,p_k)=1$. Then if $m_k$ is even then $p_k$ is odd. Hence $U(k)^{p_k}=I$, if $m_k$ is odd then either $p_k$ is odd or even. In any case $U^{2p_k}=I$. Hence, the first case follows immediately.  

If $\theta\in \{2\pi/3,-2\pi/3\}$, then $U$ has eigenvalues $-1,1,e^{\pm\iota 2\pi/3},e^{\pm\iota \pi/3}$ for all $k\in \{0,\hdots,N-1\}$ . Clearly, $lcm(2,3,6)=6$ is the period of the walk in these two cases. Finally, we shall show that any using any other coin different from the ones described in previous cases, leads to an aperiodic walk i.e. only finite period can be found using permutation coins from the class $\mathcal{X}_\theta$.Assume the walk is periodic with period $\tau$ for any coin. From theorem\ref{neccon2period}, we see for coin $C=(c_{ij})_{3\times 3}$ . Hence, $2N\frac{2\cos{\theta}+1}{3}$ We notice that from Theorem \ref{eigenX} that $\lambda_{k,3}/\lambda_{k,4}=\frac{x+1}{2}\pm\iota \frac{\sqrt{(1-x)(x+3)}}{2}$ and $\lambda_{k,5}/\lambda_{k,6}=\frac{(x-1+2x\cos{(2\pi k/N)})}{2} \pm \iota \frac{\sqrt{(1+x+2x\cos{(2\pi k/N)})(3-x-2x\cos{(2\pi k/N)})}}{2}$ . Putting $k=0$ we get $\lambda_{0,5}/\lambda_{0,6}=\frac{3x-1}{2}\pm \iota \frac{\sqrt{(3x+1)(3-3x)}}{2}$. Putting $x=\frac{2\cos{\theta}+1}{3}$ we get $\lambda_{0,5}/\lambda_{0,6}=e^{\pm \iota \theta}$. Take $\theta$ to be rational because for irrational $\theta$, $\not\exists \tau\in \mathbb{N}$ such that $e^{\pm \iota \theta\tau}=1$. Hence $\theta=\frac{m\pi}{n}$ for $gcd(m,n)=1,m\in \mathbb{N}\cup\{0\},n\in \mathbb{N}$. Then we see that $\lambda_{0,3}/\lambda_{0,4}=\cos{(\phi)}\pm \iota \sin{(\phi)}$ where \begin{equation}\label{eqnmainperiodx}
    \cos{(\phi)}=\frac{\cos{(\theta)}+2}{3}.
\end{equation}
From  Lemma\ref{neccon2period}  we get $ c_{11}e^{\iota 2\pi k/N}+2c_{22}+c_{11}e^{-\iota 2\pi k/N} \in \mathbb{Z}[\zeta_\tau].$ Further $e^{\pm\iota 2\pi k/N}\mathbb{Z}[\zeta_\tau]\subset \mathbb{Z}[\zeta_{lcm(N,\tau)}]$. Hence $
  2Nc_{22} \in \mathbb{Z}[\zeta_\tau]$ and $2Nc_{11} \in \mathbb{Z}[\zeta_{lcm(N,\tau)}].$ Now,
since $c_{11}=c_{22} $ for coins in $\mathcal{X}$ and since $e^{\pm \frac{2i\pi k}{N}}\mathbb{Z}[\zeta_\tau]\subset \mathbb{Z}[\zeta_{lcm(N,\tau)}]$, we see that $\tau=Nl$ for $l\in \mathbb{N}$. Thus 
$$Nc_{11}e^{\iota 2\pi k/N}+2Nc_{22}+Nc_{11}e^{-\iota 2\pi k/N} \in \mathbb{Z}[\zeta_{Nl}]
    \implies  Nc_{11}e^{\iota 2\pi kl/Nl}+2Nc_{22}+Nc_{11}e^{-\iota 2\pi kl/Nl} \in \mathbb{Z}[\zeta_{Nl}].$$

Consequently, $2N\frac{2\cos{\theta}+1}{3}$ is an integer and hence $\frac{2\cos{\theta}+1}{3}$ is rational. Now by Niven's theorem \cite{Niven1956}, $\theta \in \{0,\pm \pi,\pm \frac{pi}{2},\pm\frac{\pi}{3},\pm\frac{2\pi}{3}\}$. For all such $\theta$ we obtain $\cos{\phi}\in \{1,\frac{1}{2},\frac{1}{3},\frac{2}{3}\}$. However, for finite period, $\phi$ also has to be of the form $\frac{u\pi}{v}$ for $gcd(u,v)=,u\in \mathbb{N}\cup\{0\},v\in \mathbb{N}$. Following Niven's theorem again we obtain $$(\theta,\phi)=\{(0,0),(2\pi/3,\pi/3),(2\pi/3,-\pi/3),(-2\pi/3,\pi/3),(2\pi/3,-\pi/3)$$ which again gives us the coin as permutation matrices. The rest of the proof follows immediately. \hfill{$\square$}

\begin{corollary}
The proposed walk with Grover coin on $\mathrm{Cay(D_N,\{a,b\})}$ is aperiodic.    
\end{corollary}

Now we have the following results for coins in $\mathcal{Y}.$
   
    \begin{theorem}\label{eigenY} Consider the proposed quantum walks on $\mathrm{Cay(D_N,\{a,b\})}$ with coin operator $C\in \mathcal{Y}.$ Then the eigenvalues of $U(k)$ are given by \begin{eqnarray*}
     && \lambda_{k,1}=-1, \,\, \lambda_{k,2}= 1, \,\, \lambda_{k,3}/\lambda_{k,4}=\frac{(x-1)}{2} \pm \iota \frac{\sqrt{(1+x)(3-x)}}{2}, \\
     && \lambda_{k,5}/\lambda_{k,6}=\frac{(x+1+2x\cos{(2\pi k/N)})}{2} \pm \iota \frac{\sqrt{(1-x-2x\cos{(2\pi k/N)})(3+x+2x\cos{(2\pi k/N)})}}{2},
    \end{eqnarray*} with associated eigenvectors $$\ket{\nu_{k,j}}=\begin{cases}
            \bmatrix{0&0&1&0&0&0}^T  \mbox{ if } x=-1,y=0,j=1\\  
        \bmatrix{0&0&0&0&-1&1}^T  \mbox{ if } x=-1,y=0,j=2\\
        \bmatrix{0&0&0&1&0&0}^T  \mbox{ if } x=-1,y=0,j=3\\
        \bmatrix{0&0&0&0&1&1}^T  \mbox{ if } x=-1,y=0,j=4\\
        \bmatrix{0&1&0&0&0&0}^T  \mbox{ if } x=-1,y=0,j=5\\
        \bmatrix{1&0&0&0&0&0}^T  \mbox{ if } x=-1,y=0,j=6\\
        \bmatrix{0&0&-1&-1&1&1}^T   \mbox{ if } x=\frac{1}{3},y=\frac{-2}{3},j=2\\
                \bmatrix{\frac{-\lambda_{k,j}(\lambda_{k,j}(1+x+y)+y)}{-1+\lambda_{k,j}x+\lambda_{k,j}xe^{i\frac{2\pi k}{N}}+\lambda_{k,j}^2xe^{i\frac{2\pi k}{N}}}\\ \frac{-(\lambda_{k,j}(1+x+y)+y)}{-x-\lambda_{k,j}x-\lambda_{k,j}xe^{-i\frac{2\pi k}{N}}+\lambda_{k,j}^2e^{-i\frac{2\pi k}{N}}} \\ 
            \frac{\lambda_{k,j}((1+x+y)+ \lambda_{k,j}ye^{i\frac{2\pi k}{N}})}{-1+\lambda_{k,j}x+\lambda_{k,j}xe^{i\frac{2\pi k}{N}}+\lambda_{k,j}^2xe^{i\frac{2\pi k}{N}}}  \\ 
            \frac{((1+x+y)+ \lambda_{k,j}ye^{-i\frac{2\pi k}{N}})}{-x-\lambda_{k,j}x-\lambda_{k,j}xe^{-i\frac{2\pi k}{N}}+\lambda_{k,j}^2e^{-i\frac{2\pi k}{N}}}\\
            \frac{\lambda_{k,j}(-x-\lambda_{k,j}x-\lambda_{k,j}xe^{i\frac{2\pi k}{N}}+\lambda_{k,j}^2e^{i\frac{2\pi k}{N}})}{-1+\lambda_{k,j}x+\lambda_{k,j}xe^{i\frac{2\pi k}{N}}+\lambda_{k,j}^2xe^{i\frac{2\pi k}{N}}}\\ 1
            } \mbox{ otherwise, } \\
            \end{cases} $$  where $-1\leq x\leq 1/3$, $k\in\{0,1,\hdots,N-1\}.$
    \end{theorem}
    
    \pf From equation (\ref{fourierevol}), the matrix $U(k)$ is given by \begin{small}$$
    \bmatrix{xe^{-2\iota\pi k/N} & 0 & ye^{-2\iota\pi k/N}& 0&(-1-x-y)e^{-2\iota\pi k/N}&0\\0&xe^{2\iota\pi k/N} & 0 & ye^{2\iota\pi k/N}& 0&(-1-x-y)e^{2\iota\pi k/N}\\ (-1-x-y)&0&x&0&y&0\\ 0&(-1-x-y)&0&x&0&y\\ 0&y&0&(-1-x-y)&0&x\\ y&0&(-1-x-y)&0&x&0},$$ \end{small} where $x^2+y^2+xy+x+y=0,-1\leq x\leq 1/3$. The matrix has the characteristic polynomial \begin{eqnarray*}
    \chi(z) &=& z^6-2x(1+\cos{(2\pi k/N)})z^5+(2(x^2-x)\cos{(2\pi k/N)} +x^2)z^4\\ && -(2(x^2-x)\cos{(2\pi k/N)} +x^2)z^2+2x(1+\cos{(2\pi k/N)})z-1.
\end{eqnarray*}
The rest of the proof follows similar to the proof of Theorem \ref{eigenX}. \hfill{$\square$}

Now considering the one-parametric representation of the coins in $\mathcal{Y}$ given by \cite{Sarkar2020} $$
   \mathcal{Y}_{\theta} =\left\{\bmatrix{\frac{2\cos{\theta}-1}{3}  &  \frac{-1-\cos{\theta}}{3}+\frac{\sin{\theta}}{\sqrt{3}} & \frac{-1-\cos{\theta}}{3}-\frac{\sin{\theta}}{\sqrt{3}}\\\frac{-1-\cos{\theta}}{3}+\frac{\sin{\theta}}{\sqrt{3}}&\frac{2\cos{\theta}-1}{3}&\frac{-1-\cos{\theta}}{3}+\frac{\sin{\theta}}{\sqrt{3}} \\ \frac{-1-\cos{\theta}}{3}+\frac{\sin{\theta}}{\sqrt{3}} & \frac{-1-\cos{\theta}}{3}-\frac{\sin{\theta}}{\sqrt{3}}    &  \frac{2\cos{\theta}-1}{3} }|-\pi\leq x\leq \pi  \right\},
$$ we have the following result whose proof is similar to the proof of Theorem \ref{Xperiod}.

\begin{theorem}\label{Yperiod}
Let $C\in \mathcal{Y}_{\theta}.$ Then the period of the proposed walk on $\mathrm{Cay(D_N,\{a,b\})}$ is given by 
$$\begin{cases}
    lcm\{2,2p_k:1\leq k\leq N-1\},\mbox{ where }\frac{(N-2)k}{N}=\frac{m_k}{p_k},gcd(m_k,p_k)=1,\mbox{ if } C = -I(\theta= \pi)  \vspace{0.05in}\\
    6, \mbox{ if } \theta\in \{\frac{\pi}{3},\frac{-\pi}{3}\} \,\, (\mbox{permutation coins}), \\
    \infty, \mbox{ otherwise. }
\end{cases}$$
\end{theorem}

\subsection{$C\in\mathcal{Z}\cup \mathcal{W}$}

 In this section we study the periodicity property of the proposed quantum walks on $\mathrm{Cay(D_N,\{a,b\})}$ with coin operators from $\mathcal{Z}$ and $\mathcal{W}$ as described in equations (\ref{Zclass}) - (\ref{Wclass}). The one-parameter parametrization of the coins in $\mathcal{Z}$ can be obtained by setting $x=\frac{2\cos{\theta}+1}{3}$ and $y=\frac{(1-\cos{\theta})}{3}+\frac{1}{\sqrt{3}}\sin{\theta}$ and we denote the set as $\mathcal{Z}_{\theta}$, whereas setting $x=\frac{2\cos{\theta}-1}{3}$, $y=\frac{-(1+\cos{\theta})}{3}+\frac{1}{\sqrt{3}}\sin{\theta}$ the coins in $\mathcal{W}$ are parametrized and we denote the set as  $\mathcal{W}_{\theta},$ where $-\pi\leq \theta\leq \pi$ \cite{Sarkar2020}. First we have the following theorem.

\begin{theorem}\label{eigenZ}
Consider the proposed walk on $\mathrm{Cay(D_N,\{a,b\})}$ with coin operator $C\in \mathcal{W}.$ Then the eigenvalues of $U(k)$ are given by $$\lambda_{k,1}=-1, \,\,
            \lambda_{k,2}= 1, \,\,
            \lambda_{k,3}/\lambda_{k,4}=e^{\pm\iota\omega(k)}, \,\,
            \lambda_{k,5}/\lambda_{k,6}=e^{\pm\iota\eta(k)},$$ where \begin{eqnarray*}
                \cos{\omega(k)} &=& \frac{1-x-y+x\cos{\frac{2\pi k}{N}}}{2}+\frac{1}{2}\sqrt{1+2y\cos{\frac{2\pi k}{N}}+y^2-\left(x\sin{\frac{2\pi k}{N}}\right)^2}, \\
            \cos{\eta(k)} &=& \frac{1-x-y+x\cos{\frac{2\pi k}{N}}}{2}-\frac{1}{2}\sqrt{1+2y\cos{\frac{2\pi k}{N}}+y^2-\left(x\sin{\frac{2\pi k}{N}}\right)^2},
            \end{eqnarray*} $-1/3\leq x\leq 1$, $k\in\{0,1,\hdots,N-1\}$ with the corresponding eigenvectors
$$\ket{\nu_{k,j}}=\begin{cases}
            \bmatrix{1&0&0&0&0&0}^T  \mbox{ if } x=1,y=0,j=5\\  
        \bmatrix{0&1&0&0&0&0}^T  \mbox{ if } x=1,y=0,j=6\\
        \bmatrix{e^{-i\frac{2\pi k}{N}}&0&1&0&0&0}^T  \mbox{ if } x=0,y=1,j=3\\
        \bmatrix{0&e^{i\frac{2\pi k}{N}}&0&1&0&0}^T  \mbox{ if } x=0,y=1,j=4\\
        \bmatrix{-e^{-i\frac{2\pi k}{N}}&0&1&0&0&0}^T  \mbox{ if } x=0,y=1,j=5\\
        \bmatrix{0&-e^{i\frac{2\pi k}{N}}&0&1&0&0}^T  \mbox{ if } x=0,y=1,j=6\\
        \bmatrix{0&0&1&0&0&0}^T  \mbox{ if } x=0,y=0,j=2\\
        \bmatrix{0&0&0&1&0&0}^T  \mbox{ if } x=0,y=0,j=3\\
        \bmatrix{1&e^{i\frac{2\pi k}{N}}&0&0&e^{i\frac{2\pi k}{N}}&1}^T  \mbox{ if } x=0,y=0,j=4\\
                \bmatrix{\frac{\lambda_{k,j}(\lambda_{k,j}-1)(1-x-y)}{-1+\lambda_{k,j}(1-x-y)+\lambda_{k,j}xe^{i\frac{2\pi k}{N}}+\lambda_{k,j}^2ye^{i\frac{2\pi k}{N}}}\\ \frac{(\lambda_{k,j}-1)(1-x-y)}{-y-\lambda_{k,j}x-\lambda_{k,j}e^{-i\frac{2\pi k}{N}}(1-x-y)+\lambda_{k,j}^2e^{-i\frac{2\pi k}{N}}} \\ 
            \frac{\lambda_{k,j}(\lambda_{k,j}e^{i\frac{2\pi k}{N}}-1)x}{-1+\lambda_{k,j}(1-x-y)+\lambda_{k,j}xe^{i\frac{2\pi k}{N}}+\lambda_{k,j}^2ye^{i\frac{2\pi k}{N}}}  \\ 
            \frac{(\lambda_{k,j}e^{-i\frac{2\pi k}{N}}-1)x}{-y-\lambda_{k,j}x-\lambda_{k,j}e^{-i\frac{2\pi k}{N}}(1-x-y)+\lambda_{k,j}^2e^{-i\frac{2\pi k}{N}}} \\
            \frac{\lambda_{k,j}({-y-\lambda_{k,j}x-\lambda_{k,j}e^{i\frac{2\pi k}{N}}(1-x-y)+\lambda_{k,j}^2e^{i\frac{2\pi k}{N}}})}{-1+\lambda_{k,j}(1-x-y)+\lambda_{k,j}xe^{i\frac{2\pi k}{N}}+\lambda_{k,j}^2ye^{i\frac{2\pi k}{N}}}\\ 1
            }, \mbox{ otherwise. } \\
        \end{cases} $$
\end{theorem}

\pf The characteristic polynomial of $U(k)$ are \begin{eqnarray*}
    \chi(z)=z^6-2(1-x-y+x\cos{(2\pi K/N)})z^5-((4y-2y^2)\cos{(2\pi K/N)}+2y^2-1)z^4+\\\nonumber((4y-2y^2)\cos{(2\pi K/N)}+2y^2-1)z^2+2(1-x-y+x\cos{(2\pi K/N)})z-1.
\end{eqnarray*} Note that the product of eigenvalues of $U(k)$ is $-1$ . Hence, $-1$ shall always be an eigenvalue. Since all the complex eigenvalues of $U(k)$ come in conjugate pairs hence, $1$ is another eigenvalue. Hence, $\lambda_{k,1}=-1$ and $\lambda_{k,2}=1$. Then the characteristic functions of $U(k)$  can be rewritten as \begin{eqnarray}
    \chi(z)=(z^2-1)(z^4-2(1-x-y+x\cos{(\frac{2k\pi}{N})})z^3-((4y-2y^2)\cos{(\frac{2k\pi}{N})}+2y^2-2)z^2\\\nonumber-2(1-x-y+x\cos{(\frac{2k\pi}{N})})z+1)
\end{eqnarray} Assume other two eigenvalues are $e^{\pm \iota \omega(k)}$ and  $e^{\pm \iota \eta(k)}$ then we have \begin{eqnarray*}
    (z^4-2(1-x-y+x\cos{(\frac{2k\pi}{N})})z^3-((4y-2y^2)\cos{(\frac{2k\pi}{N})}+2y^2-2)z^2\\\nonumber-2(1-x-y+x\cos{(\frac{2k\pi}{N})})z+1)=(z^2-2z\cos{\omega(k)}-1)(z^2-2z\cos{\eta(k)}-1)
\end{eqnarray*} 
Comparing both sides, its easy to solve for $\cos{\omega(k)},\cos{\eta(k)}$. The rest of the proof follows easily. \hfill{$\square$}

\begin{theorem}\label{Zperiod}
The period of the proposed walk with $C\in\mathcal{Z}_{\theta}$  is given by 
$$\begin{cases}
    lcm\{2,2p_k:1\leq k\leq N-1\},\mbox{ where }\frac{2k}{N}=\frac{m_k}{p_k},gcd(m_k,p_k)=1,\mbox{ if } \theta=2\pi/3  \vspace{0.05in}\\
    4, \mbox{ if } \theta=\frac{-2\pi}{3}\\
    lcm\{4,c_kp_k:1\leq k\leq N-1\},\mbox{ where }\frac{2k}{N}=\frac{m_k}{p_k},gcd(m_k,p_k)=1, c_k=1 \mbox{ if } m_k \mbox{ is even and } \\ \hfill{c_k=2 \mbox{ if } m_k \mbox{ is odd } ,\mbox{ if } C = I(\theta= 0)} \\
    \infty, \mbox{otherwise.}
    \end{cases}$$    
\end{theorem}

\pf First we consider $\theta\in \{0,2\pi/3,-2\pi/3\}.$ The eigenvalues of $U(k)$ when $C\in \mathcal{Z}_\theta$ when $\theta=0$ are given by
$\{\pm1, e^{\pm \iota \frac{\pi}{2}},e^{\pm \iota \frac{2\pi k}{N}}\}$.The eigenvalues of $U(k)$ when $C\in \mathcal{Z}_\theta$ when $\theta=2\pi/3$ are given by
$\{\pm1, -e^{\pm \iota \frac{\pi k}{N}},e^{\pm \iota \frac{\pi k}{N}}\}$ and finally eigenvalues of $U(k)$ when $C\in \mathcal{Z}_\theta$ when $\theta=-2\pi/3$ are given by
$\{-1,1,1,1 e^{\pm \iota \frac{\pi}{2}}\}, k\in \{0,\hdots,N-1\}.$ Rest of the proof follows similar to Theorem \ref{Xperiod}.

Next let $\theta\in \notin\{0,2\pi/3,-2\pi/3\}.$ The proof follows on a similar way to theorem \ref{Xperiod}. Assume the walk is periodic with period $\tau$. From lemma\ref{neccon2period}, we see that $c_{11}e^{\iota 2\pi k/N}+2c_{22}+c_{11}e^{-\iota 2\pi k/N} \in \mathbb{Z}[\zeta_\tau]$ for a quantum coin $C=[c_{ij}],$ where $c_{11}=x,c_{22}=(1-x-y)$.
Using algebra we see that for $k=0$ then $2(1-y)\in \mathbb{Z}[\zeta_{\tau}]$. We can again see that 
    $$\sum_{k=0}^{N-1}e^{-\iota 2\pi k/N} (c_{11}e^{\iota 2\pi k/N}+2c_{22}+c_{11}e^{-\iota 2\pi k/N}) \in \mathbb{Z}[\zeta_{lcm(N,\tau)}]
     \implies Nx\in  \mathbb{Z}[\zeta_{lcm(N,\tau)}]$$
Also, $$\sum_{k=0}^{N-1} (c_{11}e^{\iota 2\pi k/N}+2c_{22}+c_{11}e^{-\iota 2\pi k/N}) \in \mathbb{Z}[\zeta_{lcm(N,\tau)}]
    \implies 2N(1-x-y)\in  \mathbb{Z}[\zeta_{\tau}].$$

Hence, $2Nx\in \mathbb{Z}[\zeta_{\tau}]$. Thus $2Nx\in \mathbb{Z}[\zeta_{lcm(N,\tau)}]\implies 2Nx\in \mathbb{Z}[\zeta_{\tau}]$. Hence, using the logic similar to theorem\ref{Xperiod}, we see that $2N\frac{2\cos{\theta}+1}{3}$ and $2N(\frac{1-\cos{\theta}}{3}-\frac{1}{\sqrt{3}}\sin{\theta})$ is an integer and by Niven's theorem, $\theta\in \{0,\pm \pi,\pm \pi/3,\pm 2\pi/3\}$. For $\theta=\pm\pi/3$, we see that $y=2/3\mbox{ or }-1/3$ and  for $k=0$, from Theorem \ref{eigenZ}, we obtain 
$$\lambda_{0,1}=-1, \, \lambda_{0,2}=1, \, \lambda_{0,3}=1, \,
    \lambda_{0,4}=1, \, \lambda_{0,5} =-y+i\sqrt{1-y^2}, \,
    \lambda_{0,6} =-y-i{\sqrt{1-y^2}}.$$
But $\arccos{(-2/3)},\arccos{(1/3)}$ is not a rational multiple of $\pi$ using Niven's theorem and Chebyschev polynomials\cite{Kreyszig1978}. For, $\theta=\pm \pi/2$, then we see that $2(1-y)=\frac{4-2\sqrt{3}}{3}\not \in \mathbb{Z}[\zeta_{\tau}]$. The rest of the proof follows immediately. \hfill{$\square$}

Now we consider coins from $C\in \mathcal{W}.$

\begin{theorem}\label{eigenW}
Consider the proposed walk on $\mathrm{Cay(D_N,\{a,b\})}$ with coin operator $C\in \mathcal{W}.$ Then the eigenvalues of $U(k)$ are given by $$\lambda_{k,1}=-1, \, \lambda_{k,2}= 1, \, \lambda_{k,3}, \lambda_{k,4}=e^{\pm\iota\omega(k)}, \,
\lambda_{k,5},\lambda_{k,6}=e^{\pm\iota\eta(k)},$$ where 
\begin{eqnarray*}
\cos{\omega(k)} &=& \frac{-1-x-y+x\cos{\frac{2\pi k}{N}}}{2}+\frac{1}{2}\sqrt{1-2y\cos{\frac{2\pi k}{N}}+y^2-\left(x\sin{\frac{2\pi k}{N}}\right)^2}, \\
\cos{\eta(k)} &=& \frac{-1-x-y+x\cos{\frac{2\pi k}{N}}}{2}-\frac{1}{2}\sqrt{1-2y\cos{\frac{2\pi k}{N}}+y^2-\left(x\sin{\frac{2\pi k}{N}}\right)^2}
\end{eqnarray*} with the corresponding eigenvectors
            $$\ket{\nu_{k,j}}=\begin{cases}
            \bmatrix{1&0&0&0&0&0}^T  \mbox{ if } x=-1,y=0,j=5\\  
        \bmatrix{0&1&0&0&0&0}^T  \mbox{ if } x=-1,y=0,j=6\\
        \bmatrix{-e^{-i\frac{2\pi k}{N}}&0&1&0&0&0}^T  \mbox{ if } x=0,y=-1,j=3\\
        \bmatrix{0&-e^{i\frac{2\pi k}{N}}&0&1&0&0}^T  \mbox{ if } x=0,y=-1,j=4\\
        \bmatrix{e^{-i\frac{2\pi k}{N}}&0&1&0&0&0}^T  \mbox{ if } x=0,y=-1,j=5\\
        \bmatrix{0&e^{i\frac{2\pi k}{N}}&0&1&0&0}^T  \mbox{ if } x=0,y=-1,j=6\\
        \bmatrix{0&0&1&0&0&0}^T  \mbox{ if } x=0,y=0,j=1\\
        \bmatrix{0&0&0&1&0&0}^T  \mbox{ if } x=0,y=0,j=2\\
        \bmatrix{1&e^{i\frac{2\pi k}{N}}&0&0&e^{i\frac{2\pi k}{N}}&1}^T  \mbox{ if } x=0,y=0,j=3\\
                \bmatrix{\frac{-\lambda_{k,j}(\lambda_{k,j}+1)(1+x+y)}{1+\lambda_{k,j}(1+x+y)-\lambda_{k,j}xe^{i\frac{2\pi k}{N}}+\lambda_{k,j}^2ye^{i\frac{2\pi k}{N}}}\\ \frac{-(\lambda_{k,j}+1)(1+x+y)}{y-\lambda_{k,j}x+\lambda_{k,j}e^{-i\frac{2\pi k}{N}}(1+x+y)+\lambda_{k,j}^2e^{-i\frac{2\pi k}{N}}} \\ 
            \frac{\lambda_{k,j}(\lambda_{k,j}e^{i\frac{2\pi k}{N}}+1)x}{1+\lambda_{k,j}(1+x+y)-\lambda_{k,j}xe^{i\frac{2\pi k}{N}}+\lambda_{k,j}^2ye^{i\frac{2\pi k}{N}}}  \\ 
            \frac{(\lambda_{k,j}e^{-i\frac{2\pi k}{N}}+1)x}{y-\lambda_{k,j}x+\lambda_{k,j}e^{-i\frac{2\pi k}{N}}(1+x+y)+\lambda_{k,j}^2e^{-i\frac{2\pi k}{N}}} \\
            \frac{\lambda_{k,j}({y-\lambda_{k,j}x+\lambda_{k,j}e^{i\frac{2\pi k}{N}}(1+x+y)+\lambda_{k,j}^2e^{i\frac{2\pi k}{N}}})}{1+\lambda_{k,j}(1+x+y)-\lambda_{k,j}xe^{i\frac{2\pi k}{N}}+\lambda_{k,j}^2ye^{i\frac{2\pi k}{N}}}\\ 1
            } \mbox{ otherwise, } \\
        \end{cases} $$ where $-1\leq x\leq 1/3$, $k\in\{0,1,\hdots,N-1\}$.
\end{theorem}

\pf  The characteristic polynomial of $U(k)$ are \begin{eqnarray*}
    \chi(z)=z^6-2(-1-x-y+x\cos{(2\pi K/N)})z^5-((4y-2y^2)\cos{(2\pi K/N)}+2y^2-1)z^4+\\\nonumber((-4y-2y^2)\cos{(2\pi K/N)}+2y^2-1)z^2+2(-1-x-y+x\cos{(2\pi K/N)})z-1
\end{eqnarray*} when the coin from $\mathcal{W}$ is taken. The rest follows similar to the proof of Theorem \ref{eigenZ}. $\square$

\begin{theorem}\label{Wperiod}
Let $C\in \mathcal{W}_{\theta}.$ Then the period of the proposed walk on $\mathrm{Cay(D_N,\{a,b\})}$ is given by
$$\begin{cases}
    lcm\{2,2p_k:1\leq k\leq N-1\},\mbox{ where }\frac{2k}{N}=\frac{m_k}{p_k},gcd(m_k,p_k)=1,\mbox{ if } \theta=2\pi/3  \vspace{0.05in}\\
    4, \mbox{ if } \theta=\frac{-2\pi}{3}\\
    lcm\{4,2p_k:1\leq k\leq N-1\},\mbox{ where} \frac{2k}{N}=\frac{m_k}{p_k},gcd(m_k,p_k)=1,\mbox{ if } C = I(\theta= 0) \\
    \infty, \mbox{otherwise.}
    \end{cases}$$
\end{theorem}

\pf The eigenvalues of $U(k)$ when $C\in \mathcal{W}_\theta$ when $\theta=0$ are given by
$\{\pm1, e^{\pm \iota \frac{\pi}{2}},-e^{\pm \iota \frac{2\pi k}{N}}\}$.The eigenvalues of $U(k)$ when $C\in \mathcal{W}_\theta$ when $\theta=2\pi/3$ are given by
$\{\pm1, -e^{\pm \iota \frac{\pi }{2}},-e^{\pm \iota \frac{\pi k}{N}}\}$ and finally eigenvalues of $U(k)$ when $C\in \mathcal{Z}_\theta$ when $\theta=-2\pi/3$ are given by
$\{1,-1,-1,-1 e^{\pm \iota \frac{\pi}{2}}\}, k\in \{0,\hdots,N-1\}$. Rest of the proof follows similar to the proof of Theorem \ref{Xperiod}. \hfill{$\square$}

\begin{remark}
We mention that the periodicity of lively quantum walks on cycle graphs $C_N$ (Cayley graphs corresponding to $\Z_N$), as studied in \cite{Sarkar2020}, is finite for coins that can be permutation matrix or certain non-permutation matrices belonging to the classes $\mathcal{X},\mathcal{Y},\mathcal{Z},\mathcal{W}$.

Now, from the above theorems we observe that the periodicity of the proposed quantum walks on $\mathrm{Cay(D_N,\{a,b\})}$ is finite only for the coins that are of the form $\pm P,$ where $P$ denotes a permutation matrix belonging to the said classes. This distinguishes the commutativity aspect of the underlying group for the Cayley graph in the quantum walk model. 
\end{remark}

\section{Localization properties of the proposed quantum walks}\label{sec:4}

In this section, we perform a numerical study to investigate localization property of the proposed quantum walk model on $\mathrm{Cay(D_N,\{a,b\})}.$ We proceed by deriving the formula of time-averaged probability of finding the walker at any vertex $(s,r)$, $s\in\{0,1\},$ $r\in\{0,\hdots, N-1\}.$  

Recall that we denote $(\lambda_{k,j},\ket{\omega_{k,j}})$ as the eigenpairs of $U(k)$, $k\in\{0,\hdots, N-1\},$ $j\in\{1,\hdots,6\}$ with $\braket{\omega_{k,j}|\omega_{k,j}}=1,$ $\ket{\omega_{k,j}}=[\omega_{k,j,1}, \omega_{k,j,2}, \hdots, \omega_{k,j,6}]^T\in\C^6.$ Then, defining \begin{eqnarray}
    \ket{{\omega}(0,r)}_{k,j} &=& [\omega_{k,j,1}, 0,\omega_{k,j,3}, 0, \omega_{k,j,5}, 0]^T \\
    \ket{{\omega}(1,r)}_{k,j} &=& [0,\omega_{k,j,2},0, \omega_{k,j,4}, 0, \omega_{k,j,6}]^T,  
\end{eqnarray} and utilizing equation (\ref{eqn:fts}) we obtain
\begin{equation}
 \overline{P(s,r,T)}=\frac{1}{T}\sum_{t=0}^{T-1}\frac{1}{N^2}\sum_{k,k'=0}^{N-1}\sum_{j,j'=1}^6\lambda_{k,j}^t\overline{\lambda_{k',j'}^t}e^{\frac{2\iota \pi (k-k')r}{N}}\braket{\omega_{k,j}|\Psi(k,0)}\braket{\Psi(k',0)|\omega_{k',j'}}\braket{\omega(s,r)_{k',j'}|\omega(s,r)_{k,j}},
\end{equation} $s\in\{0,1\},$ $r\in\{0,\hdots,N-1\}.$ Then we say the walk localizes at $(s,r)$ if $\overline{P(s,r,T)} >0$ for a large value of $T$ (in principle, $T\rightarrow \infty$) and for a given initial position along with an initial state.

Note that if $r=0$ and the walker starts from $(0,r)$ or $(1,r)$ and if $T\rightarrow \infty$, then we see from \cite{Mandal2023} that $$\lim_{T\rightarrow\infty}\frac{1}{T}\sum_{t=0}^{T-1}\lambda_{k,j}^t\overline{\lambda_{k',j'}^t}=\begin{cases}
    0 \mbox{ if } \lambda_{k,j}\neq \lambda_{k',j'}\\
    1, \mbox{ otherwise. }
\end{cases}$$ Hence, we obtain \begin{equation}
     \overline{P(s,0,T)} =\frac{1}{N^2}\sum_{k=0}^{N-1}\sum_{j=1}^6|\braket{\omega_{k,j}|\Psi(k,0)}|^2 \braket{\omega(s,0)_{k,j}|\omega(s,0)_{k,j}},
\end{equation} $s\in\{0,1\}.$

Now we provide numerical simulation results of the time-averaged probability of the proposed walks for several coin states belonging to $\mathcal{X}_{\theta},$ $\mathcal{Y}_{\theta},$ $\mathcal{Z}_{\theta}$ and $\mathcal{W}_{\theta}$ with a variety of different initial positions, initial states, and total number of vertices. All numerical simulations have been performed on MATLAB 2019a on  a computer with 16 GB RAM, 1.5GHZ processor.

\subsection{Time-averaged probability}

\begin{figure}[H]
    \centering
    \subfigure[$\ket{\psi_0}=\ket{0}_3\ket{1}_2\ket{0}_N$, $C\in \mathcal{X}_\theta, \theta=\pi$]{\includegraphics[height=3.5 cm,width=7 cm]{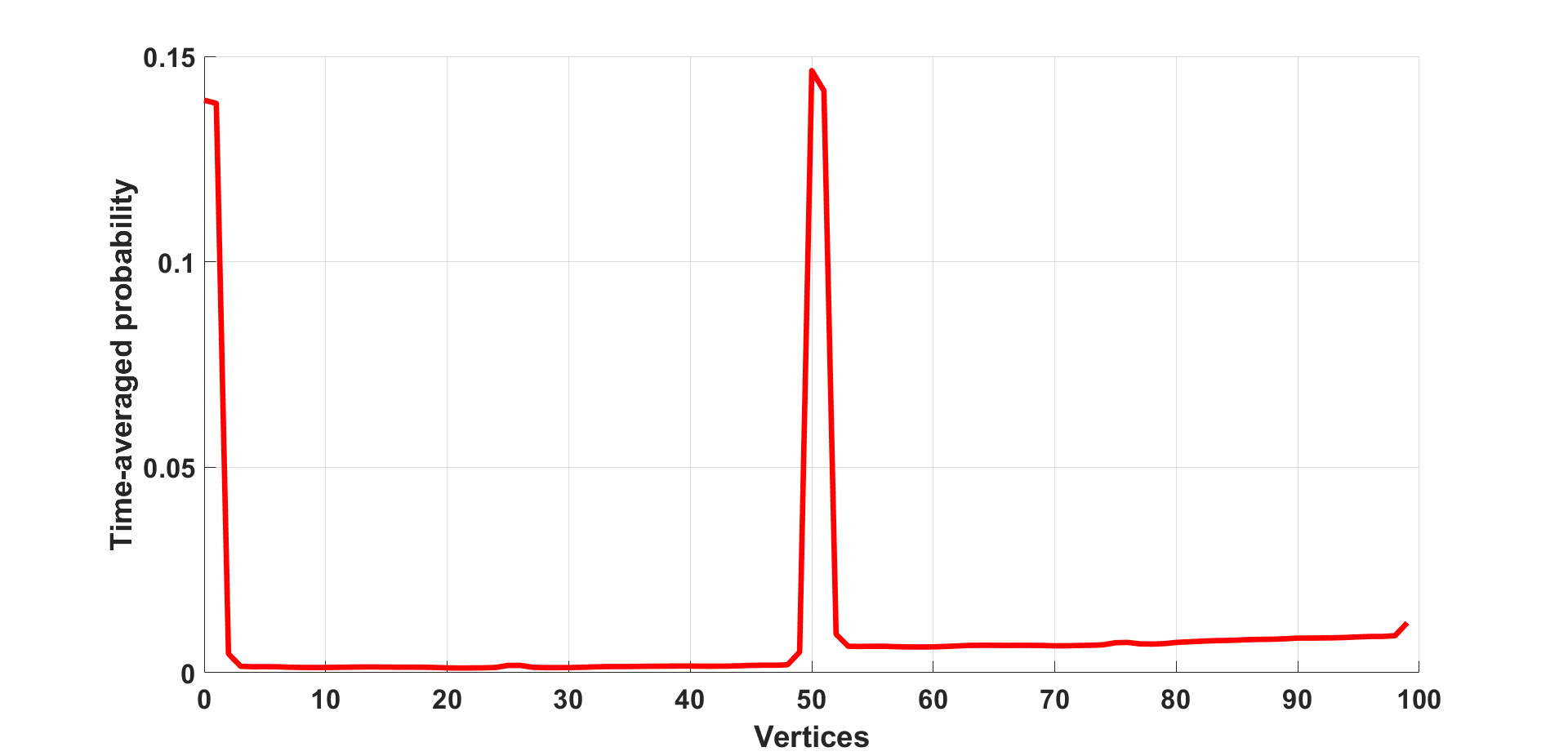}}
     \subfigure[$\ket{\psi_0}=\ket{1}_3\ket{1}_2\ket{0}_N$, $C\in \mathcal{X}_\theta, \theta=\pi$]{\includegraphics[height=3.5 cm,width=7 cm]{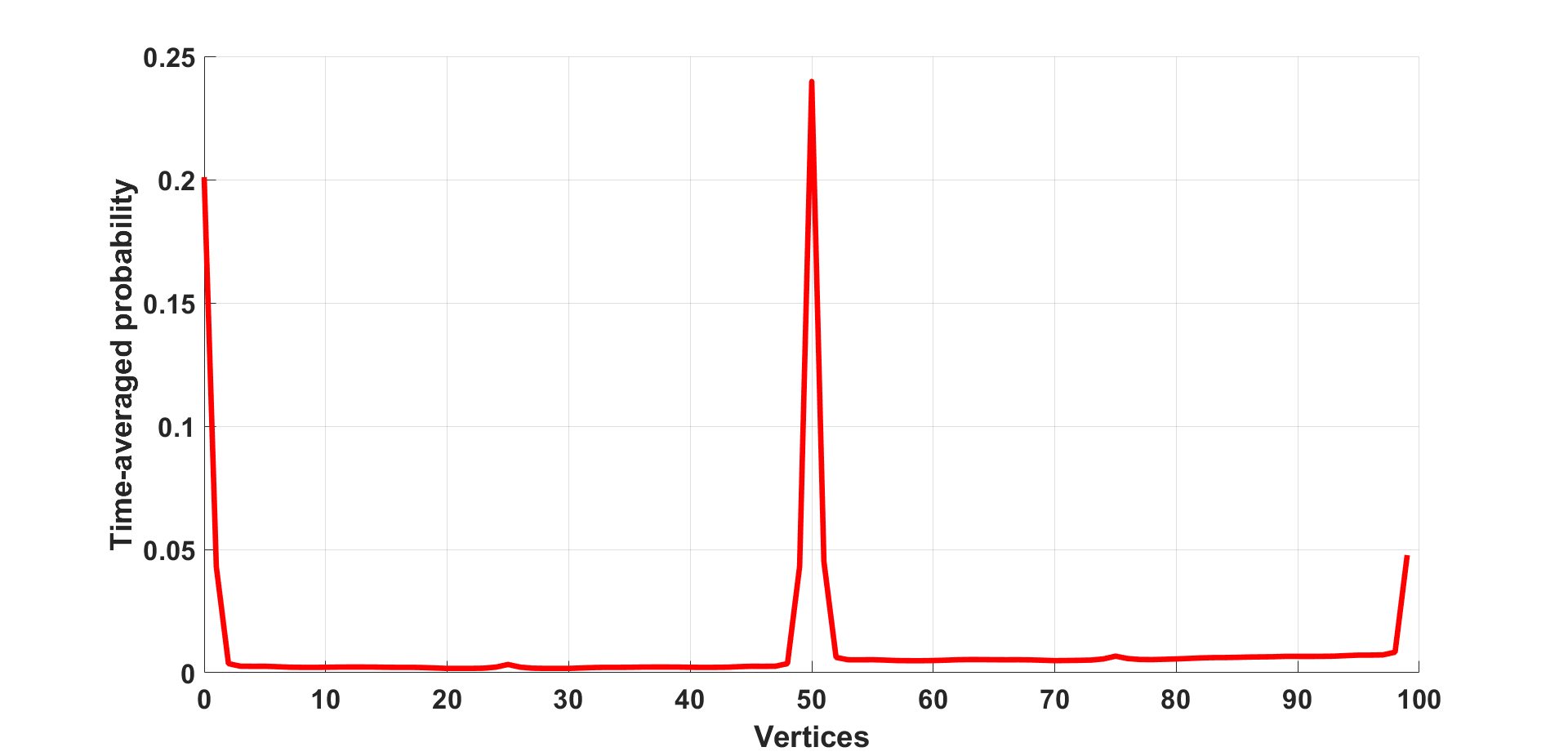}}\\
     \subfigure[$\ket{\psi_0}=\ket{2}_3\ket{1}_2\ket{0}_N$, $C\in \mathcal{X}_\theta, \theta=\pi$]{\includegraphics[height=3.5 cm,width=7 cm]{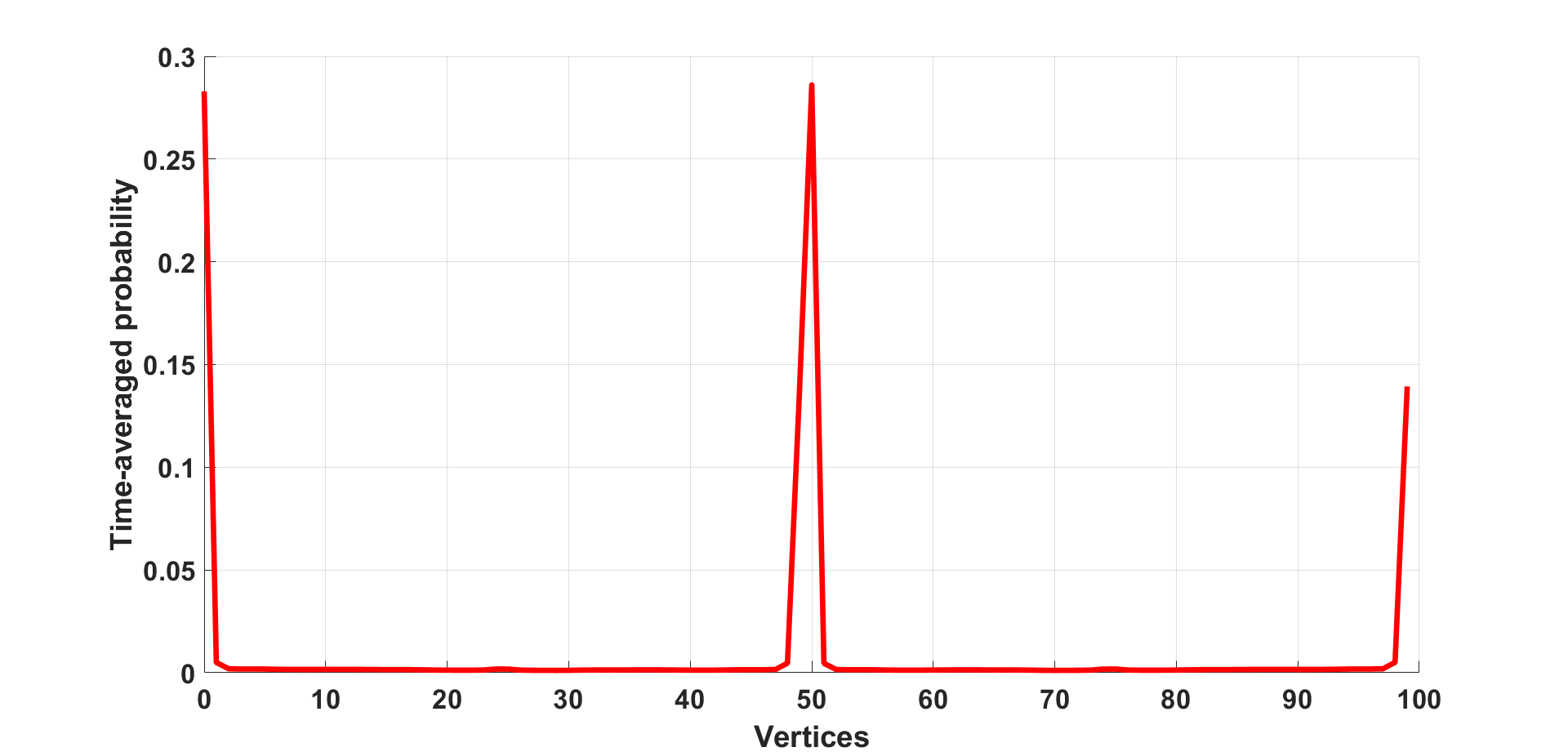}}
     \subfigure[$\ket{\psi_0}\frac{1}{\sqrt{3}}(\ket{0}_3+\ket{1}_3+\ket{2}_3)\ket{1}_2\ket{0}_N$,$C\in \mathcal{X}_\theta, \theta=\pi$ ]{\includegraphics[height=3.5 cm,width=7 cm]{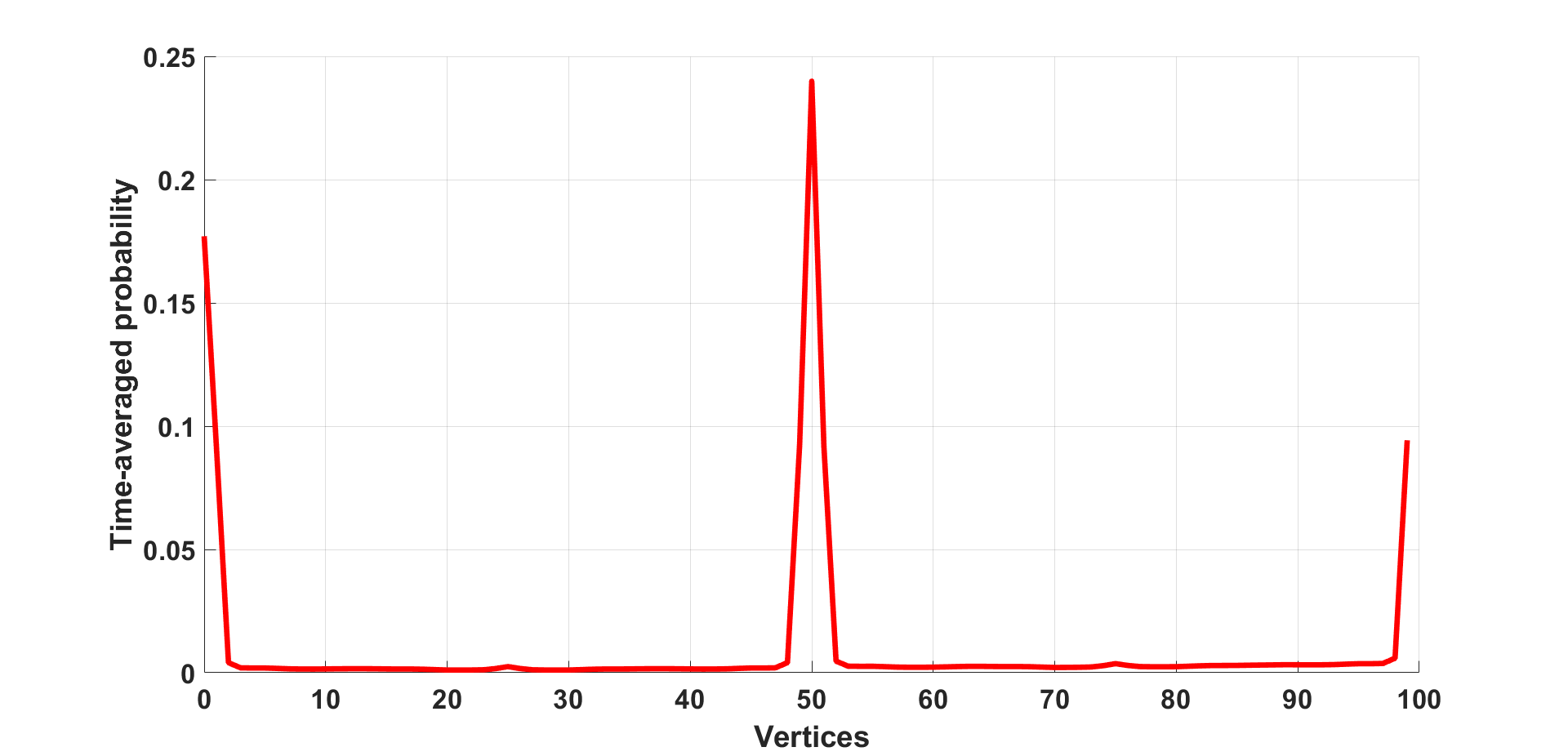}}
     \subfigure[$\ket{\psi_0}=\ket{0}_3\ket{1}_2\ket{0}_N$,$C\in \mathcal{Y}_\theta, \theta=\pi/2$ ]{\includegraphics[height=3.5 cm,width=7 cm]{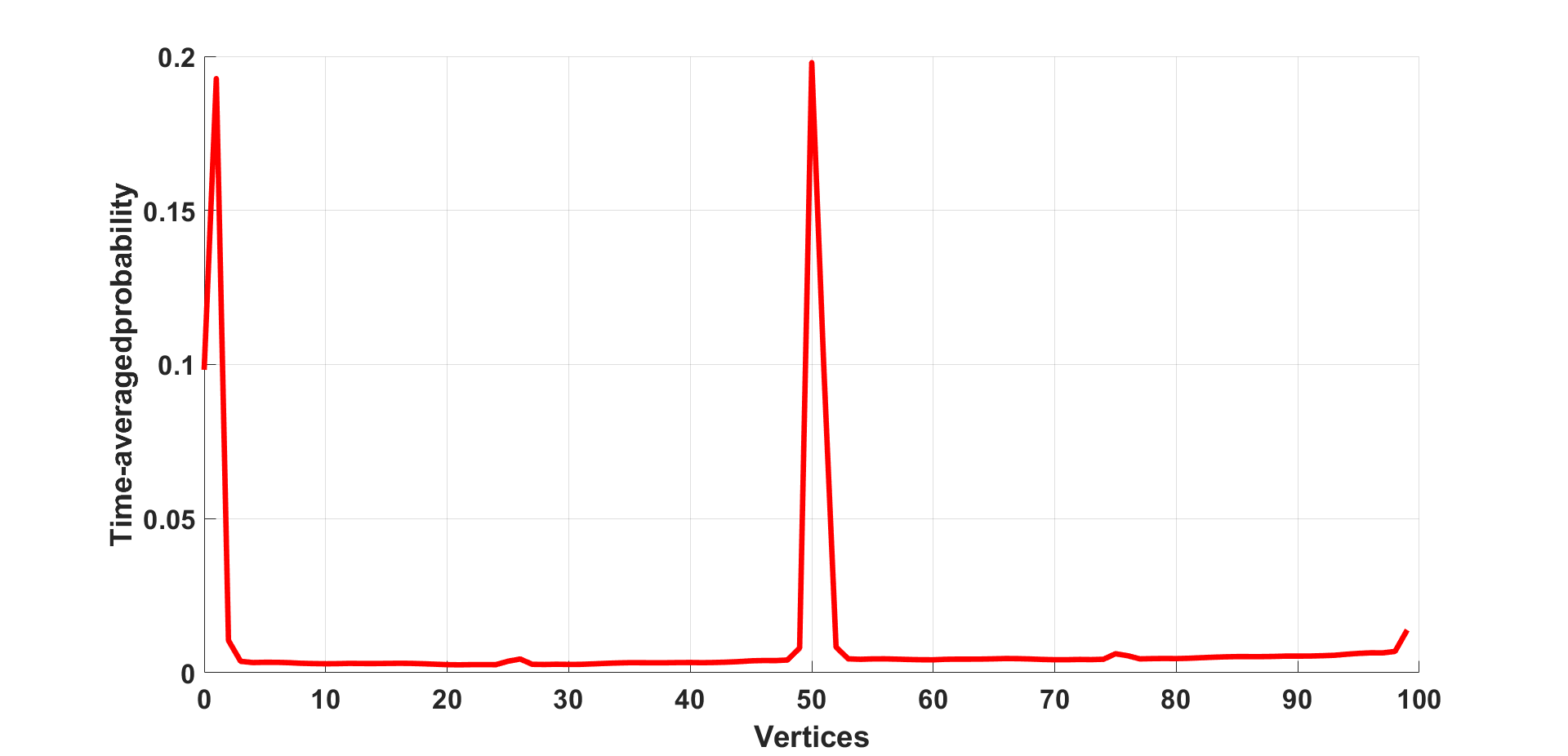}}
     \subfigure[$\ket{\psi_0}=\ket{1}_3\ket{1}_2\ket{0}_N$,$C\in \mathcal{Y}_\theta, \theta=\pi/2$]{\includegraphics[height=3.5 cm,width=7 cm]{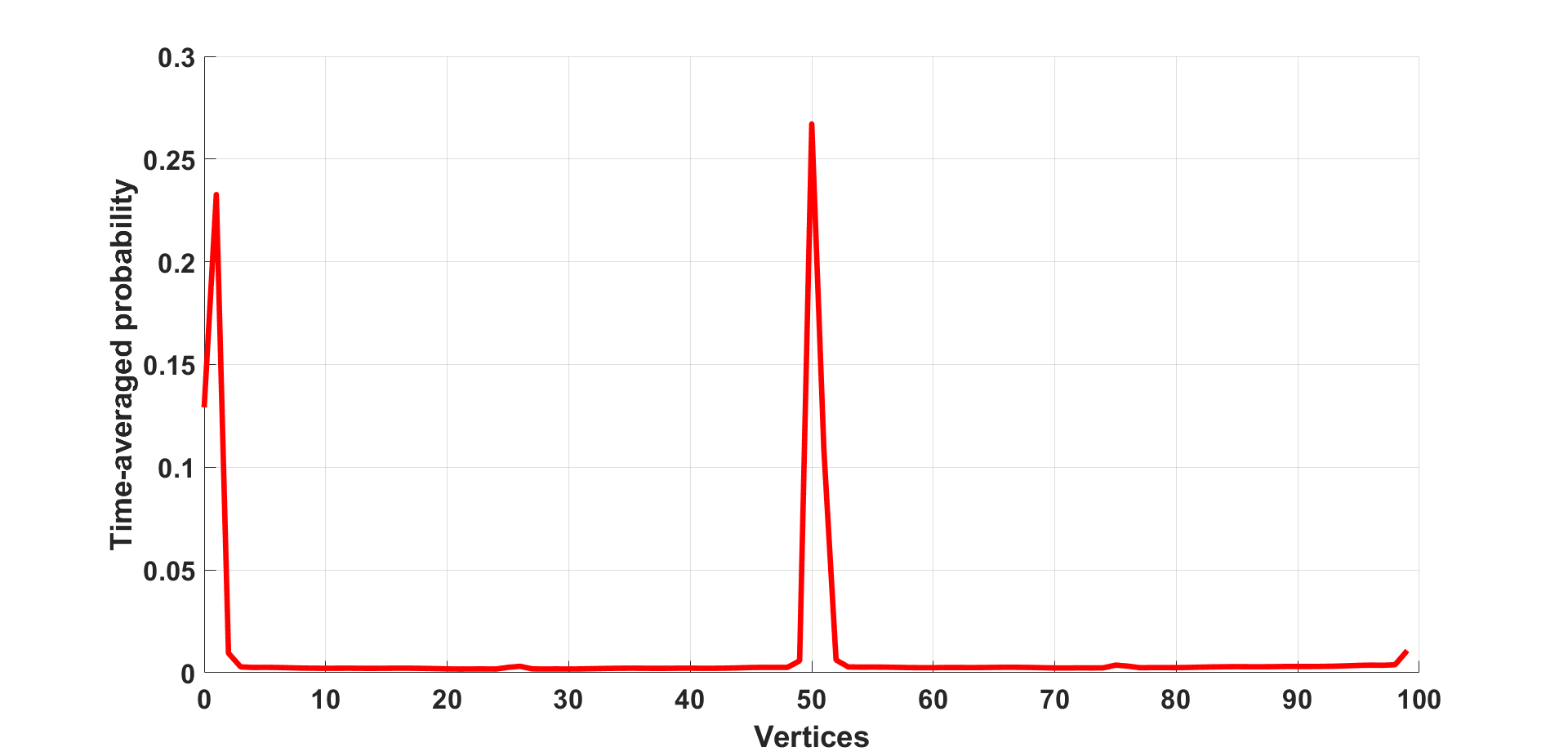}}\\
     \subfigure[$\ket{\psi_0}=\ket{2}_3\ket{1}_2\ket{0}_N$,$C\in \mathcal{Y}_\theta, \theta=\pi/2$]{\includegraphics[height=3.5 cm,width=7 cm]{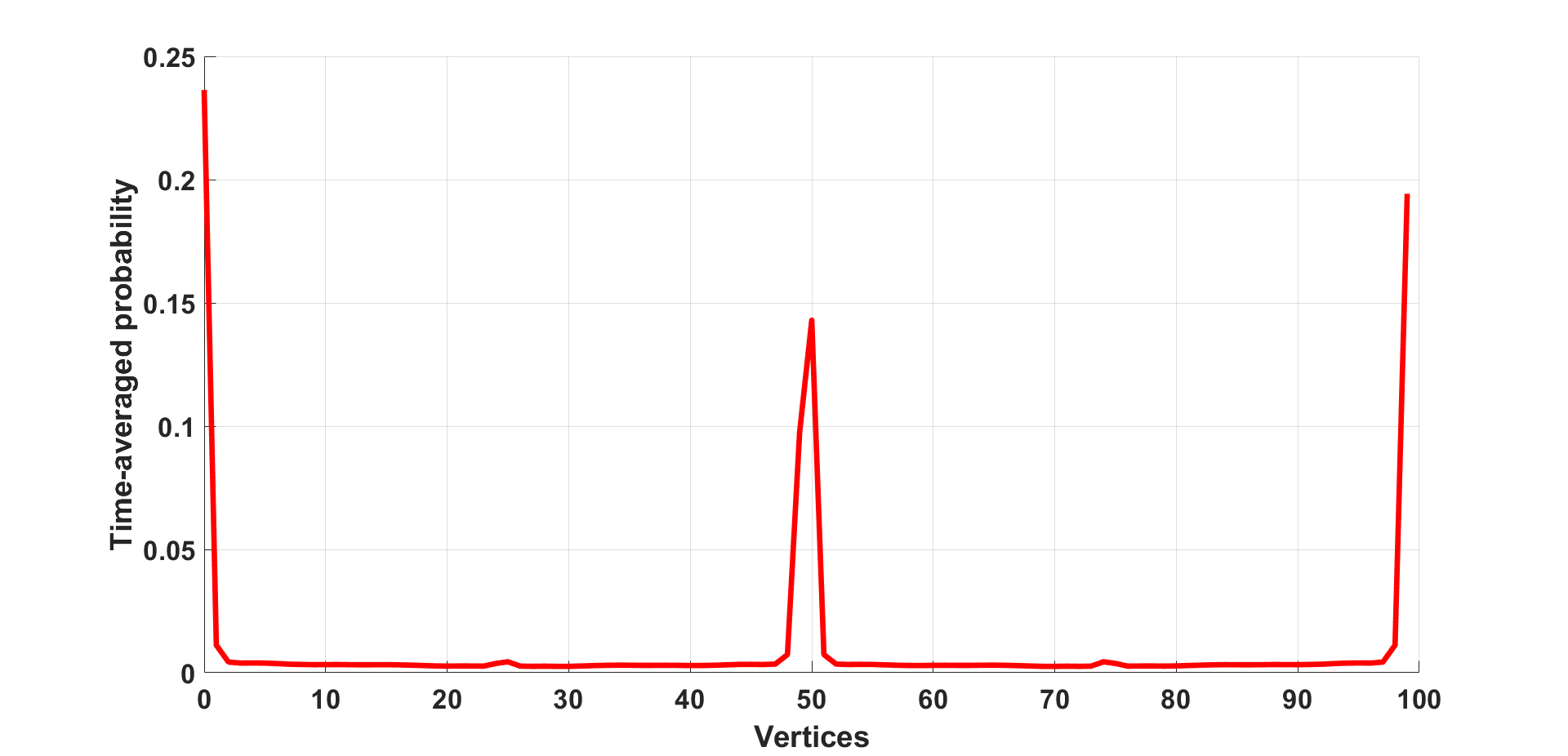}}
     \subfigure[$\ket{\psi_0}\frac{1}{\sqrt{3}}(\ket{0}_3+\ket{1}_3+\ket{2}_3)\ket{1}_2\ket{0}_N$,$C\in \mathcal{Y}_\theta, \theta=\pi/2$ ]{\includegraphics[height=3.5 cm,width=7 cm]{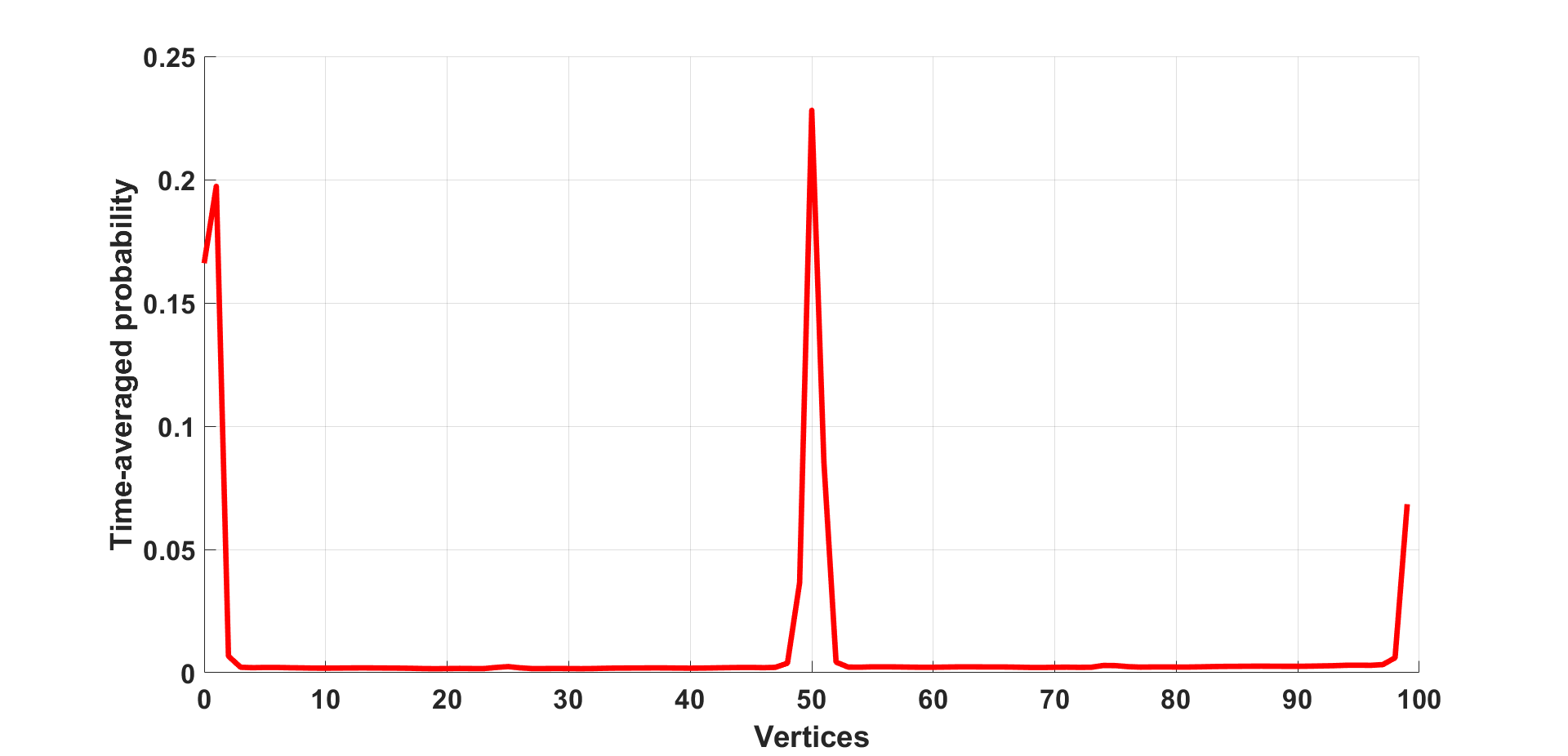}}
    \caption{The time-averaged probability curve for $T=300$ of the proposed walk on $\mathrm{Cay(D_{50},\{a,b\})}$ with the Grover coin and initial position $(1,0)$ and coins taken from $\mathcal{X}_\theta,\mathcal{Y}_\theta$.}\label{fig:timeavD50X180}
\end{figure}

\begin{figure}[H]
    \centering
    \subfigure[$\ket{\psi_0}=\ket{0}_3\ket{1}_2\ket{0}_N$,$C\in \mathcal{Z}_\theta, \theta=\pi/3$]{\includegraphics[height=3.5 cm,width=7 cm]{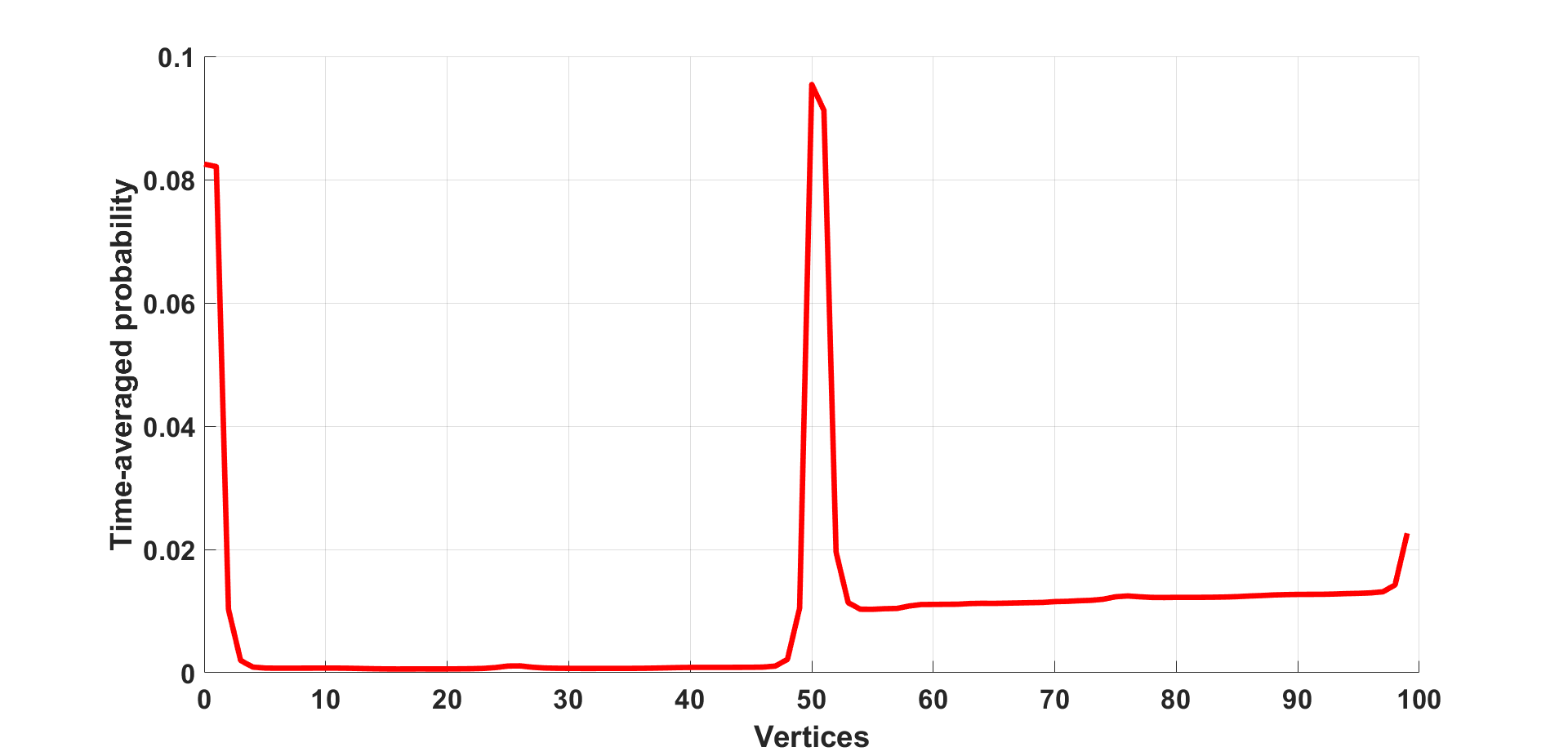}}
     \subfigure[$\ket{\psi_0}=\ket{1}_3\ket{1}_2\ket{0}_N$,$C\in \mathcal{Z}_\theta, \theta=\pi/3$]{\includegraphics[height=3.5 cm,width=7 cm]{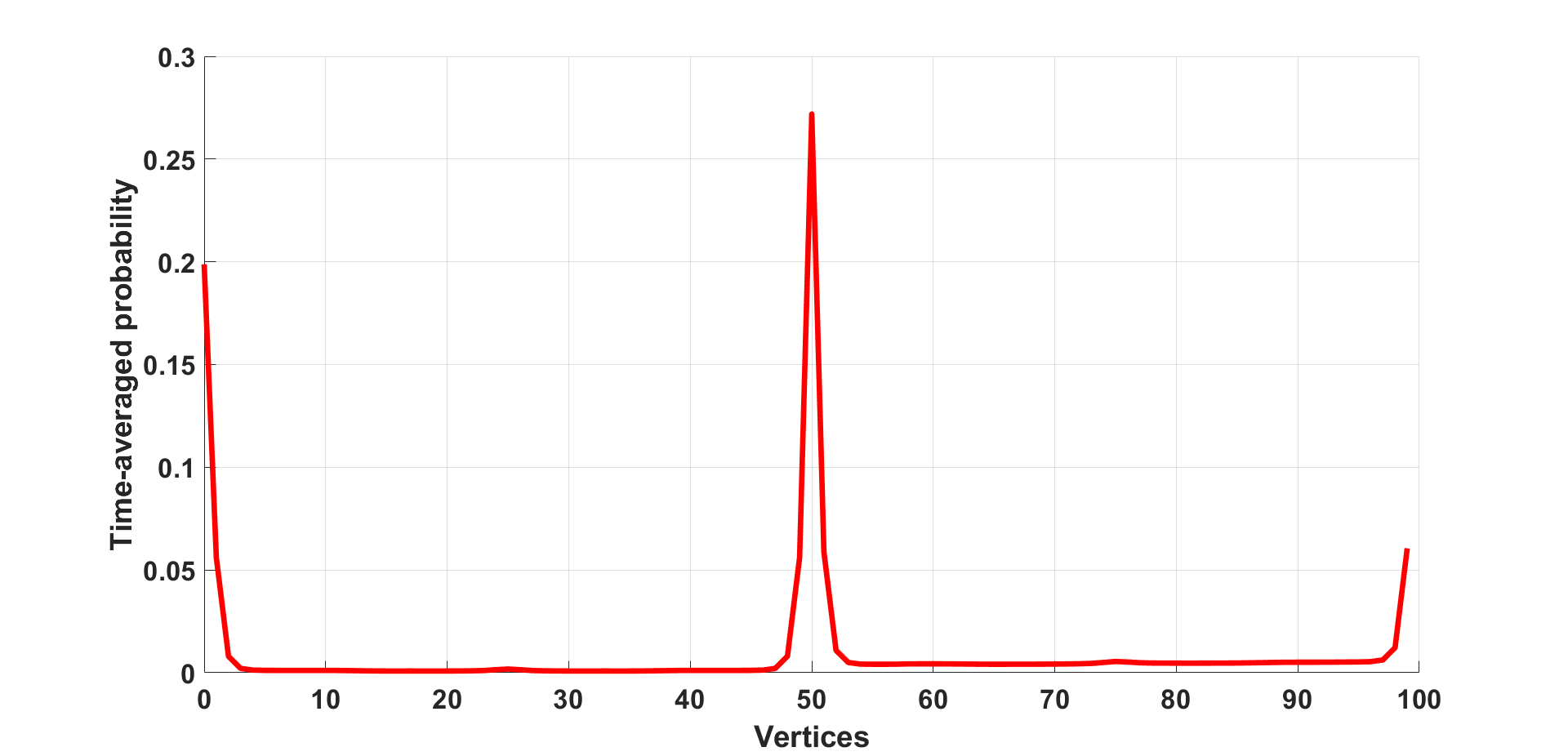}}\\
     \subfigure[$\ket{\psi_0}=\ket{2}_3\ket{1}_2\ket{0}_N$,$C\in \mathcal{Z}_\theta, \theta=\pi/3$]{\includegraphics[height=3.5 cm,width=7 cm]{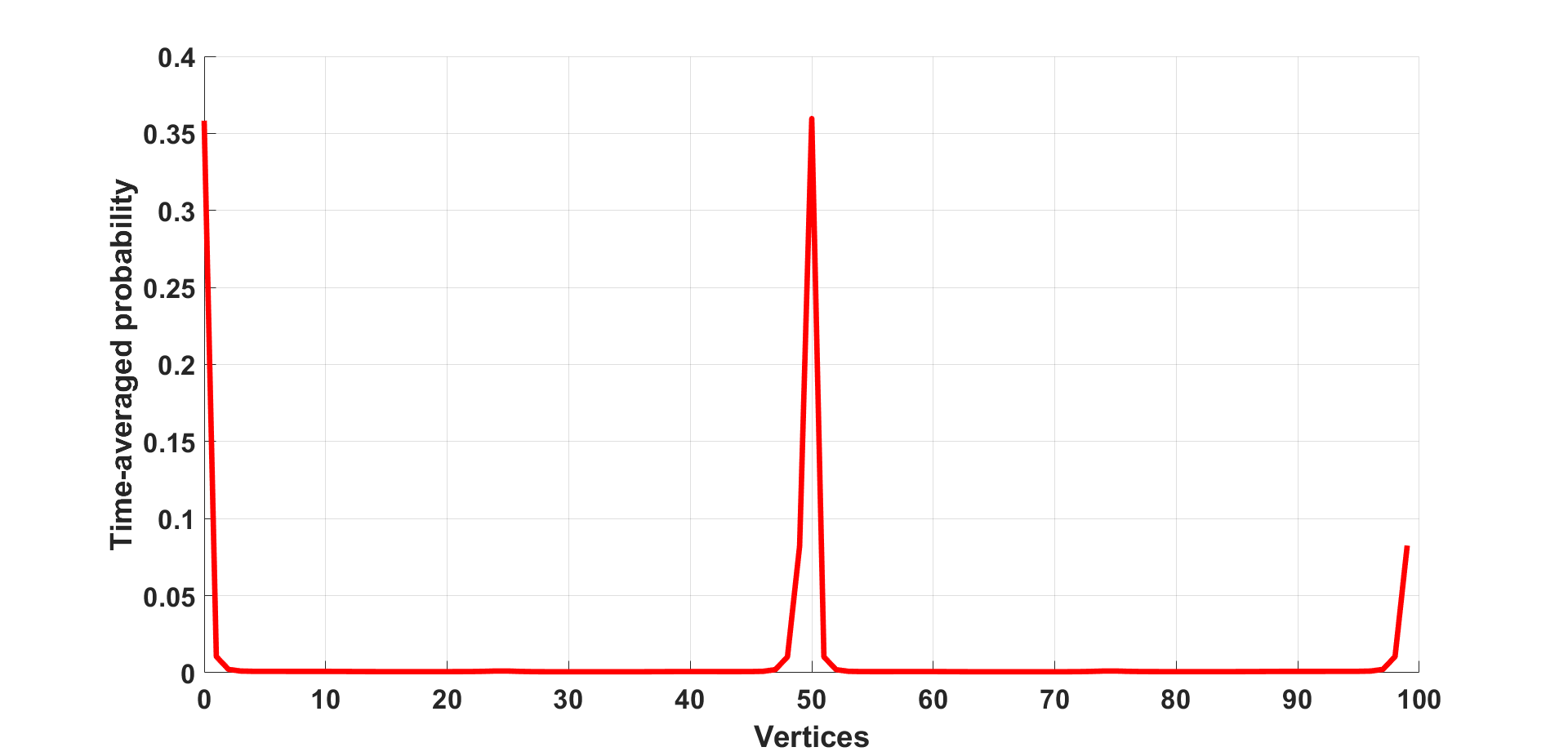}}
     \subfigure[$\ket{\psi_0}=\frac{1}{\sqrt{3}}(\ket{0}_3+\ket{1}_3+\ket{2}_3)\ket{1}_2\ket{0}_N$,$C\in \mathcal{Z}_\theta, \theta=\pi/3$ ]{\includegraphics[height=3.5 cm,width=7 cm]{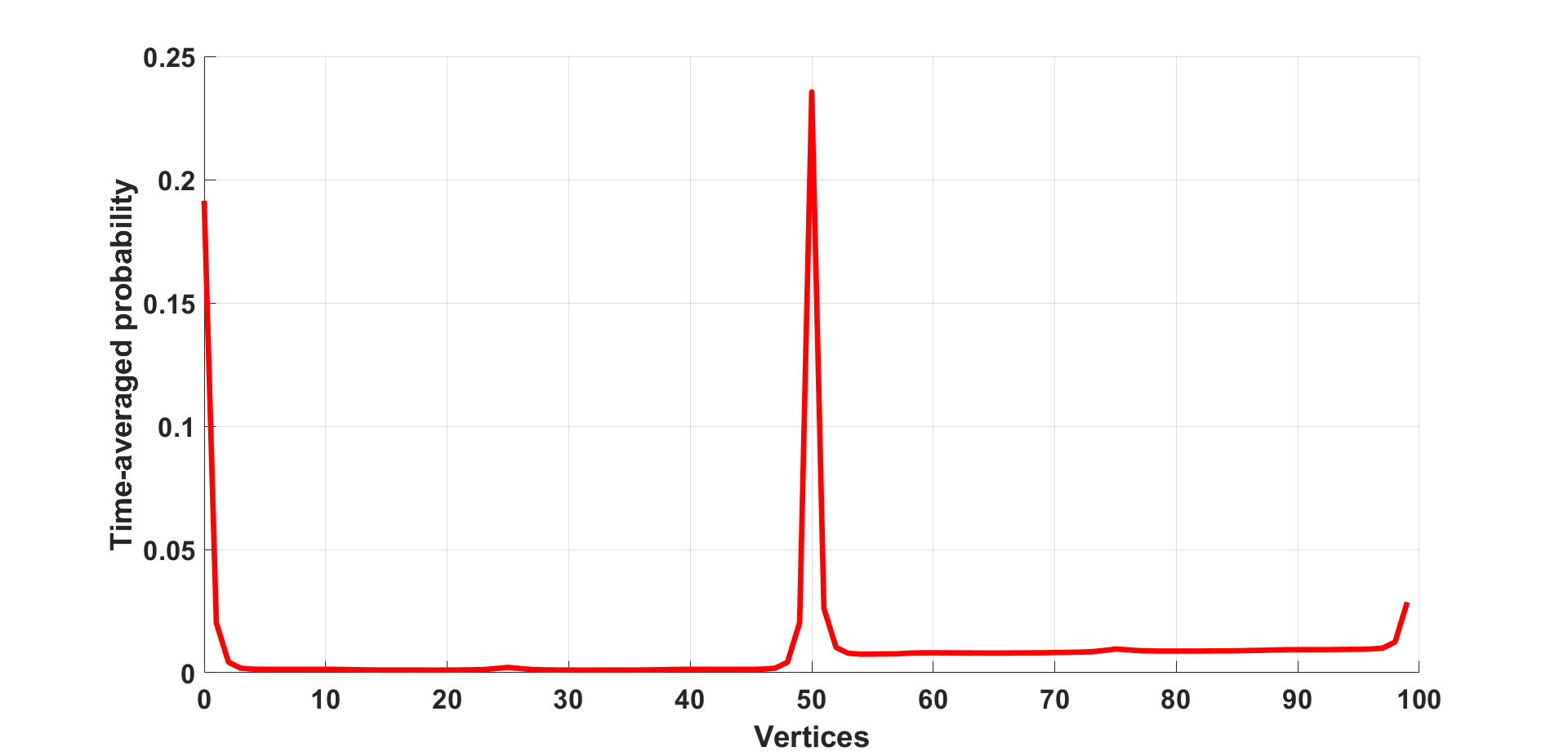}}
   \subfigure[$\ket{\psi_0}=\ket{0}_3\ket{1}_2\ket{0}_N$,$C\in \mathcal{W}_\theta, \theta=-\pi/4$]{\includegraphics[height=3.5 cm,width=7 cm]{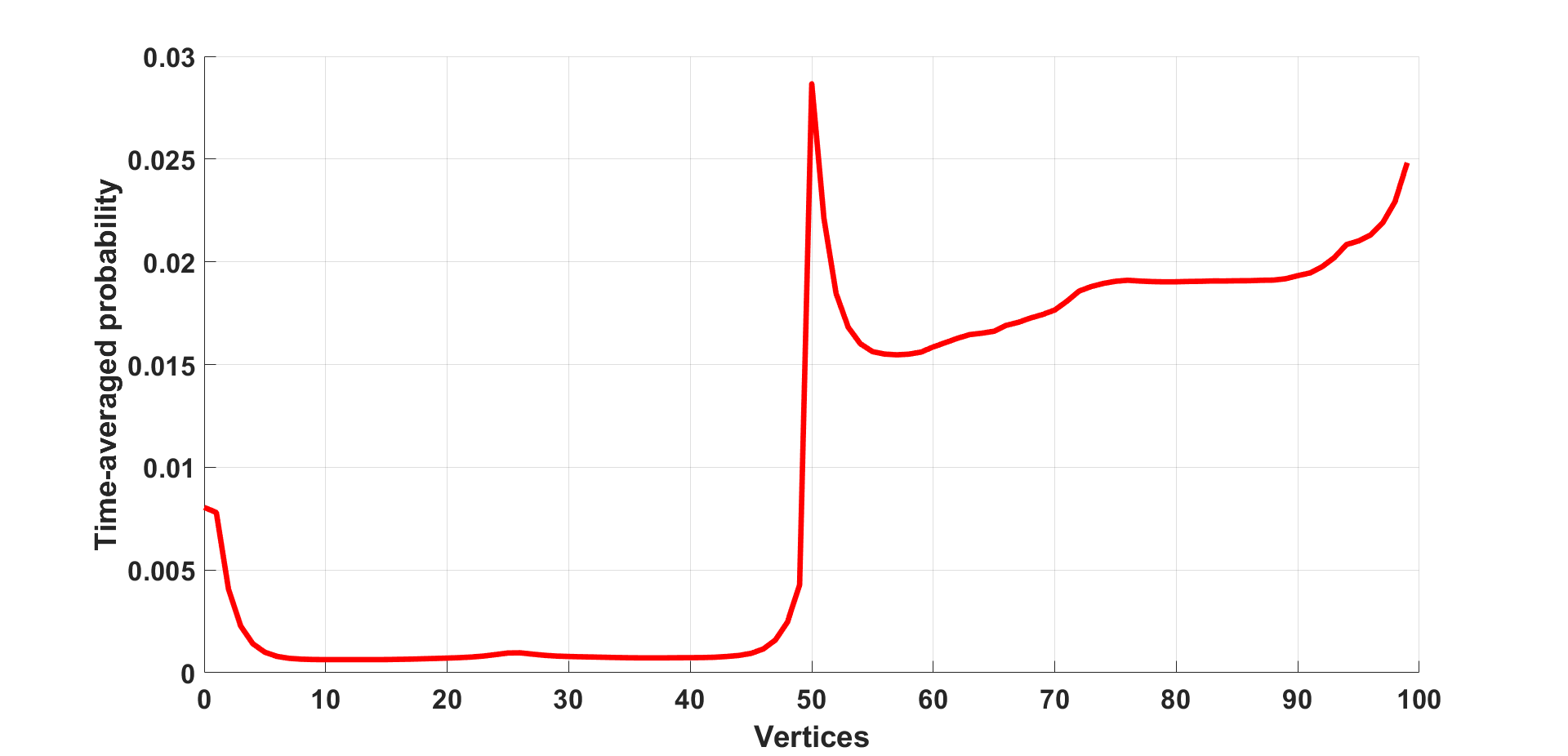}}
     \subfigure[$\ket{\psi_0}=\ket{1}_3\ket{1}_2\ket{0}_N$,$C\in \mathcal{W}_\theta, \theta=-\pi/4$]{\includegraphics[height=3.5 cm,width=7 cm]{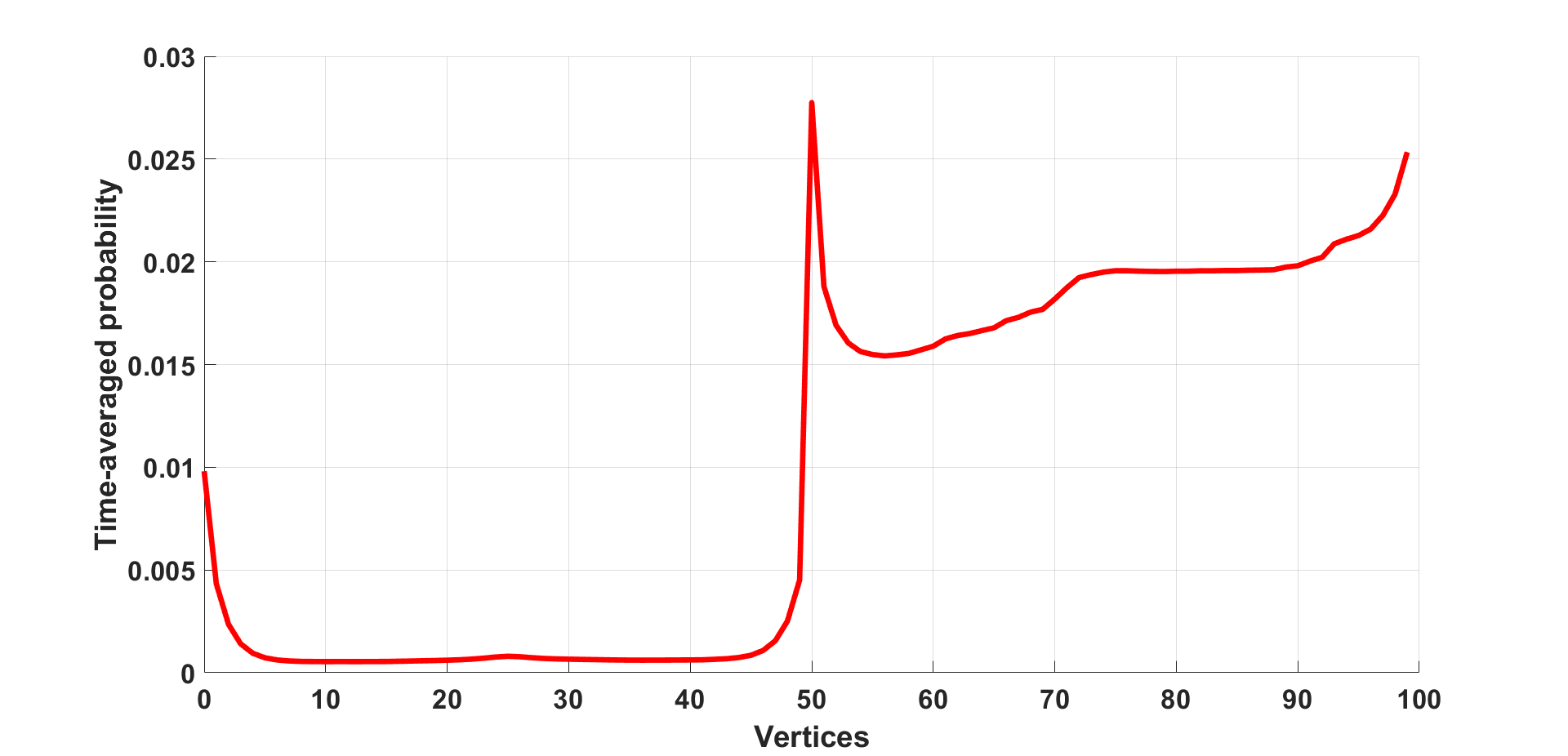}}\\
     \subfigure[$\ket{\psi_0}=\ket{2}_3\ket{1}_2\ket{0}_N$,$C\in \mathcal{W}_\theta, \theta=-\pi/4$]{\includegraphics[height=3.5 cm,width=7 cm]{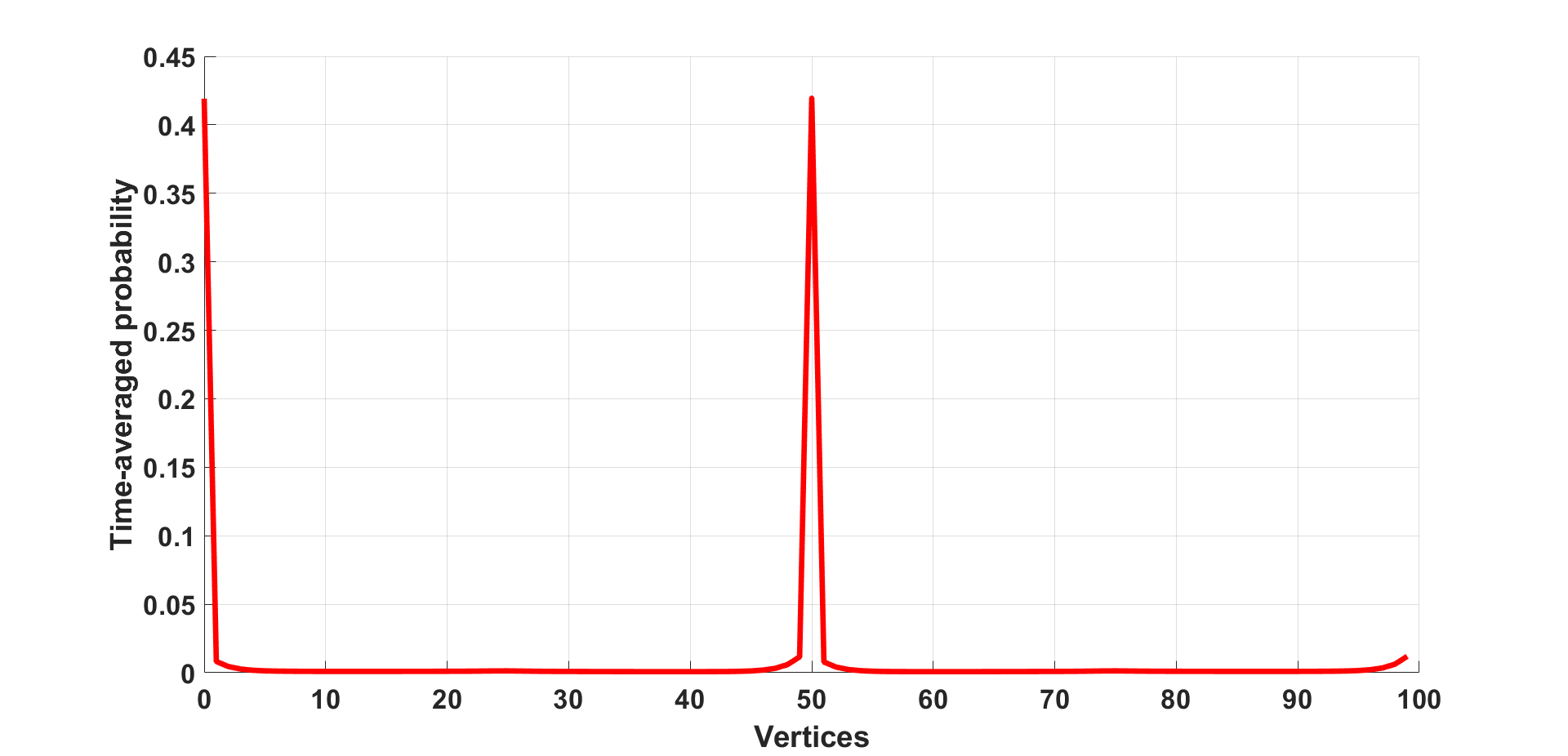}}
     \subfigure[$\ket{\psi_0}=\frac{1}{\sqrt{3}}(\ket{0}_3+\ket{1}_3+\ket{2}_3)\ket{1}_2\ket{0}_N$,$C\in \mathcal{W}_\theta, \theta=-\pi/4$ ]{\includegraphics[height=3.5 cm,width=7 cm]{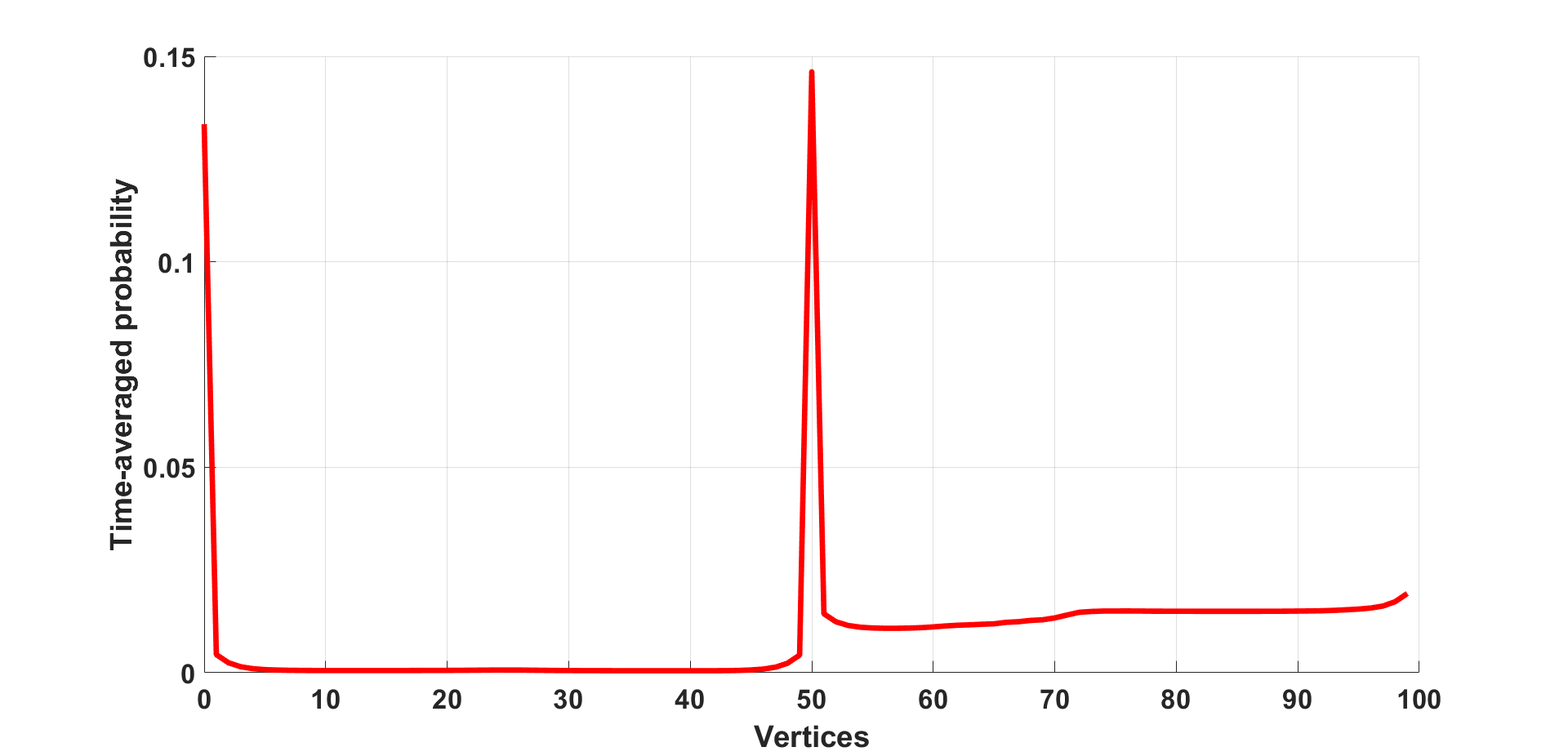}}
    \caption{The time-averaged probability curve for $T=300$ of the proposed walk on $\mathrm{Cay(D_{50},\{a,b\})}$ with the Grover coin and initial position $(1,0)$ and coins taken from $\mathcal{Z}_\theta,\mathcal{W}_\theta$.}\label{fig:timeavD50Z60}
\end{figure}

In Figure \ref{fig:timeavD50X180}, we plot the time-averaged probability of the proposed walk with initial position $(1,0)$ at all the vertices of $\mathrm{Cay(D_N,\{a,b\})}$ for $N=50$, for different initial coin states when $T$ and the coin $C\in \mathcal{X}_\theta$ with $\theta=\pi$ i.e. the Grover coin $\mathsf{G}$ and $\mathcal{Y}_\theta$ with $\theta=\pi/2$. The probability curve shows that the walk localizes in all the cases and the probability values is maximum at the starting position and its reflection vertex. This observation is also reported in \cite{Liu2021}. A similar phenomena is also observed in Figure \ref{fig:timeavD50Z60},  taking coins from $\mathcal{Z}_\theta$ and $\mathcal{W}_\theta$ for $\theta=\pi/3$ and $-\pi/4$ respectively.

\begin{figure}[H]
    \centering
    \subfigure[$\overline{P(0,0,500)}, C\in \mathcal{X}_\theta$]{\includegraphics[height=3.5 cm,width=8 cm]{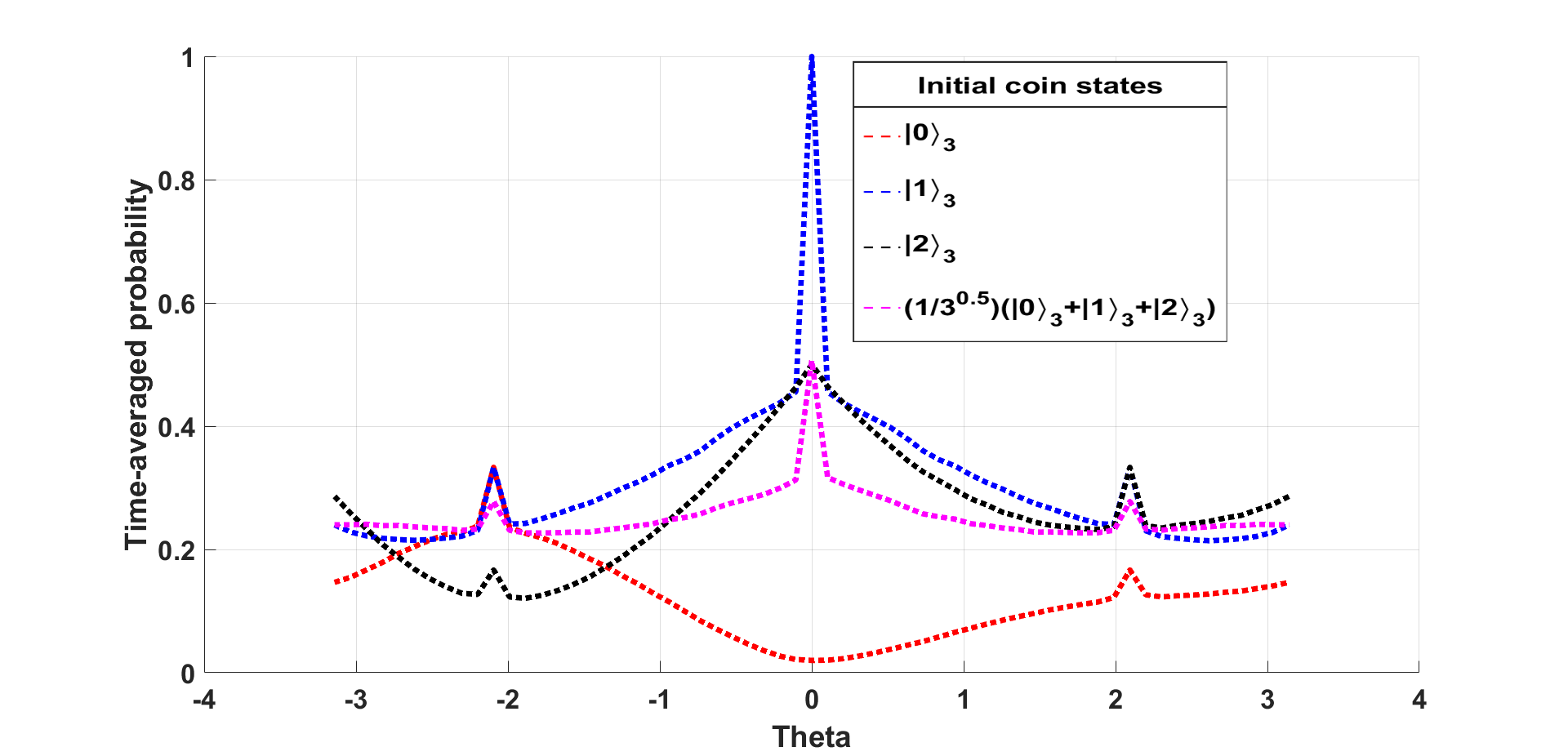}}
     \subfigure[$\overline{P(0,1,500)}, C\in \mathcal{X}_\theta$]{\includegraphics[height=3.5 cm,width=8 cm]{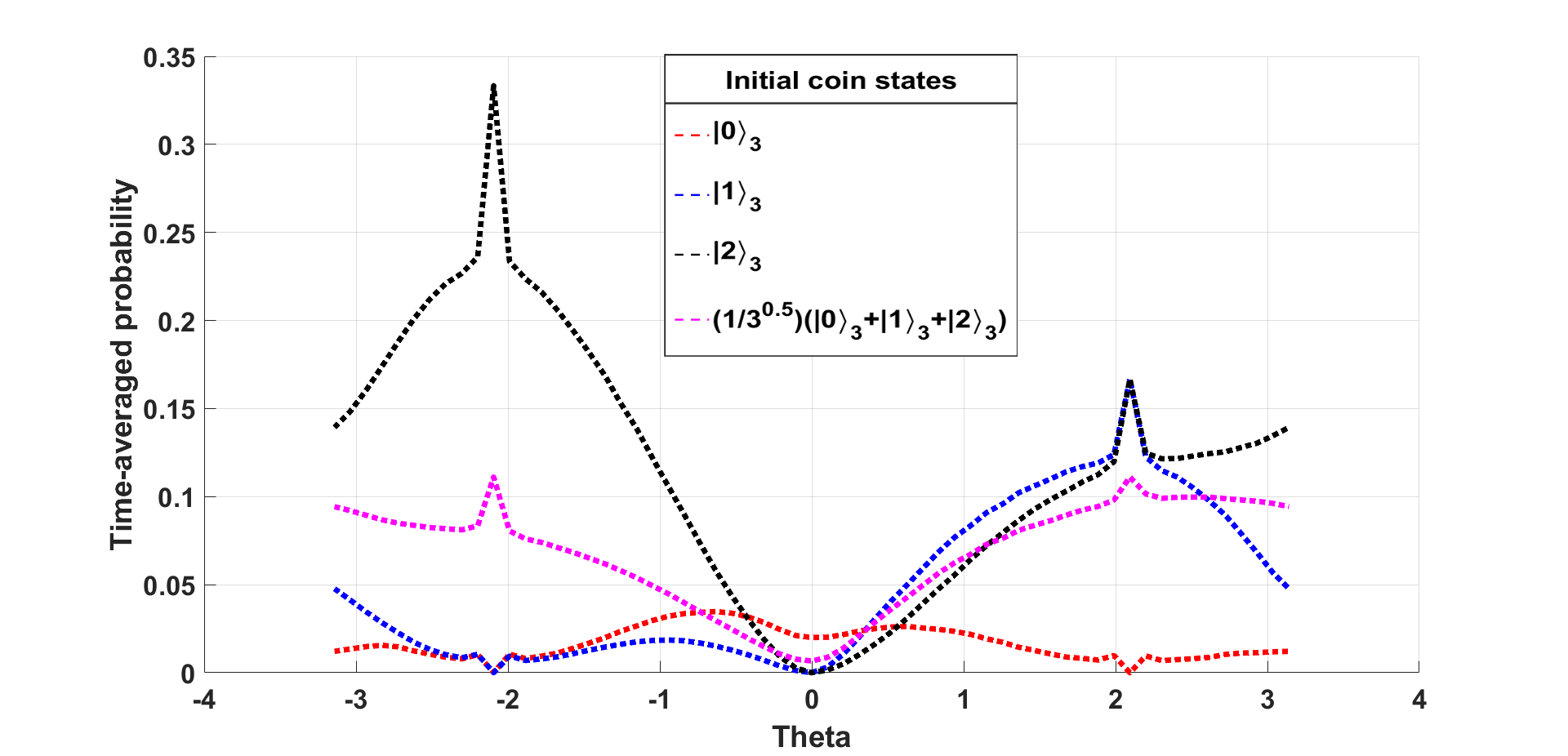}}\\
    \subfigure[$\overline{P(1,0,500)}, C\in \mathcal{X}_\theta$]{\includegraphics[height=3.5 cm,width=8 cm]{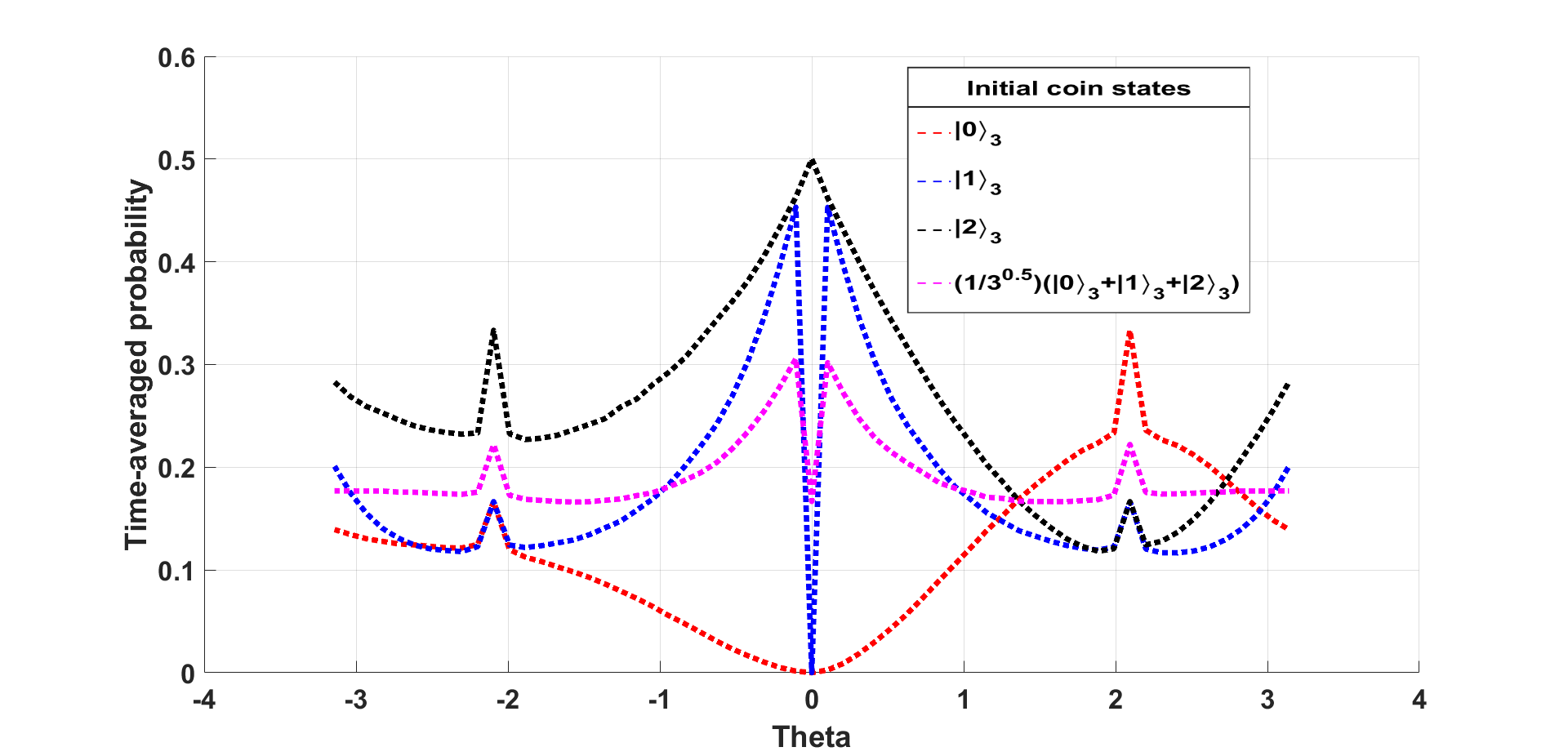}}
    \subfigure[$\overline{P(0,0,500)}, C\in \mathcal{Y}_\theta$]{\includegraphics[height=3.5 cm,width=8 cm]{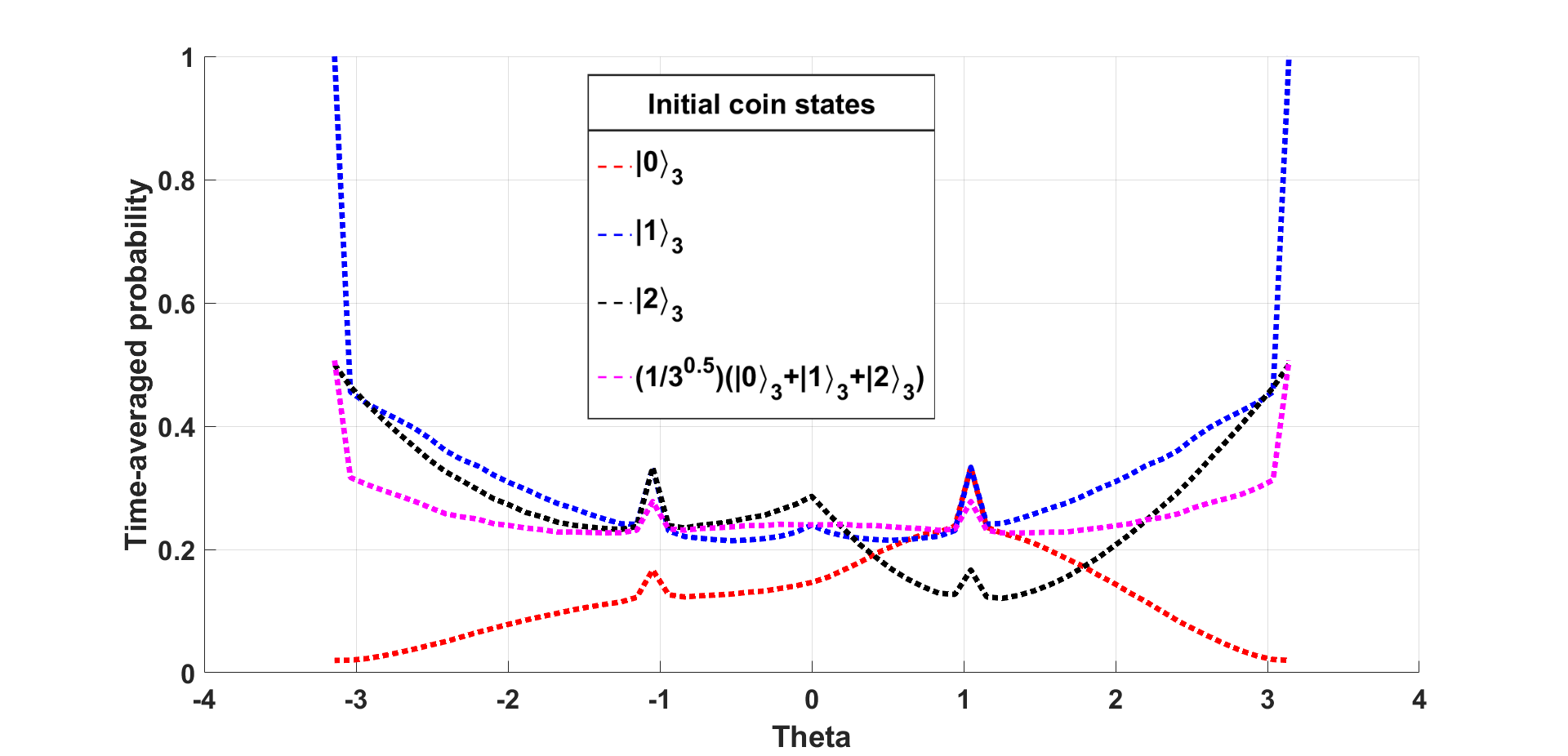}}
     \subfigure[$\overline{P(0,1,500)}, C\in \mathcal{Y}_\theta$]{\includegraphics[height=3.5 cm,width=8 cm]{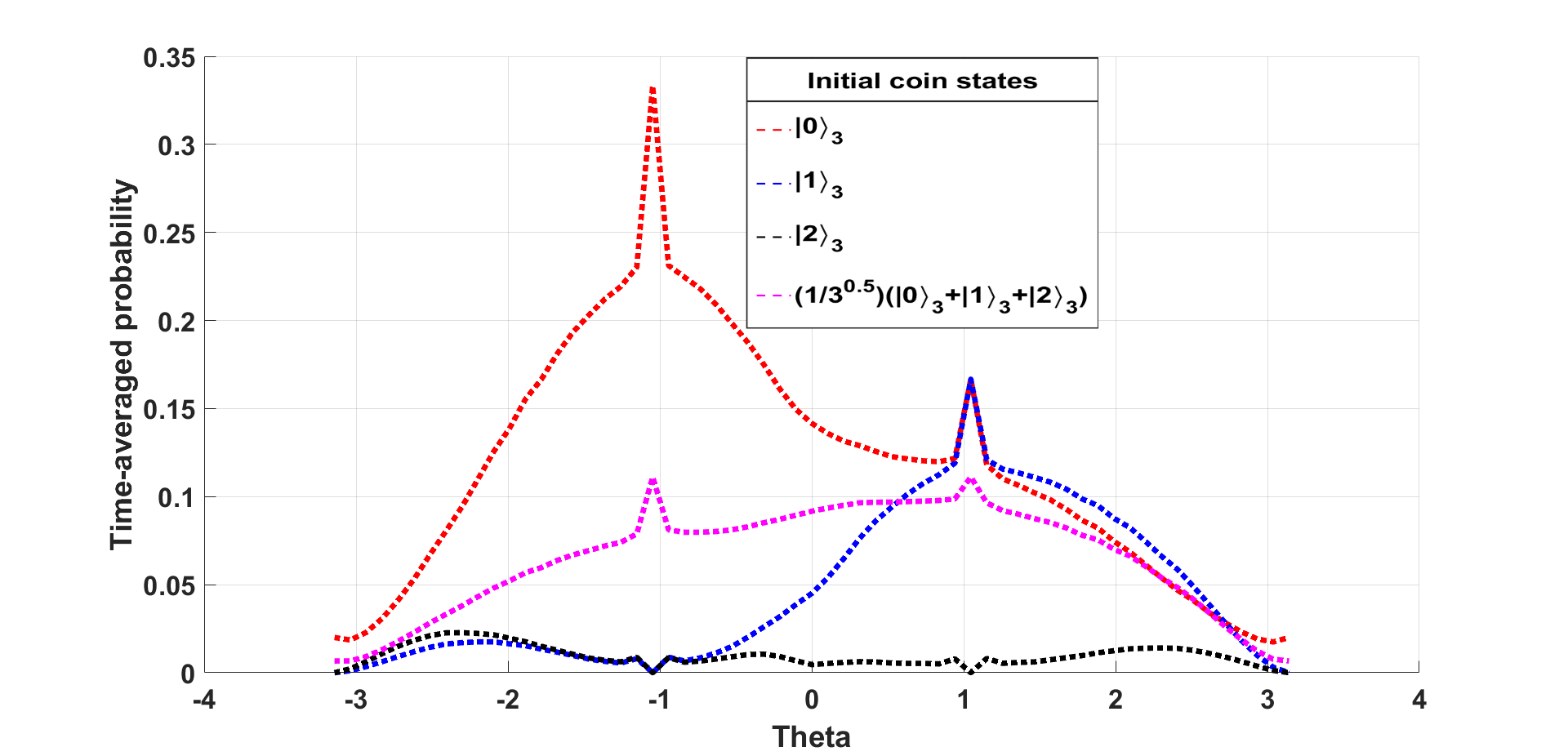}}     \subfigure[$\overline{P(1,0,500)}, C\in \mathcal{Y}_\theta$]{\includegraphics[height=3.5 cm,width=8 cm]{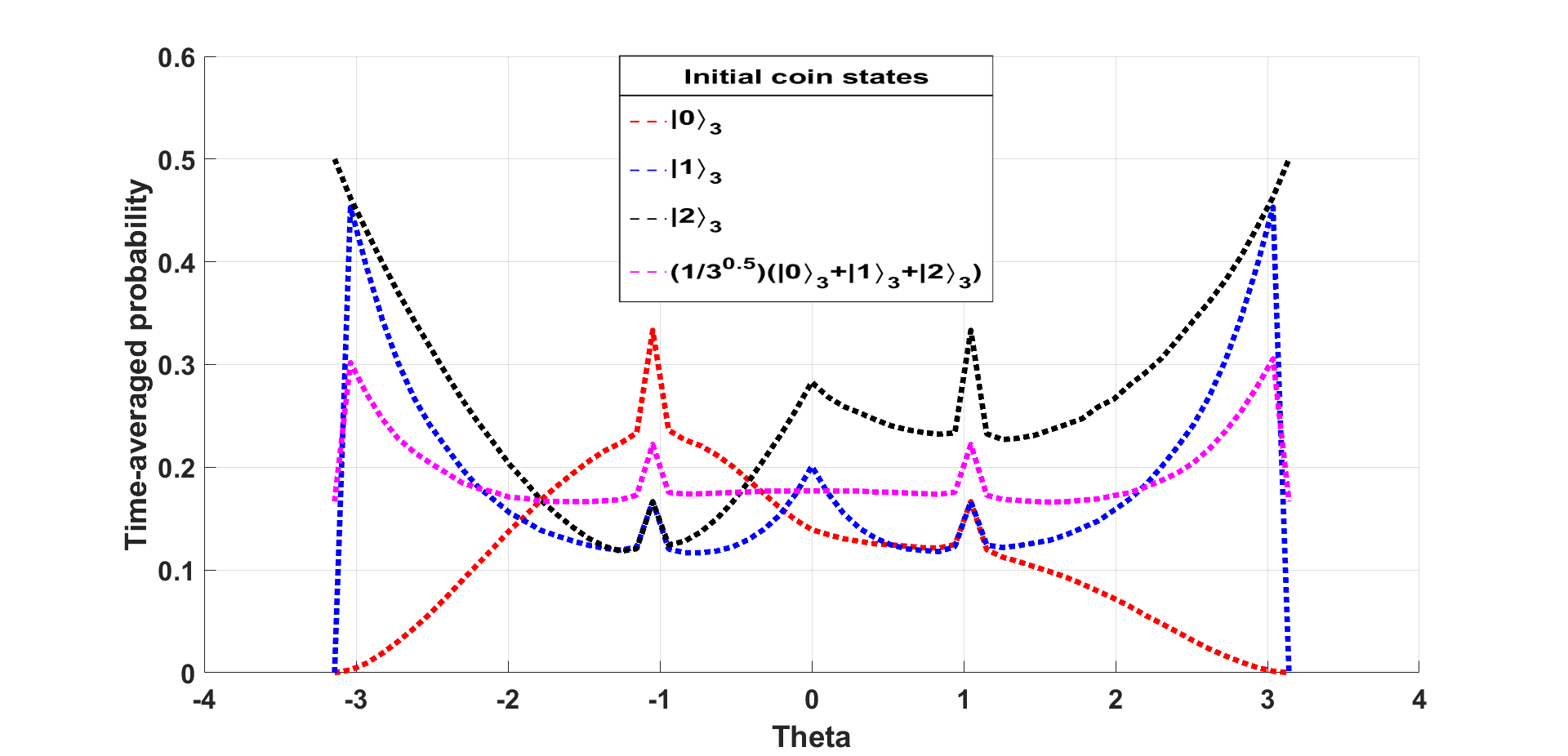}}

    \caption{ Time averaged probability of the walk  at different vertices on $\mathrm{Cay(D_{50},\{a,b\})}$ when the coin $C\in \{\mathcal{X}_\theta,\mathcal{Y}_\theta\}$. The time steps are taken up to $500$.}\label{fig:ThetavariesX}
\end{figure}

Now, in Figure \ref{fig:ThetavariesX} we plot the time-averaged probability at various positions of the walk on $\mathrm{Cay(D_{50},\{a,b\})}$ when the coin is taken from the classes $\mathcal{X}_\theta,\mathcal{Y}_\theta$ for several values of $\theta$ obtained by discretizing the interval $[-\pi,\pi]$ into $60$ equidistant points for various initial coin states. The initial position of the walker is taken to be $(0,0)$. We observe that the probability value is positive with the peaks and crests of time-averaged probability found mainly at permutation matrices.  For the coins taken from $\mathcal{X}_\theta$, the large and smaller sharp peaks (indicating maxima) along with the lower troughs (indicating minima) are found at $\theta\in \{0,\pm\frac{2\pi}{3}\}$ i.e. when the coin is a permutation matrix. For the coins from $\mathcal{Y}_\theta$, such maxima and minima are found at $\theta\in \{\pm \frac{\pi}{3},\pm \pi\}$ which again gives permutation matrices.

\begin{figure}[H]
    \centering
    \subfigure[$\overline{P(0,0,500)}, C\in \mathcal{Z}_\theta$]{\includegraphics[height=3.5 cm,width=8 cm]{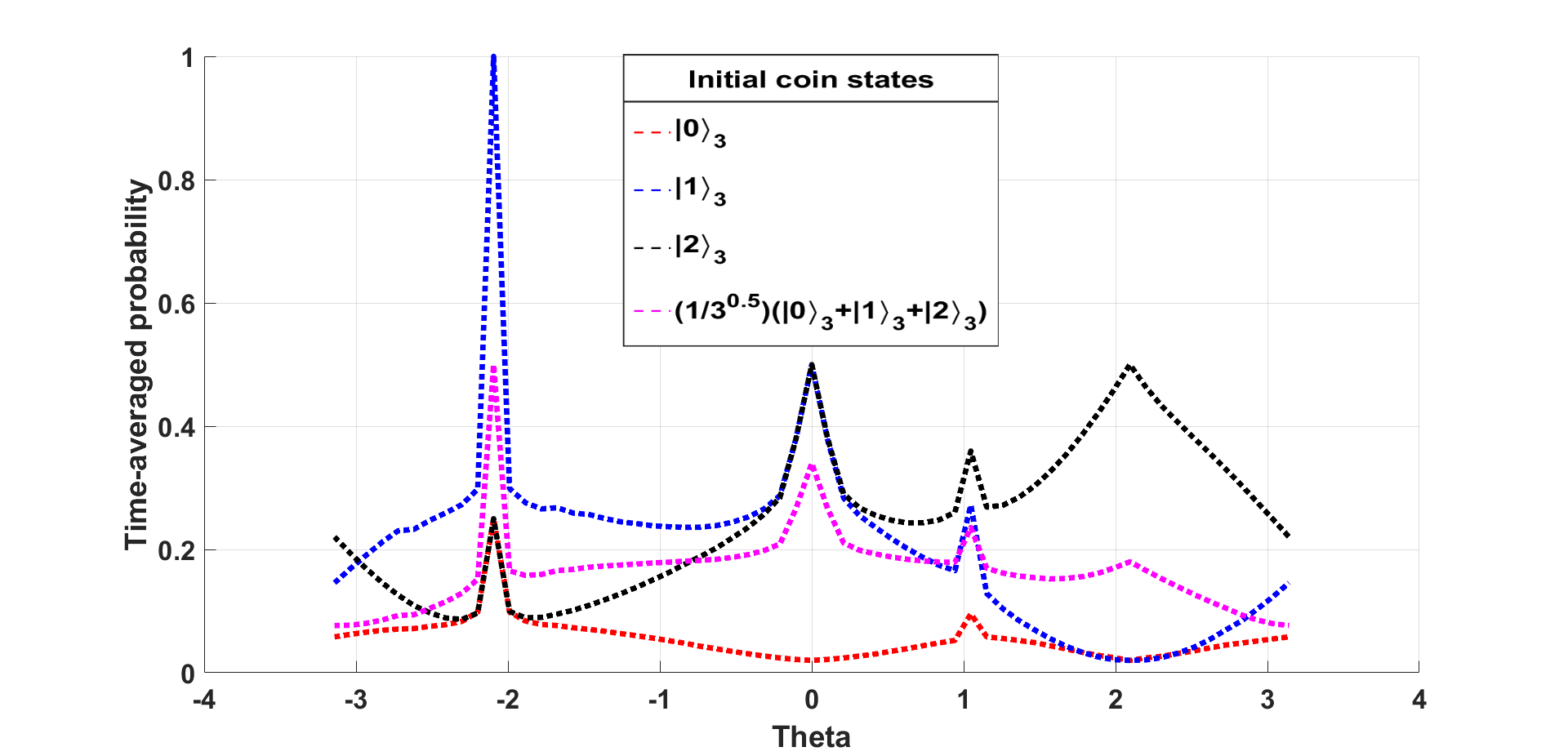}}
     \subfigure[$\overline{P(0,1,500)}, C\in \mathcal{Z}_\theta$]{\includegraphics[height=3.5 cm,width=8 cm]{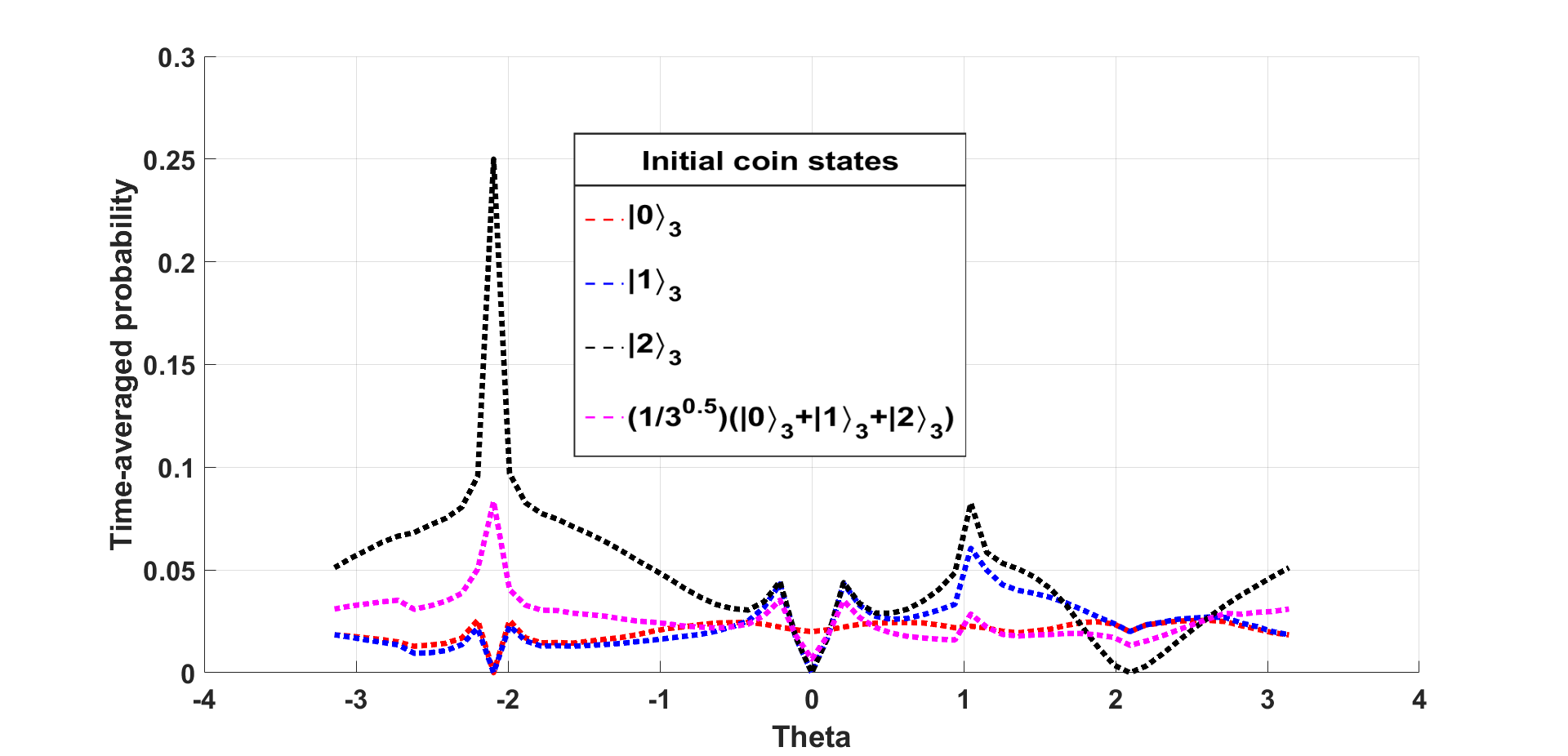}}\\
     \subfigure[$\overline{P(1,0,500)}, C\in \mathcal{Z}_\theta$]{\includegraphics[height=3.5 cm,width=8 cm]{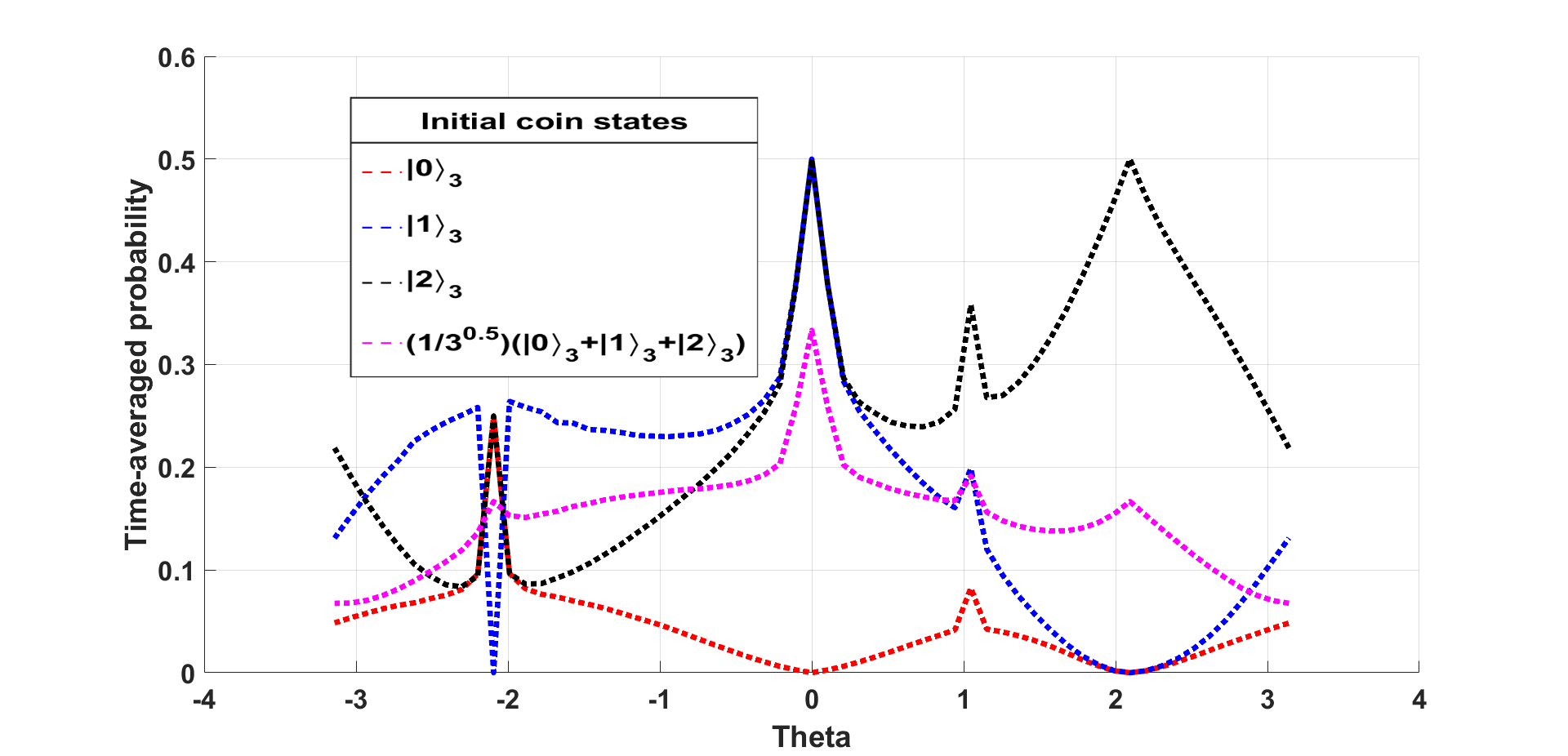}}
         \subfigure[$\overline{P(0,0,500)}, C\in \mathcal{W}_\theta$]{\includegraphics[height=3.5 cm,width=8 cm]{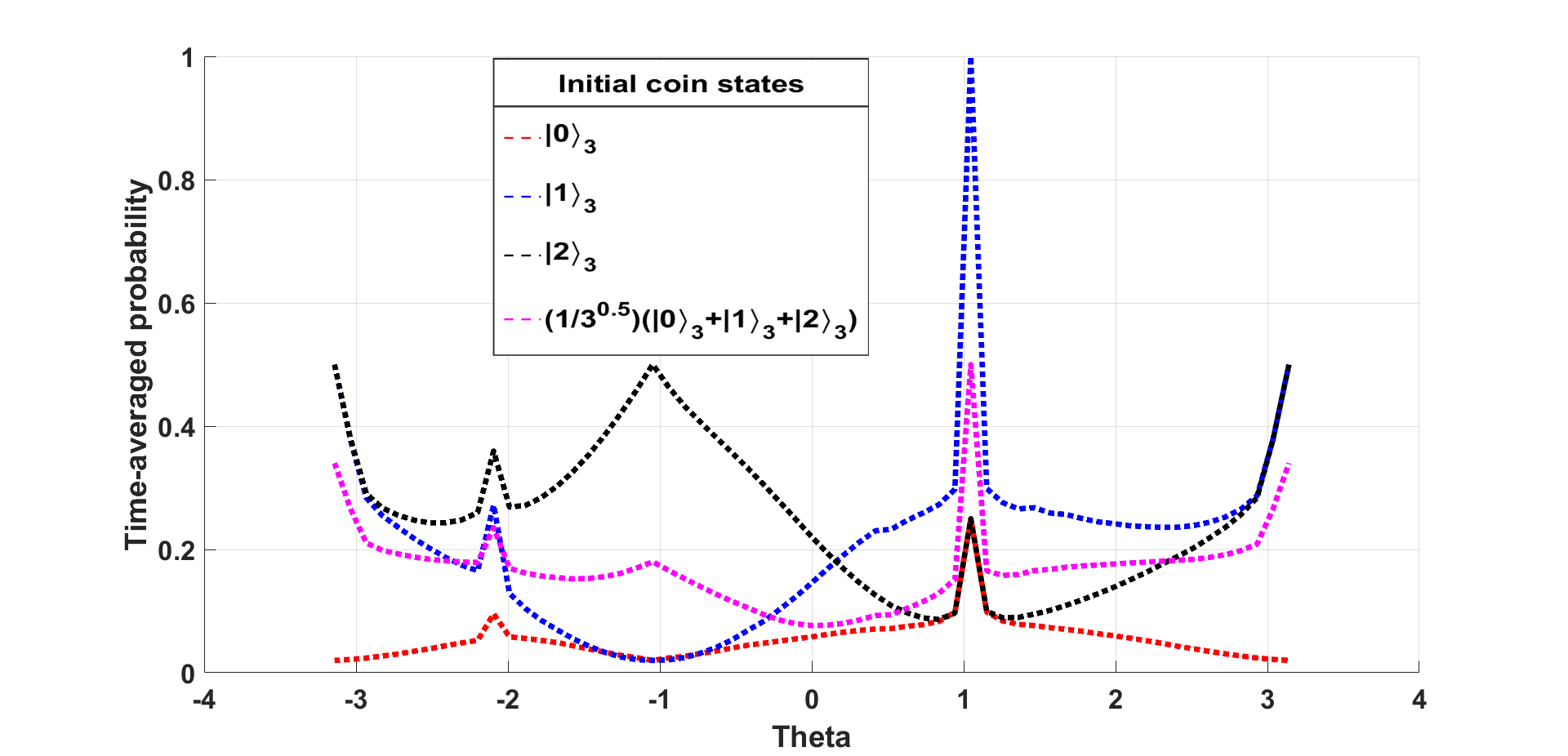}}
     \subfigure[$\overline{P(0,1,500)}, C\in \mathcal{W}_\theta$]{\includegraphics[height=3.5 cm,width=8 cm]{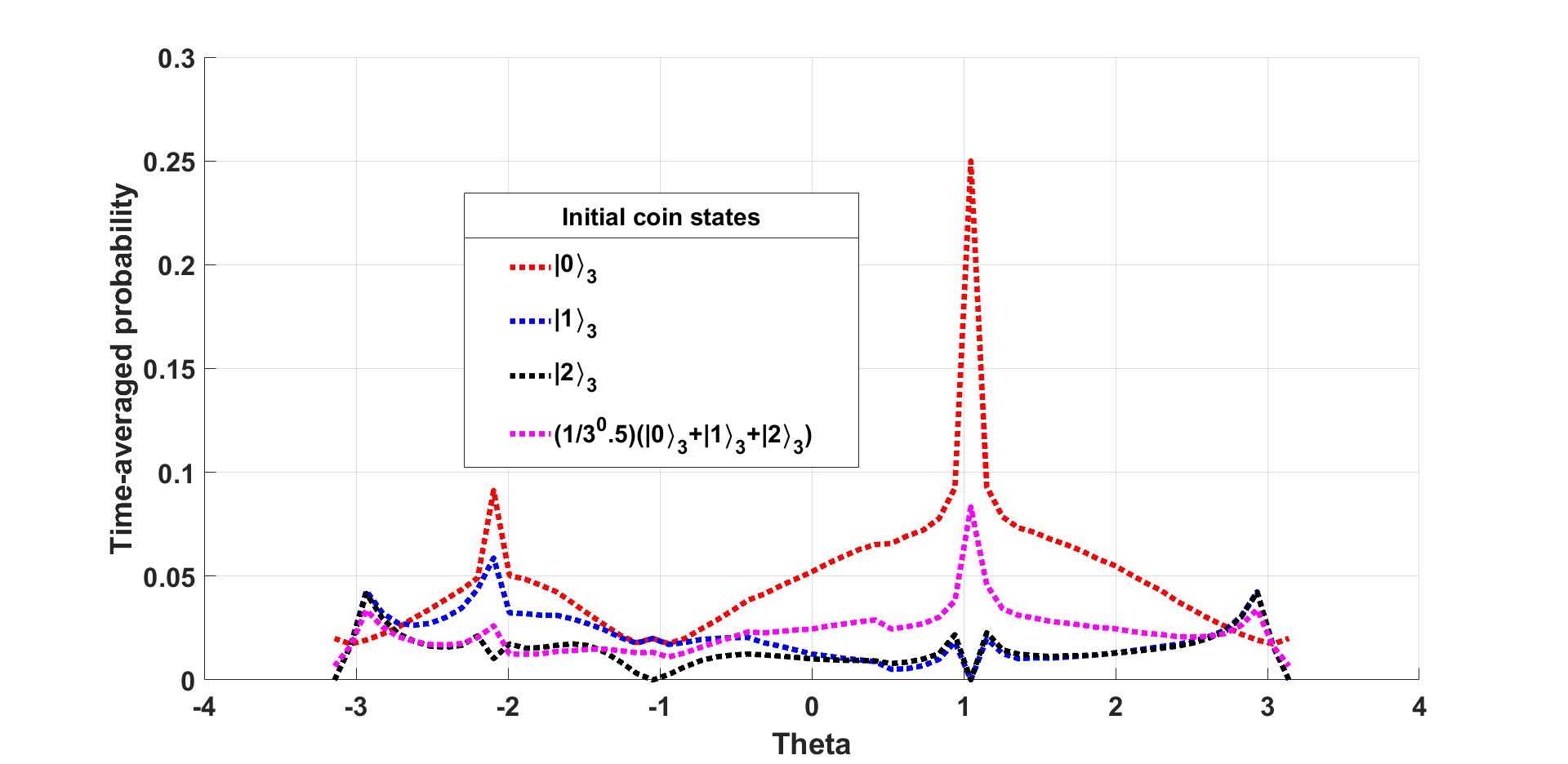}}
     \subfigure[$\overline{P(1,0,500)}, C\in \mathcal{W}_\theta$]{\includegraphics[height=3.5 cm,width=8 cm]{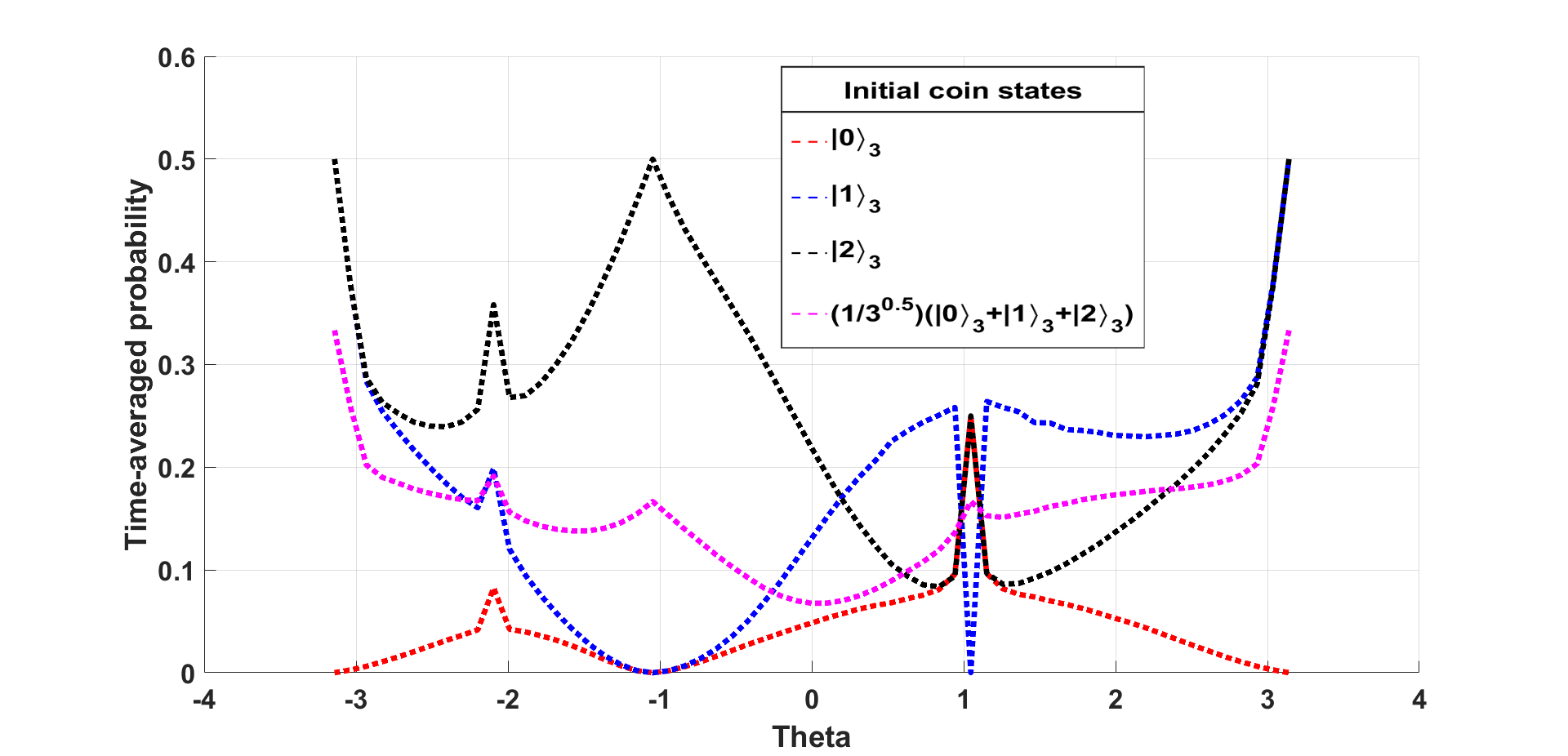}}

    \caption{ Time averaged probability of the walk  at different vertices on $\mathrm{Cay(D_{50},\{a,b\})}$ when the coin $C\in \{\mathcal{Z}_\theta,\mathcal{W}_\theta\}$. The time steps are taken up to $500$.}\label{fig:ThetavariesZ}
\end{figure}

Then in Figure \ref{fig:ThetavariesZ} we plot the time-averaged values similar to the Figure \ref{fig:ThetavariesX} with initial position being at $(0,0)$. For the coins from $\mathcal{Z}_\theta$, the points of extrema are found for $\theta\in \{0,\pm\frac{2\pi}{3}\}$ or in other words, permutation matrices and similarly for the coins from $\mathcal{W}_\theta$ maxima and minima are found at $\theta\in \{\pm \frac{\pi}{3},\pm \pi\}$. In all of the cases, we see that the walk either does not show localization at the the said vertices mentioned or has the highest probability of localization in case of permutation matrices.

\begin{figure}[H]
    \centering
     \subfigure[$C\in \mathcal{X}_\theta,\theta=\pi$]{\includegraphics[height=3.5 cm,width=8 cm]{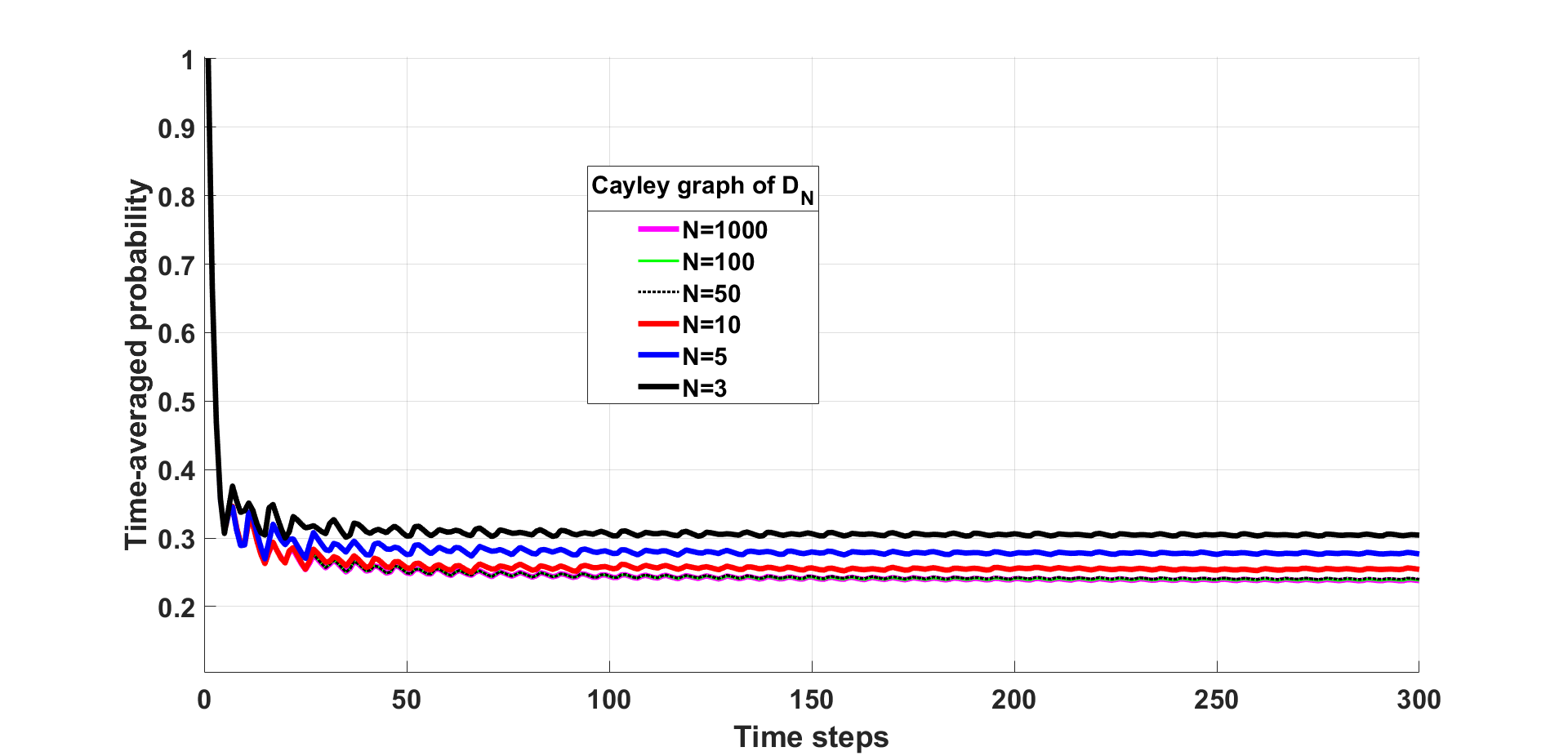}}
   \subfigure[$C\in \mathcal{X}_\theta,\theta=\pi/2$]{\includegraphics[height=3.5 cm,width=8 cm]{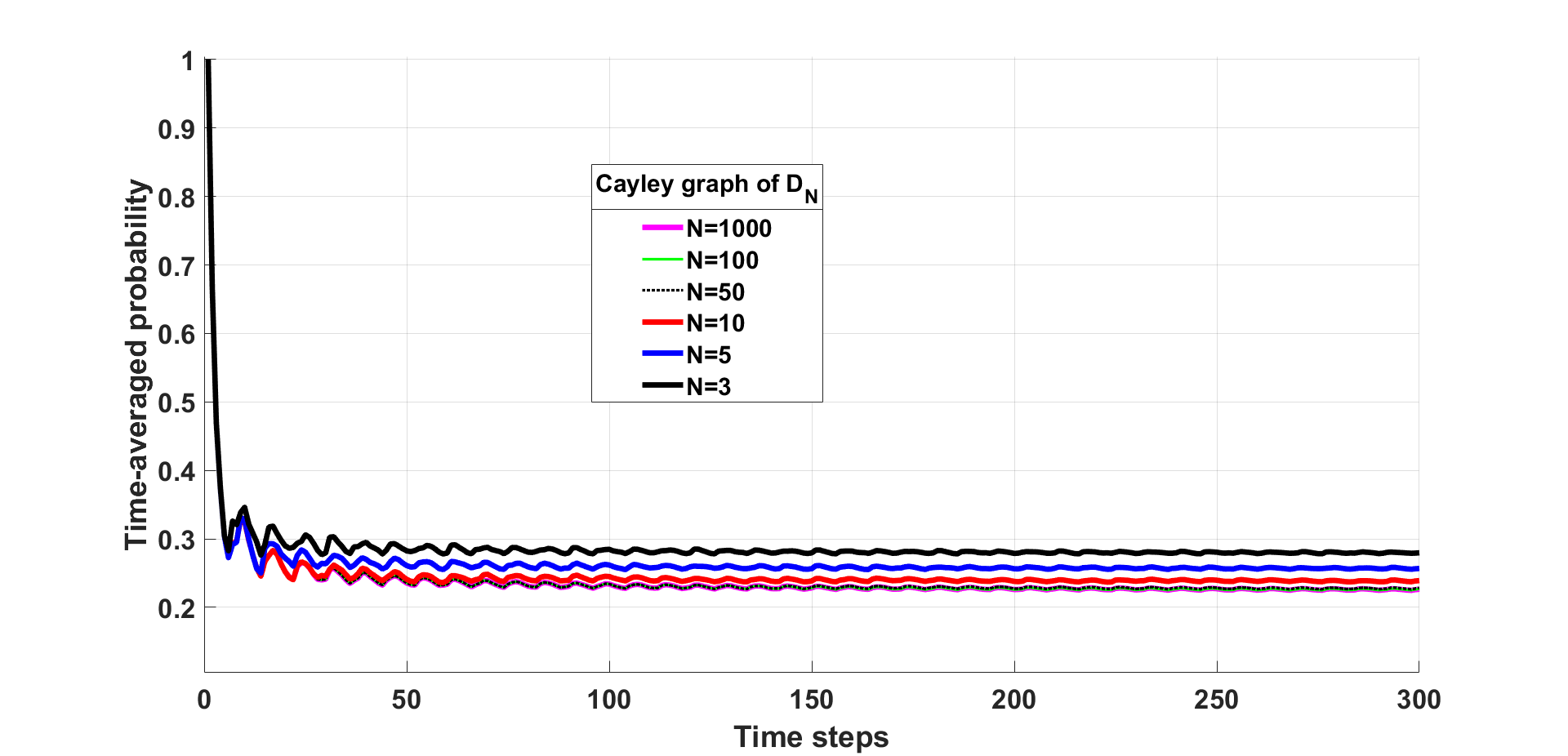}}
   \subfigure[$C\in \mathcal{Y}_\theta,\theta=\pi/5$]{\includegraphics[height=3.5 cm,width=8 cm]{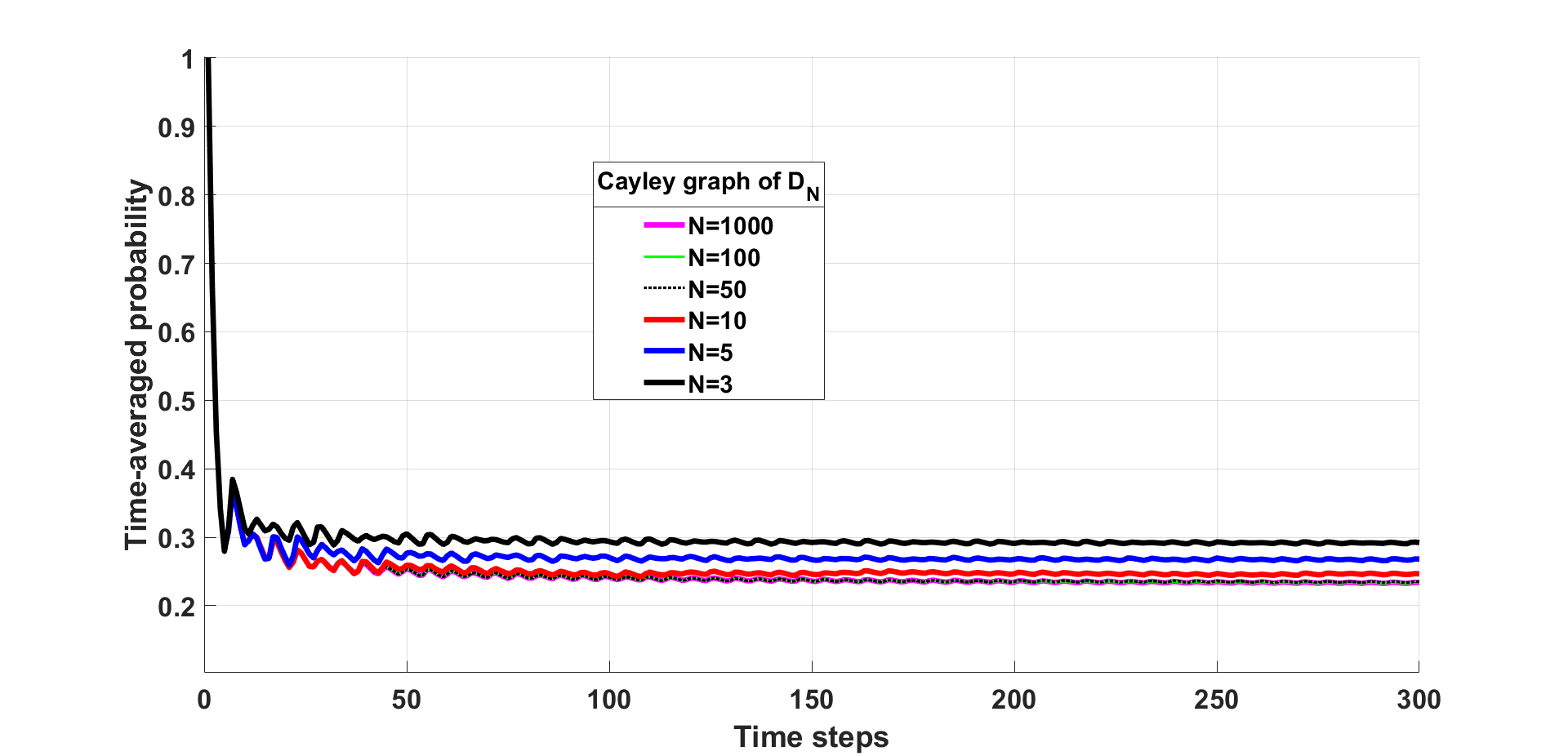}}
   \subfigure[$C\in \mathcal{Y}_\theta,\theta=-\pi/4$]{\includegraphics[height=3.5 cm,width=8 cm]{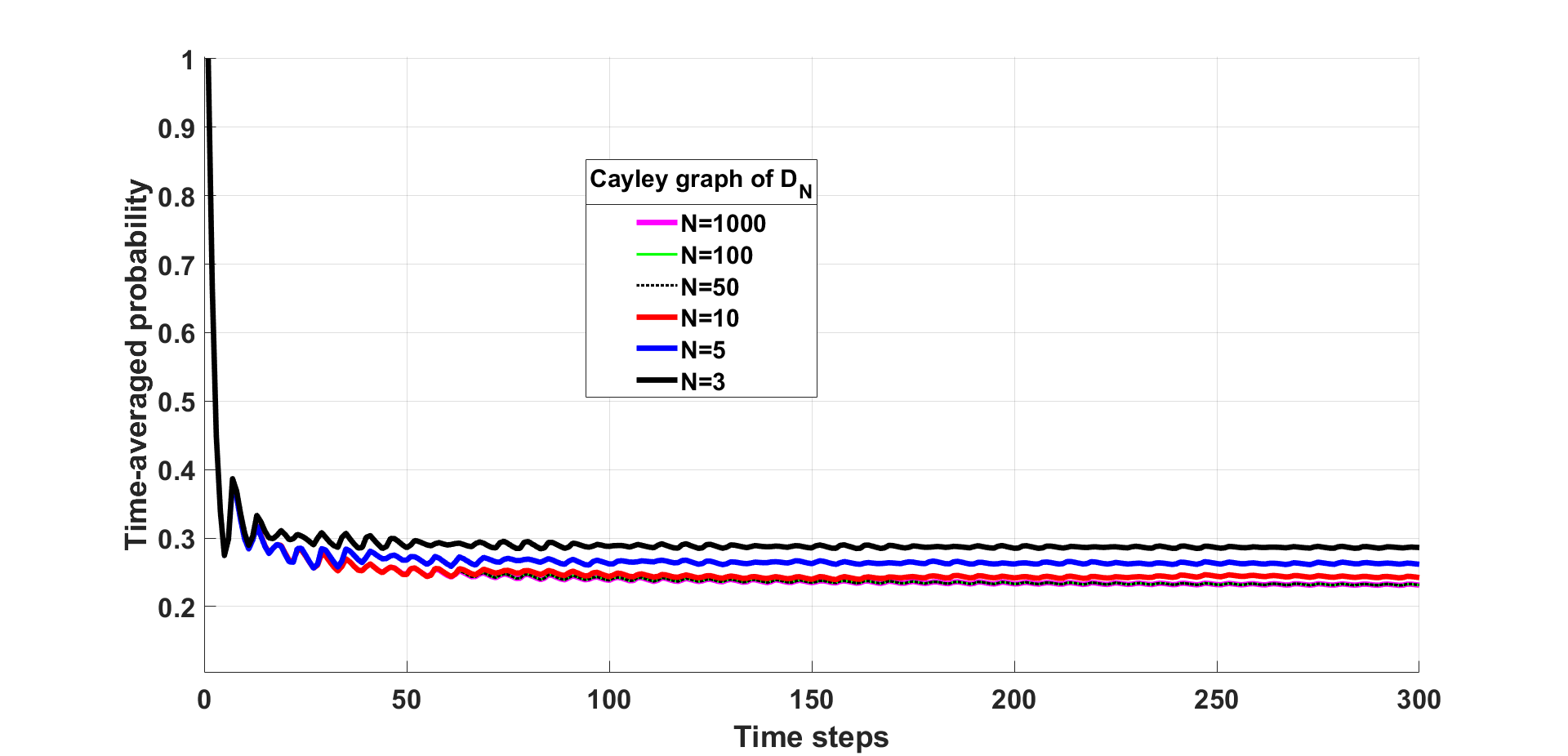}}
    \subfigure[$C\in \mathcal{Z}_\theta,\theta=\pi$]{\includegraphics[height=3.5 cm,width=8 cm]{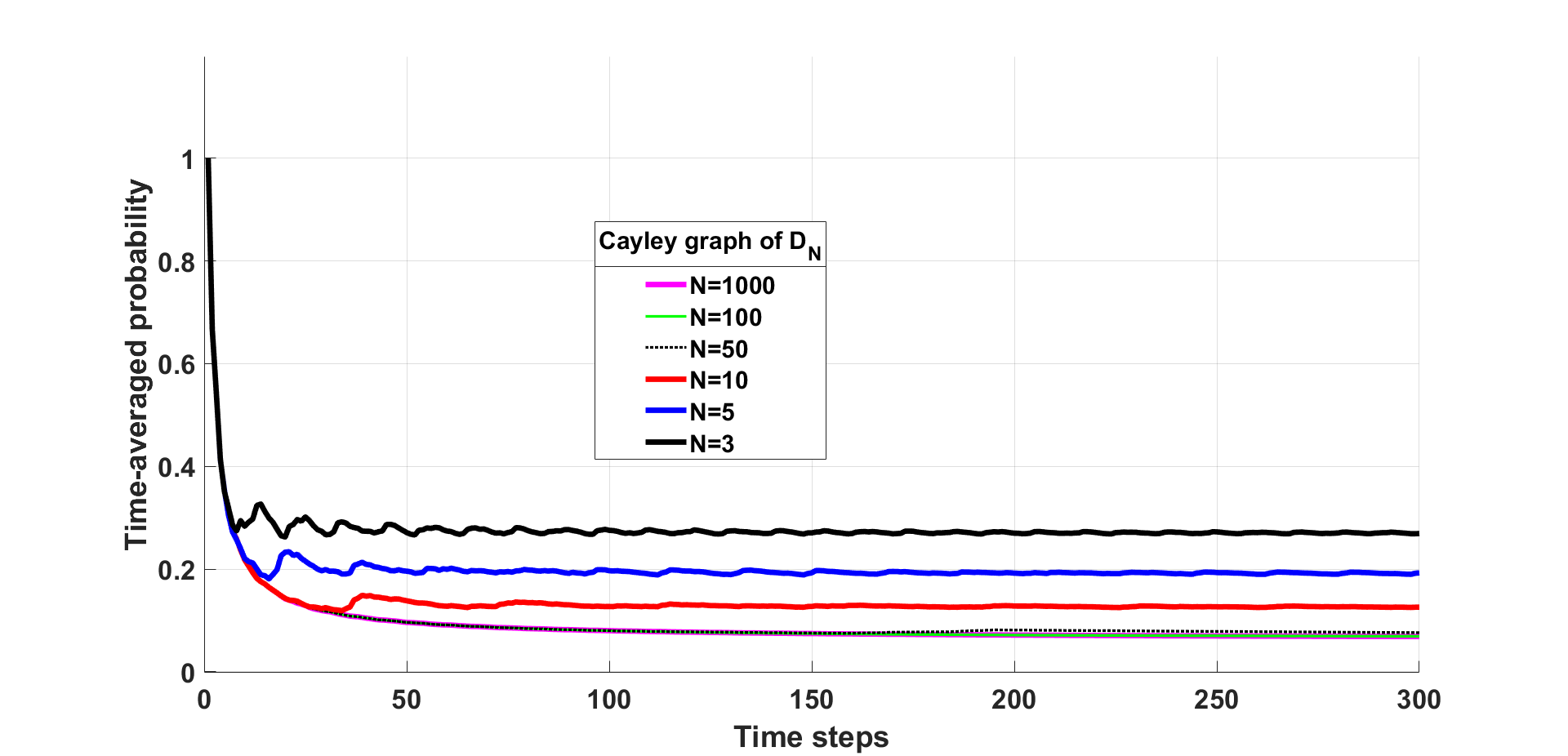}}
    \subfigure[$C\in \mathcal{W}_\theta,\theta=\pi/4$]{\includegraphics[height=3.5 cm,width=8 cm]{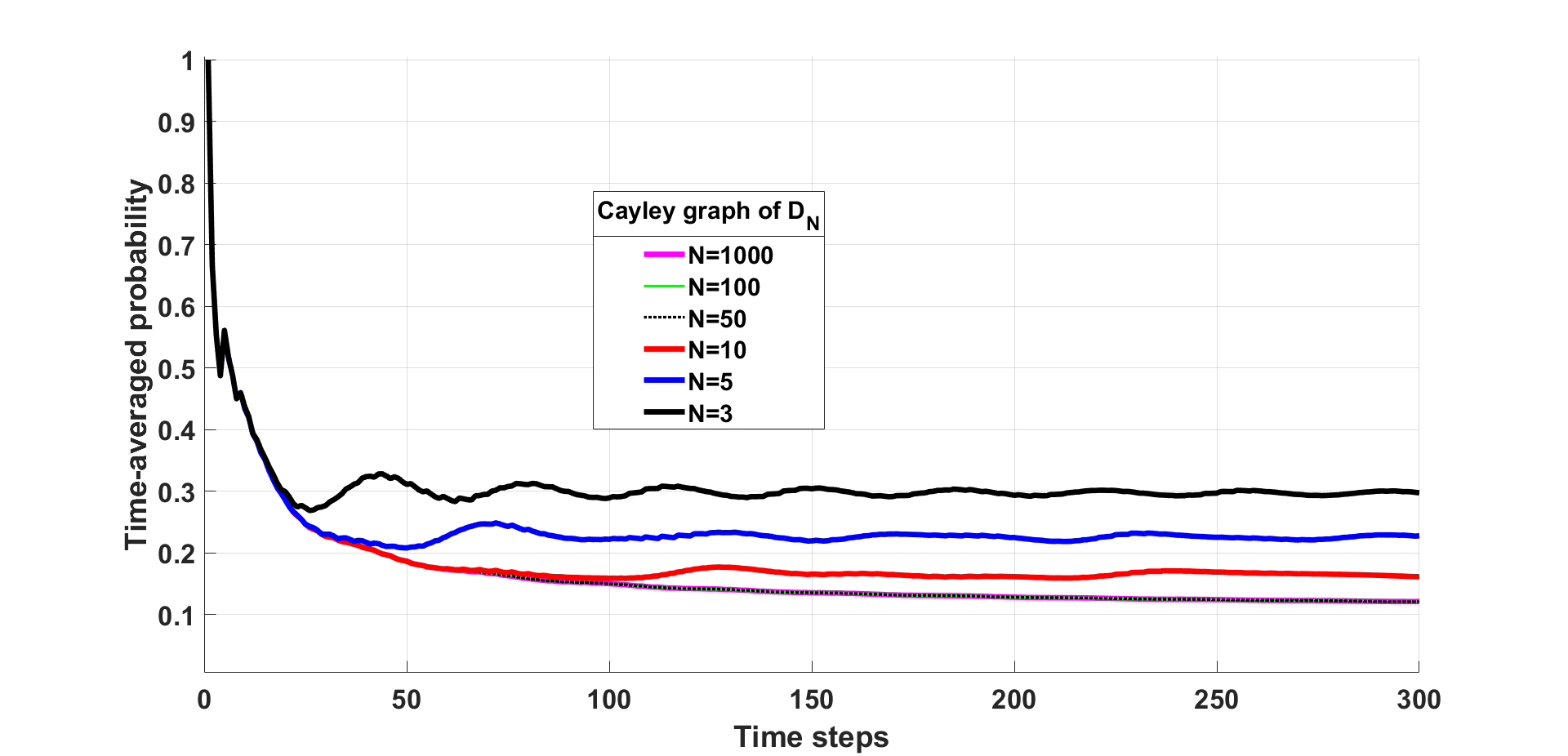}}
    \caption{ Time averaged probability for $\mathrm{Cay(D_N,\{a,b\})}$ taking coins from the classes $,\mathcal{X}_\theta,\mathcal{Y}_\theta,\mathcal{Z}_\theta,\mathcal{W}_\theta$ with initial position $(1,0)$ and initial coin state $\frac{1}{\sqrt{3}}(\ket{0}_3+\ket{1}_3+\ket{2}_3)$. The time step is taken up to $300$.  }\label{fig:NvariesX180}
\end{figure}

 %In fact for $\mathcal{Z}_\theta,\mathcal{W}_\theta$, we have seen that the walk does not localize at the reflection vertex of the starting vertex unlike the the coin classes $\mathcal{X}_\theta,\mathcal{Y}_\theta$. It is interesting to see that for Grover coins. the walk localizes at the starting vertex and its reflection unlike these coin classes that does not contain the Grover matrix.\par We shall also observe how the time-averaged probability at the starting vertex depends on the size of the underlying Cayley graph in the quantum walk.
In Figure \ref{fig:NvariesX180}, we observe the time-averaged probability of finding a particle at the starting vertex $(1,0)$ for $\mathrm{Cay(D_N,\{a,b\})}$ where $N$ varies with the initial coin state $\frac{1}{\sqrt{3}}(\ket{0}_3+\ket{1}_3+\ket{2}_3)$. The coins have been taken from the class $\mathcal{X}_\theta$ where $\theta={\pi}$ i.e. $C=\mathsf{G}$ and $\theta=\pi/2$ i.e. $\sqrt{\mathsf{G}}$. For coins in $\mathcal{Y}_\theta$, we have taken two non-trivial coins viz. $\theta=\pi/5$,$\theta=-\pi/4$. For coins belonging to $\mathcal{Z}_\theta$, $\theta=\pi$ i.e. a permutation of $\mathsf{G}$ and for coins from $\mathcal{W}_\theta$, $\theta=\pi/4$ is chosen as a non-trivial example. The time steps are taken up to $300$ and we see how the time-averaged probability converges. Note that the choice of the coin is arbitrary since similar phenomenon is observed for other coins as well.

%It is to be noticed that existence of degenerate eigenvalues like $1$ or $-1$ is a necessary condition for localization of the quantum walk\cite{Inui2004} and for our classes of coins, such degenerate eigenvalues exist. For permutation matrices however, it is observed that the time-averaged probabilities achieve local maxima or minima for canonical basis states. This may be attributed to the fact that the permutation coin matrix just changes one basis state to another rather than producing a linear superposition of all canonical basis states like other non-permutation operators.

\noindent{\bf Conclusion}
We have studied the periodicity and localization properties of discrete-time quantum walks on Cayley graphs corresponding to Dihedral groups with the coin operators that are (real) linear sum of permutation matrices of order $3$. It is surprising to observe that the walk does not exhibit periodicity for Grover coin, in contrast to several walks on different graphs as observed in literature. Indeed, the walks are periodic only for coins of the form $\pm P,$ where $P$ denotes a permutation matrix. Finally, we perform an extensive numerical study for analyzing localization of these walks and establish that the walks localize for several coins for a variety of Cayley graphs of different sizes. In particular, the time-averaged probability maximizes or minimizes for permutation coins when the initial coin state is one of the canonical coin states. We also notice an inverse correlation with the time-averaged probability of the quantum walk at the starting vertex and the size of the graph i.e. for smaller sized graphs, the time-averaged probability for finding the walker at the initial vertex was found to be greater than that of the larger sized graphs. \\

% In this paper, we have generalized the notion of discrete-time quantum walks on dihedral groups by using quantum coins as a linear sum of permutation matrices which are in turn also permutative (all rows are permutations of the first row). We analyize the periodicity of the walk for several classes of such coin operators viz. $\mathcal{X},\mathcal{Y},\mathcal{Z},\mathcal{W}$ and we have shown that the quantum walk remains aperiodic unless the quantum coin belongs to the class of permutations. Further, we have investigated the localization property of the walks for varying sizes of the underlying graph along with different initial states and we have seen for several quantum coins that the walk localizes at the starting vertex. Though, deriving an analytical expression for long-time limit in such a walk is cumbersome, we theorize that the walk mainly localizes due to the presence of degenerate eigenvalues $\pm 1$. We also notice an inverse correlation with the time-averaged probability of the quantum walk at the starting vertex and the size of the graph meaning for smaller sized graphs, the time-averaged probability for finding the walker at the initial vertex was greater than that of larger sized graphs. In future, the authors aim to provide a quantum circuit for this DTQW walk model in order to validate the results derived here.  

\noindent{\bf Acknowledgement}
Rohit Sarma Sarkar acknowledges support through Prime Minister's Research Fellowship (PMRF), Government of India.

\end{document}